\documentclass[trackchanges, twocolumn]{aastex701}

\newcommand{\pkg}[1]{\texttt{#1}}

\usepackage{chngcntr}
\usepackage{amssymb}
\usepackage{amsmath}
\usepackage{float}
\usepackage{hyperref}
\usepackage{subfigure}
\usepackage{rotating}
\usepackage{xcolor}
\hypersetup{
    colorlinks,
    linkcolor={blue!50!black},
    citecolor={blue!50!black},
    urlcolor={blue!80!black}
}

\usepackage{graphicx}
\usepackage{verbatim}
\usepackage{placeins}
\usepackage{multirow}
\usepackage{tabularx}
\usepackage{url}
\usepackage{enumitem}

\shortauthors{Khadir et al.}

\begin{document}
\title{Search for a Dispersion Measure Excess Toward Nearby Galaxy Halos Using Fast Radio Bursts from the Second CHIME/FRB Baseband Catalog}

\author[orcid=0009-0001-2196-8251
,sname='Khadir']{Affan Khadir}
\affiliation{Department of Physics, McGill University, 3600 rue University, Montréal, QC H3A 2T8, Canada}
\affiliation{Trottier Space Institute, 3550 rue University, Montréal, QC H3A 2A7, Canada}
\email[show]{affan.khadir@mail.mcgill.ca}  

\author[orcid=0000-0002-8897-1973, sname='Pandhi']{Ayush Pandhi} 
\affiliation{Department of Physics, McGill University, 3600 rue University, Montréal, QC H3A 2T8, Canada}
\affiliation{Trottier Space Institute, 3550 rue University, Montréal, QC H3A 2A7, Canada}
\email{ayush.pandhi@mcgill.ca}

\author[orcid=0000-0003-2317-1446, gname=Jason,sname=Hessels]{Jason Hessels}
\affiliation{Department of Physics, McGill University, 3600 rue University, Montréal, QC H3A 2T8, Canada}
\affiliation{Trottier Space Institute, 3550 rue University, Montréal, QC H3A 2A7, Canada}
\affiliation{Anton Pannekoek Institute for Astronomy, University of Amsterdam, Science Park 904, 1098 XH Amsterdam, The Netherlands}
\affiliation{ASTRON, Netherlands Institute for Radio Astronomy, Oude Hoogeveensedijk 4, 7991 PD Dwingeloo, The Netherlands}
\email{jason.hessels@mcgill.ca}

\author[orcid=0000-0002-4209-7408]{Calvin Leung}
\affiliation{Miller Institute for Basic Research, University of California, Berkeley, CA 94720, United States}
\affiliation{Department of Astronomy, University of California, Berkeley, CA 94720, United States}
\email{calvin_leung@berkeley.edu}

\author[orcid=0000-0002-3382-9558]{B. M. Gaensler}
\affiliation{Department of Astronomy and Astrophysics, University of California, Santa Cruz, 1156 High Street, Santa Cruz, CA 95060, USA}
\affiliation{Dunlap Institute for Astronomy and Astrophysics, 50 St. George Street, University of Toronto, ON M5S 3H4, Canada}
\affiliation{David A. Dunlap Department of Astronomy and Astrophysics, 50 St. George Street, University of Toronto, ON M5S 3H4, Canada}
\email{gaensler@ucsc.edu}

\author[0000-0003-4690-2774]{Ludwig M. Böss}
\email{lboess@uchicago.edu}
\affiliation{Department of Astronomy and Astrophysics, The University of Chicago, William Eckhart Research Center, 5640 S. Ellis Ave. Chicago, IL 60637, USA}

\author[orcid=0000-0001-6422-8125 ]{Amanda M. Cook}
\email{amanda.cook@mail.mcgill.ca}
\affiliation{Department of Physics, McGill University, 3600 rue University, Montréal, QC H3A 2T8, Canada}
\affiliation{Trottier Space Institute, 3550 rue University, Montréal, QC H3A 2A7, Canada}
\affiliation{Anton Pannekoek Institute for Astronomy, University of Amsterdam, Science Park 904, 1098 XH Amsterdam, The Netherlands}
\email{alice.curtin@mcgill.ca}

\author[orcid=0000-0002-8376-1563]{Alice P. Curtin}
\affiliation{Department of Physics, McGill University, 3600 rue University, Montréal, QC H3A 2T8, Canada}
\affiliation{Trottier Space Institute, 3550 rue University, Montréal, QC H3A 2A7, Canada}
\affiliation{Anton Pannekoek Institute for Astronomy, University of Amsterdam, Science Park 904, 1098 XH Amsterdam, The Netherlands}
\email{alice.curtin@mcgill.ca}  

\author[]{Klaus Dolag}
\email{dolag@usm.lmu.de}
\affiliation{Universitäts-Sternwarte, Fakultät für Physik, Ludwig-Maximilians-Universität München, Scheinerstr.1, 81679 München, Germany}
\affiliation{Max-Planck-Institut für Astrophysik, Karl-Schwarzschild-Straße 1, 85741 Garching, Germany}

\author[orcid=0009-0009-0938-1595]{Naman Jain}
\affiliation{Department of Physics, McGill University, 3600 rue University, Montréal, QC H3A 2T8, Canada}
\affiliation{Trottier Space Institute, 3550 rue University, Montréal, QC H3A 2A7, Canada}
\email{naman.jain@mail.mcgill.ca}  

\author[orcid=0000-0001-9345-0307]{Victoria Kaspi}
\affiliation{Department of Physics, McGill University, 3600 rue University, Montréal, QC H3A 2T8, Canada}
\affiliation{Trottier Space Institute, 3550 rue University, Montréal, QC H3A 2A7, Canada}
\email{victoria.kaspi@mcgill.ca}

\author[orcid=0009-0004-4176-0062]{Afrokk Khan}
\affiliation{Department of Physics, McGill University, 3600 rue University, Montréal, QC H3A 2T8, Canada}
\affiliation{Trottier Space Institute, 3550 rue University, Montréal, QC H3A 2A7, Canada}
\email{afrasiyab.khan@mcgill.ca}

\author[0000-0003-2116-3573]{Adam E. Lanman}
\email{alanman@mit.edu}
\affiliation{Department of Physics, Massachusetts Institute of Technology, 77 Massachusetts Ave, Cambridge, MA 02139, USA}
\affiliation{MIT Kavli Institute for Astrophysics and Space Research, Massachusetts Institute of Technology, 77 Massachusetts Ave, Cambridge, MA 02139, USA}

\author[orcid=0000-0002-4279-6946]{Kiyoshi W. Masui}
\email{kmasui@mit.edu}
\affiliation{MIT Kavli Institute for Astrophysics and Space Research, Massachusetts Institute of Technology, 77 Massachusetts Ave, Cambridge, MA 02139, USA}
\affiliation{Department of Physics, Massachusetts Institute of Technology, 77 Massachusetts Ave, Cambridge, MA 02139, USA}

\author[orcid=0000-0002-2551-7554]{Daniele Michilli}
\affiliation{Laboratoire d'Astrophysique de Marseille, Aix-Marseille Univ., CNRS, CNES, Marseille, France}
\email{danielemichilli@gmail.com}

\author[0000-0002-0940-6563]{Mason Ng}
\affiliation{Department of Physics, McGill University, 3600 rue University, Montréal, QC H3A 2T8, Canada}
\affiliation{Trottier Space Institute, 3550 rue University, Montréal, QC H3A 2A7, Canada}
\email{mason.ng@mcgill.ca}  

\author[0000-0002-8912-0732]{Aaron B. Pearlman}
\altaffiliation{NASA Hubble Fellow}
\affiliation{MIT Kavli Institute for Astrophysics and Space Research, Massachusetts Institute of Technology, 77 Massachusetts Ave, Cambridge, MA 02139, USA}
\affiliation{Department of Physics, McGill University, 3600 rue University, Montréal, QC H3A 2T8, Canada}
\affiliation{Trottier Space Institute, 3550 rue University, Montréal, QC H3A 2A7, Canada}
\email{aaron.b.pearlman@mit.edu}

\author[orcid=0009-0008-2000-6959]{Sachin Pradeep E.~T. }
\affiliation{Department of Physics, McGill University, 3600 rue University, Montréal, QC H3A 2T8, Canada}
\affiliation{Trottier Space Institute, 3550 rue University, Montréal, QC H3A 2A7, Canada}
\email{sachin.pradeepetakkepravanthulicheri@mail.mcgill.ca}

\author[orcid=0000-0002-8823-8835]{Aniket Prasad}
\affiliation{Laboratoire d'Astrophysique de Marseille, Aix-Marseille Univ., CNRS, CNES, Marseille, France}
\email{aniket.prasad@lam.fr}

\author[0000-0002-4623-5329]{Mawson Sammons}
\email{mawson.sammons@lam.fr}
\affiliation{Laboratoire d'Astrophysique de Marseille, Aix-Marseille Univ., CNRS, CNES, Marseille, France}

\author[orcid=0000-0002-7374-7119]{Paul Scholz}
\email{pscholz@yorku.ca}
\affiliation{Department of Physics and Astronomy, York University, 4700 Keele Street, Toronto, ON MJ3 1P3, Canada}

\author[orcid=0000-0002-6823-2073]{Kaitlyn Shin}
\email{kaitshin@caltech.edu}
\affiliation{Division of Physics, Mathematics, and Astronomy, California Institute of Technology, Pasadena, CA 91125, USA}

\author[orcid=0000-0002-2088-3125]{Kendrick M. Smith}
\email{kmsmith@perimeterinstitute.ca}
\affiliation{Perimeter Institute for Theoretical Physics, 31 Caroline Street N, Waterloo, ON N25 2YL, Canada}

\author[orcid=0000-0002-2307-2432]{Jenny G. Sorce}
\email{jenny.sorce@univ-lille.fr}
\affiliation{Univ. Lille, CNRS, Centrale Lille, UMR 9189 CRIStAL, F-59000 Lille, France}
\affiliation{Universit\'e Paris-Saclay, CNRS, Institut d'Astrophysique Spatiale, 91405, Orsay, France}

\author[orcid=0000-0002-1491-3738]{Haochen Wang}
\email{hcwang96@mit.edu}
\affiliation{Department of Physics, Massachusetts Institute of Technology, 77 Massachusetts Ave, Cambridge, MA 02139, USA}
\affiliation{MIT Kavli Institute for Astrophysics and Space Research, Massachusetts Institute of Technology, 77 Massachusetts Ave, Cambridge, MA 02139, USA}

\author[orcid=0009-0001-3334-9482]{Michele Woodland}
\email{miwoodla@ucsc.edu}
\affiliation{Department of Astronomy and Astrophysics, University of California, Santa Cruz, 1156 High Street, Santa Cruz, CA 95060, USA}

\begin{abstract}
Fast radio bursts (FRBs) probe foreground ionized baryons via their dispersion measures (DM). Using 1748 sub-arcminute-localized sources from the Second CHIME/FRB Baseband Catalog, we conduct a statistical stacking study of the circumgalactic media (CGM) of nearby galaxies by analyzing the DMs of FRBs whose sightlines intersect foreground halos. For galaxy halos within 50\,Mpc, we measure a mean DM excess of $119\pm44$\,pc~cm$^{-3}$ ($p = 0.0053; \sim 2.6\sigma$) within one virial radius. Expanding to 150\,Mpc (628 intersections), we find a tentative decline in the mean DM as a function of impact parameter ($p = 0.0046; \sim 2.6\sigma$), which is predominantly driven by typical Milky Way (MW)-like galaxies ($10 \leq \log(M_*/M_\odot)\leq 11$). We use this DM profile to constrain parameters of the CGM electron density distribution, $n_e^\mathrm{CGM}(r) \propto r^{-3\beta}$, finding an extrapolated central electron density in the range of $(1.7-21.1) \times 10^{-3}\mathrm{\,cm}^{-3}$, with a radial power-law index of $\beta = 0.68^{+0.15}_{-0.16}$, consistent with constraints from X-ray stacking studies of MW-like galaxies. At very low impact parameters, we detect a significant DM deficit likely due to cross-matching FRBs that originate from galaxies within our galaxy sample. By deriving the DM-galaxy cross-correlation function, we show that this deficit matches theoretical predictions if 20 FRBs originate from galaxies within 150 Mpc, indicating that CHIME/FRB detects a higher fraction of low-$z$ FRBs than previously thought. We forecast that $10^4$ FRBs with similar localization errors will provide a 5-$\sigma$ detection of the CGM, and that FRBs at intermediate redshifts ($0.4\lesssim z_\mathrm{FRB}\lesssim 0.75$) are ideal for studying intervening CGMs.

\end{abstract}

\keywords{\uat{Radio bursts}{1339} --- \uat{Circumgalactic medium}{1879} --- \uat{Galaxies}{573} --- \uat{Astrostatistics}{1882}}


\section{Introduction} 


Fast radio bursts \citep[FRBs;][]{2007Sci...318..777L} are incredibly useful probes of intervening plasma as they experience a variety of propagation effects whilst they traverse through astrophysical plasma. As FRBs propagate through plasma, they undergo a dispersive delay \citep[][]{1968Natur.217..709H} such that lower frequencies arrive at a later time, and their transient nature allows us measure this dispersive delay. This time delay is proportional to the dispersion measure (DM), which is given by the integral of the free electron density along the observer's line of sight (LOS):
\begin{align}
    \mathrm{DM} = \int_0^{z} \frac{n_e(z')}{\mathrm{cm^{-3}}}\frac{1}{1+z'}\frac{dl}{dz'\mathrm{pc}}dz', 
    \label{eq:dm}
\end{align}
where $n_e$ is the free electron density in cm$^{-3}$ and $z$ is the redshift of the source. Since $n_e$ traces the density of plasma, an FRB's DM informs us about the total ionized plasma content between the observer and the source. Because the vast majority of the baryonic matter in the Universe is thought to exist as ionized plasma \citep[e.g.,][]{1998ApJ...503..518F, 1999ApJ...514....1C}, and since FRBs are the only known sources originating outside the Local Group with DMs, they serve as unique probes of the baryonic distribution of the Universe. {In addition to dispersion, FRBs also undergo other propagation effects, such as Faraday rotation \citep[e.g.,][]{2023ApJ...957L...8S, 2024ApJ...968...50P}, and scattering and scintillation \citep{2025A&A...700A..99P}, which can inform us about the magnetism \citep[e.g.,][]{2016ApJ...824..105A} and turbulence \citep[e.g.,][]{2020ApJ...898L..48X} of plasma environments in the Universe, respectively. }

{Thus, FRBs are capable of probing low-density plasma ($n_e\sim10^{-6}-10^{-5}$\,cm$^{-3}$) in the intergalactic medium \citep[IGM; e.g.,][]{2007ApJ...655..831T}, which is unconstrained by other observational techniques like X-rays or absorption lines \citep[][]{2012ApJ...759...23S}.  \citet[][]{2020Natur.581..391M} found that the cosmic component of the DM was consistent with the prediction from combining CMB-derived cosmological parameters \citep[e.g.,][]{2016A&A...594A..13P} and the expected diffuse baryon fraction from \citet{2004ApJ...616..643F} using 8 localized FRBs. Thus, FRB DMs completed the baryon census and confirmed the existence of the so-called ``missing baryons''. FRB DMs are now being used to partition the missing baryons along different components of the cosmic web \citep[e.g., diffuse IGM versus galaxy halos;][]{2020ApJ...901..134S, 2022ApJ...928....9L, 2025NatAs...9.1226C, 2024ApJ...973..151K, 2026PhRvL.137d1001W}. }



Particularly, FRB DMs are a good tool to study the circumgalactic medium (CGM), which is a reservoir for the diffuse baryons in the cosmic web. The CGM is the gas surrounding galaxies, the vast majority ($\gtrsim 99\%$) of which is ionized \citep[][]{2014ApJ...792....8W}. The properties of this gas are driven by processes such as cooling and accretion \citep[][]{2025A&A...697A.121B}, and heating by active galactic nucleus (AGN) feedback \citep[][]{2015MNRAS.448..895S}.  Thus, constraining the distribution of baryons in the CGM (such as those traced through free electrons) is an important factor in understanding feedback processes and galaxy evolution. 
The CGM of galaxies is thought to consist of the cool phase \citep[$T\sim10^4$\,K; ][]{2016ApJS..226...25L}, the warm-hot phase \citep[$10^4\, \mathrm{K}\lesssim T \lesssim 10^5\,\mathrm{K}$; ][]{2024A&A...681A..78L}, and the hot phase \citep[$T \sim 10^6$\,K; ][]{2019ApJ...882L..23D}. For more detailed reviews of the CGM, we refer the reader to \citet[][]{2017ARA&A..55..389T, 2026enap....4..370C} and the references therein. 

The hot phase occupies the largest volume and mass fraction \citep[$\gtrsim99\%$; e.g.,][]{2019MNRAS.490.3234N, 2020MNRAS.491.5056L} of the CGM; despite this, it remains poorly constrained due to its low density \citep[e.g.,][]{2017ApJ...835...52F}, which makes detecting it in X-ray emission difficult. Because of this, X-ray studies of the CGM of single galaxies have been limited to the CGM of the Milky Way \citep[MW; e.g.,][]{2023ApJ...952...41B, 2024A&A...681A..78L}. To address this low X-ray photon count problem with single sources, statistical studies combining the emission of multiple galaxies as a function of virial radius have been recently conducted to constrain the baryon distribution in the inner CGM $\lesssim 100$\,kpc \citep[e.g.,][]{2024A&A...690A.267Z}. However, due to sensitivity limits, X-ray observations are unable to extend effectively to $\sim 100$\,kpc due to the low density of plasma at these radii. Additionally, while X-ray emission observations capture light along the entire LOS, their signal is heavily biased toward the dense, inner regions of the halo due to their $n_e^2$ dependence. In contrast, FRB DMs scale linearly with $n_e$, making them uniquely sensitive to the diffuse, extended gas in the outer CGM. Thus, the distribution of baryons in the hot phase remains poorly understood and can be constrained with FRB DMs.

There have been a few studies of single halos with FRBs, many of which have focused on the MW and M31 halos \citep[e.g.,][]{2019Sci...366..231P, 2023ApJ...946...58C, 2025ApJ...991L..25L, 2025ApJ...993..221A, 2026arXiv260216781M, 2026arXiv260223749K}. However, the studies constraining extragalactic halos are mostly limited in statistical power due to the low number of sources ($\sim10^1$ FRBs). \citet[][]{2014ApJ...780L..33M} first suggested improving the signal-to-noise (S/N) and constraining power in FRB studies of foreground galaxy halos by stacking FRB DMs as a function of impact parameter to foreground galaxies.  In an early work, \citet[][]{2022NatAs...6.1035C} conducted such a stacked study of FRB DMs as a function of impact parameter for very nearby galaxies with luminosity distances ($D_L$) less than 40\,Mpc using 474 FRBs from the CHIME/FRB Intensity Catalog~1 \citep[][]{2021ApJS..257...59C}. They found 26 intersections of FRBs within a projected radial distance of 200\,kpc of foreground halos, with the intersecting population having a mean excess DM of $>90$\,pc~cm$^{-3}$ compared to the non-intersecting population at the 95\% confidence level. However, their results are limited by the FRB positional uncertainty {($>$ hundreds of arcseconds)} of their data points, along with the low number of intersections. {\citet[][]{2023ApJ...945...87W} expanded on the work of \citet[][]{2022NatAs...6.1035C} by stacking FRBs from CHIME/FRB out to 80 Mpc for galaxies with halo masses of $11 \leq \log (M_\mathrm{halo}/M_\odot)\leq 13$, and introducing an exponential down-weighting for high-DM FRBs to mitigate the impacts of high-DM outliers, and find a marginal detection at the $1-2~\sigma$ level}. 

The FRB studies of CGMs described so far have faced at least one of two major hurdles: poor localizations and a low number of FRBs. The poor localizations make it hard to confidently determine if an FRB is intersecting foreground halos, which have typical angular sizes of $50'' - 35'$ (assuming a $\sim 100$\,kpc radius at distances of $10-500$\,Mpc). For FRBs that are intersecting an intervening halo, it becomes challenging to determine an accurate impact parameter. The low number of FRBs also introduces statistical noise into the measured CGM contribution due to variance in local host contributions and in the electron distribution across the IGM, which can significantly decrease the significance and precision of measurements. To overcome these hurdles, a large catalog of precisely localized FRBs is needed. 

 The Canadian Hydrogen Intensity Mapping Experiment/Fast Radio Burst project \citep[CHIME/FRB;][]{2018ApJ...863...48C} has found the vast majority of FRB sources, discovering over 3500 unique FRB sources to date \citep[][]{2026ApJS..283...34C} due to its wide field of view, which scans the entire northern sky every 24 hours. For high S/N bursts, CHIME/FRB saves the raw channelized voltage (``baseband'') data from all of CHIME's antennas, allowing us to achieve smaller localization errors of $<1'$ \citep[][]{2021ApJ...910..147M, 2024ApJ...969..145C}; see Section \ref{sec:baseband_data}. In this paper, we use the CHIME/FRB Baseband Catalog~2 (CHIME/FRB Collaboration et al. in prep.), which has over 1600 unique FRBs {with typical localization errors on the order of tens of arcseconds, to conduct the largest statistical stacking study of free electrons in foreground galaxy CGMs to date (by sample size of localized FRBs). In particular, we search for a mean DM excess in FRBs intersecting nearby ($D_L\leq 50$\,Mpc) massive ($10\leq\log(M_*/M_\odot)\leq 11.5$) galaxy halos. We also attempt to radially map the distribution of baryons in the CGM by cross-matching FRB positions and DMs with galaxy halos in mass bins of 0.5 dex (covering a mass range of $9.5\leq \log(M_*/M_\odot)\leq 11.5$) out to a luminosity distance of 150\,Mpc. Thus, we obtain some of the first statistical constraints on the mass and density distribution of the hot gas in the inner CGM using FRBs, where processes such as baryonic feedback are crucial and remain relatively unconstrained. }   

This paper is structured as follows: Section~\ref{sec:methods} describes the galaxy catalogs and the FRB catalog used, as well as the methods carried out to prepare these catalogs for cross-correlation; Section~\ref{sec:results} presents the results of cross-correlating FRB positions with galaxy positions; and Section~\ref{sec:discussion} provides discussion and interpretation on the implications of our cross-correlation results for the distribution of baryons in galaxy halos. Throughout this study, we do not consider potential host galaxy halos to be intervening. We also assume the Planck2018 cosmological parameters \citep[][]{2020A&A...641A...6P}: $H_0 = 67.4\,\mathrm{km~s^{-1}~Mpc^{-1}},~\Omega_{m, 0} = 0.31,~\Omega_{\Lambda, 0} = 0.69$.   

\section{Methods}
\label{sec:methods}
{Throughout our analysis, we use FRBs from the CHIME/FRB Baseband Catalog 2. To search for excess DM in FRBs intersecting galaxy halos, we cross-match FRB positions with that of galaxies from the 50~Mpc Galaxy Catalog \citep[50MGC;][]{2024AJ....167...31O}. To radially map the distribution of baryons in CGMs, we employ the REGALADE \citep[][]{2026A&A...706A.284T} galaxy catalog. Below, we describe these data and the pre-processing steps to prepare these catalogs for our analysis.}

\subsection{CHIME/FRB Baseband Catalog~2}
\label{sec:baseband_data}
The first CHIME/FRB Baseband Catalog \citep{2024ApJ...969..145C} provided baseband data for 140 FRBs, along with higher time resolution compared to the intensity data, for some of the 536 bursts from the CHIME/FRB Intensity Catalog~1 \citep{2021ApJS..257...59C}. With baseband data, we can form offline beams at higher angular scales, which in turn enables further refinement of the FRB position \citep[$\sim 500\times$ smaller sky area than intensity data uncertainties;][]{2021ApJ...910..147M}. The average positional uncertainty of an FRB in this catalog is at the sub-arcminute level. Since the baseband data has a higher time resolution (2.56\,$\mu$s) it also enables improved DM measurements for the FRB sources. This catalog has now been updated with an order of magnitude more bursts in the second CHIME/FRB Baseband Catalog (``BaseCat2", hereafter) that consists of 2086 bursts, 1748 of which are unique sources (\citeauthor{2026arXiv260508410C} \citeyear{2026arXiv260508410C}{; }CHIME/FRB Collaboration et al. 2026, in prep.) detected between 9 December 2018 and 15 September 2023. All of these bursts have a S/N ratio greater than 12 (S/N $>10$ for subsequent bursts of known repeaters). 

\begin{figure}[!htb]
    \centering
    \includegraphics[width=0.47\textwidth]{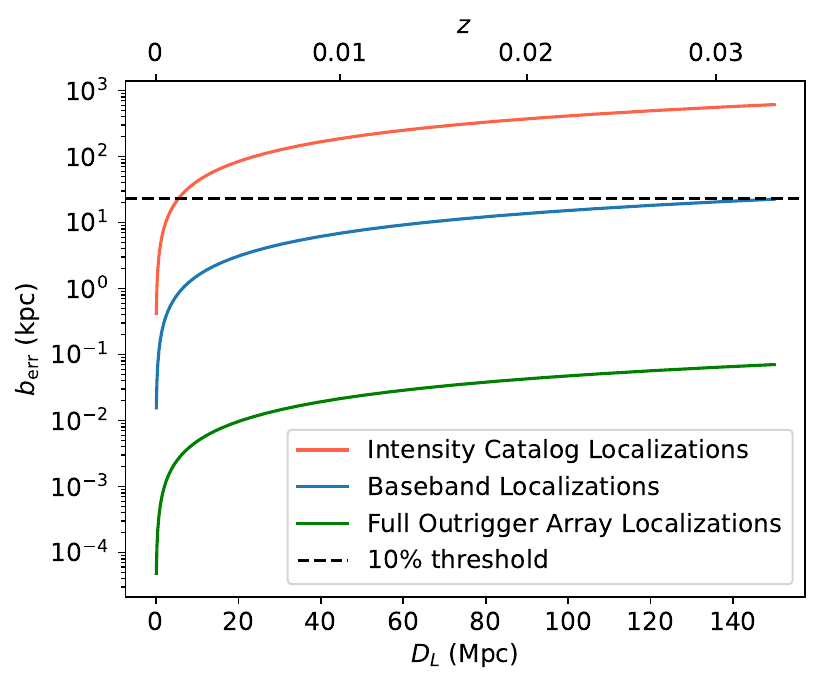}
    \caption{The median uncertainty on foreground galaxy impact parameter for the different types of localizations that can be obtained from the CHIME/FRB instrument, as a function of luminosity distance to the foreground galaxy. The localization uncertainties for the intensity ($\theta_{\mathrm{err}}^{\mathrm{med}} = 0.19^\circ$), baseband ($\theta_\mathrm{err}^\mathrm{med} = 32^{\prime\prime}$), and full Outrigger array ($\theta_\mathrm{err}^\mathrm{med} = 100$\, mas) localizations are presented in orange, blue and green, respectively. The dotted line represents the 10\% error threshold for typical MW-like galaxies.}
    \label{fig:1}
\end{figure}

Because of the high S/N of the bursts, we obtain a relatively small median $1\sigma$ uncertainty on the localization of bursts in BaseCat2 of $32^{\prime\prime}$. We consider a measurement of the impact parameter ($b$) to intervening galaxies along the LOS to be reliable if the fractional error $b/r_\mathrm{vir} \lesssim 10\%$, where $r_\mathrm{vir}$ is the virial radius. Thus, thanks to the high localization accuracy, which enables low impact parameter uncertainties, we can reliably measure $b$ out to $D_L\sim 150$\,Mpc; at this distance, the median angular localization error corresponds to an uncertainty in transverse scale of 23\,kpc, which is on the order of $\sim 10\%$ of $r_\mathrm{vir}$ for Milky Way-sized galaxies \citep{2019A&A...621A..56P}. In Figure~\ref{fig:1} we show the median error on the derived FRB impact parameter to intervening galaxies at redshift $z$ for typical CHIME/FRB sources with (i) only intensity catalog localizations, (ii) baseband localizations, and (iii) full CHIME/FRB Outrigger localizations. Currently, both the intensity data and the Outrigger are non-ideal for a statistical stacking study of the CGM; the intensity data suffers from poor localizations, and the Outrigger data suffers from a low number of bursts. Thus, currently, baseband data is perfect for CGM studies, as it has sufficiently good localizations and also a large number of FRBs.

Of the 2086 total bursts in BaseCat2, there are 411 that originate from 73 sources that produce repeat bursts \citep[][]{2019ApJ...885L..24C, 2023ApJ...947...83C, 2026arXiv260508410C}. For repeaters we use the localization from the burst with the smallest uncertainty in CHIME/FRB baseband data or, where available, position uncertainties measured with Very Long Baseline Interferometry \citep[i.e., for FRBs~20180916B, 20201124A, 20200120E, 20220912A, 20190208A, 20190417A, 20220529A;][]{2020Natur.577..190M, 2022ApJ...927L...3N, 2022Natur.602..585K, 2024MNRAS.529.1814H, 2024ApJ...977L...4H, 2026ApJ...996L..16M, 2026Sci...391..280L}, many of which are localized as part of the PRECISE project \citep[][]{2022evlb.confE..35M}.  We exclude FRB~20200120E from our analysis, as it is located in a globular cluster in M81 and there are no intervening galaxy halos between us and this source \citep[][]{2022Natur.602..585K}.  We also remove 26 FRBs that are thought to be either embedded or background to galaxy clusters, based on the results of \citet[][]{2026arXiv260810358S}, as DM contributions from the hot cluster gas can dominate over any CGM DM contributions; for further discussion of this, see Section~\ref{sec:environment}.

{For all the bursts in BaseCat2}, we use the DMs obtained by maximizing the coherent power using the  \pkg{DM\_phase}\footnote{\url{https://github.com/danielemichilli/DM_phase}} algorithm \citep[][]{2019ascl.soft10004S}. There were a few events for which issues were faced with running \pkg{DM\_phase} due to either a low S/N in the beamformed baseband data or due to  \pkg{DM\_phase} not converging. In these cases, we use DMs obtained by incoherently maximizing the S/N across the full burst. We note that the choice of either DM will not significantly affect the results of our analysis as, on average, the two DMs differ by $< 5$\,pc~cm$^{-3}$ \citep[e.g., Figure~3 of][]{2020ApJ...896L..40P}. 

After obtaining accurate DMs for all the sources in the catalog, we also correct for the MW contribution to obtain the extragalactic DM contribution:
\begin{align}
    \rm{DM_{egal}} = DM  - DM_{MW},
\end{align}
where $\mathrm{DM_{MW}}$ is the MW DM contribution, which we assume is composed of a disk and a halo component: 
\begin{align}
    \rm{DM_{MW}} = \rm{DM_{MW,~disk}} + \rm{DM_{MW,~halo}}.
\end{align}
Here, we assume the NE2025 model for the MW disk \citep[][]{2026ApJ..1002....3O}, and obtain $\rm{DM_{\rm{MW, ~disk}}}$ by integrating out to 50\,kpc. For the halo, we assume the best-fit analytical model provided in Equation~8 of \citet[][YT2020]{2020ApJ...888..105Y}. {YT2020 is Galactic latitude dependent and predicts a mean halo DM of 43\,pc~cm$^{-3}$, which is consistent with the often assumed constant value of $\sim30$\,pc~cm$^{-3}$ \citep[e.g.,][]{2015MNRAS.451.4277D, 2019MNRAS.490....1A, 2022ApJ...927...35C, 2023ApJ...946...58C}}. 

\subsection{Foreground galaxy halo catalogs}
We conduct a two-pronged analysis: first, measure the effect of intervening galaxy CGMs on an FRB's DM, and then to probe the radial distribution of gas in the CGM \citep[e.g.,][]{2019MNRAS.485..648P, 2024A&A...690A.267Z}. For these analyses, we define two different galaxy halo data sets below.  


\subsubsection{Detecting the effect of galaxy CGMs on FRB DMs}
\label{sec:cgm_det}

Since the gas in galaxy halos will contribute to the measured DM of a background FRB, we expect that FRBs traversing through galaxy halos have higher DMs on average. To measure the DM contribution of galaxy halos, we compare the DMs of FRBs whose sightlines intersect halos of known galaxies with those that do not. This requires a control sample (the ``non-intersecting" sample) that is not contaminated by FRBs whose sightlines cross halos of galaxies missing from the catalog, since such contamination biases the measured DM excess low. Furthermore, having a larger maximum distance would increase the number of FRBs originating from galaxies within our galaxy catalog, which would also bias our DM measurements to the lower end (see Section \ref{sec:contact}). For these reasons, we need a high completeness for typical MW-like galaxies; the redshift completeness of local galaxy catalogs for $L^*$ systems decreases as we go out to higher luminosity distances, dropping to below 80\% past a redshift of 0.01 \citep[][]{2018ApJ...860...22K, 2020ApJ...895...32F}. Therefore, for this component of the analysis, we restrict ourselves to a maximum luminosity distance of 50\,Mpc ($z \approx 0.0117$). 


Because we restrict ourselves to a maximum luminosity distance of 50\,Mpc for this analysis, we use 50MGC \citep[][]{2024AJ....167...31O}. 50MGC is a compilation of galaxy information from HyperLeda \citep[][]{2014A&A...570A..13M}, NASA Sloan Atlas\footnote{\url{ http://nsatlas.org}}, and the Catalog of Local Volume Galaxies \citep[LVG;][]{2013AJ....145..101K}. The catalog contains a sample of 15424 nearby galaxies, with distances obtained from the best estimate among redshift-based distances from the Cosmicflows-3 Distance-Velocity Calculator \citep[][]{2020AJ....159...67K}, redshift independent distances from the LVG catalog and the NASA/IPAC Extragalactic Database distances \citep[][]{2017AJ....153...37S}, and they assigned a distance of $16.5\pm 1.1\,$Mpc for galaxies in the Virgo cluster \citep[][]{2007ApJ...655..144M}. In all, they have a median fractional error of 0.18 in their distances. 

Of the 15424 galaxies, 11740 of the galaxies in the catalog have stellar masses determined. The stellar masses were estimated using the photometric colors of the galaxies. Specifically, \citet[][]{2024AJ....167...31O} used the empirical relationship between the $(g-i)$ colors of the galaxies and the stellar mass to luminosity in the $i$ band ratio ($M_*/L_i$) from Equation 7 of \citet[][]{2011MNRAS.418.1587T}.
In our cross-matching, we need to ensure that the FRB sightline passes through the CGM of the galaxy and, therefore, need an estimate of the radius of the CGM from the stellar mass. For this, we use the semi-empirical relation obtained by \citet[][]{2013ApJ...773...16Z}:
\begin{align}
    r_\mathrm{vir} = 200~\mathrm{kpc} \left(\frac{M_*}{10^{10.3}M_\odot}\right)^{1/5}. 
    \label{eq:vir}
\end{align}
In obtaining this relationship, they assumed a power-law relationship between the stellar mass and the virial mass of $M_* \propto M_\mathrm{vir}^{5/3}$. Additionally, they also assumed a spherical halo collapse and obtained the critical overdensity for collapse from the empirical fits of \citet[][]{1998ApJ...495...80B}. We note here that Equation~\ref{eq:vir} was specifically applied to galaxies that have a median redshift of 0.1. However, we assert that this is acceptable for our purposes, as there are no large evolutionary differences between galaxies at these low redshifts \citep[$z < 0.2$; e.g.,][]{2012MNRAS.427.1238C, 2014ARA&A..52..415M}. Additionally, we follow the framework established by \citet[][]{2020ApJ...893...82F, 2022ApJ...928...37F} by assuming that $r_\mathrm{CGM} = 1.1 r_\mathrm{vir}$. Because we require a stellar mass to obtain the CGM radius, we restrict our analysis to the 11740 galaxies that have stellar masses listed in the 50MGC catalog. 

We also restrict our sample to galaxies that have a stellar mass greater than $10^{10}~M_\odot$, as these are known to have large reservoirs of hot ionized gas \citep[e.g.,][]{2024A&A...690A.268Z}, and so will have the most detectable impact on a background FRB's DM. Additionally, we constrain the analysis to a maximum stellar mass of $10^{11.5}~M_\odot$; above this mass, halos approach the galaxy group and cluster regime \citep[e.g.,][]{2018MNRAS.475..648P, 2025A&A...702A..22C, 2025MNRAS.537.3954M}. These mass limits correspond to the following CGM radii: $174\,\mathrm{kpc} \lesssim r_\mathrm{CGM} \lesssim 347$\,kpc. 

\begin{figure}[!htb]
    \centering
    \includegraphics[width=0.47\textwidth]{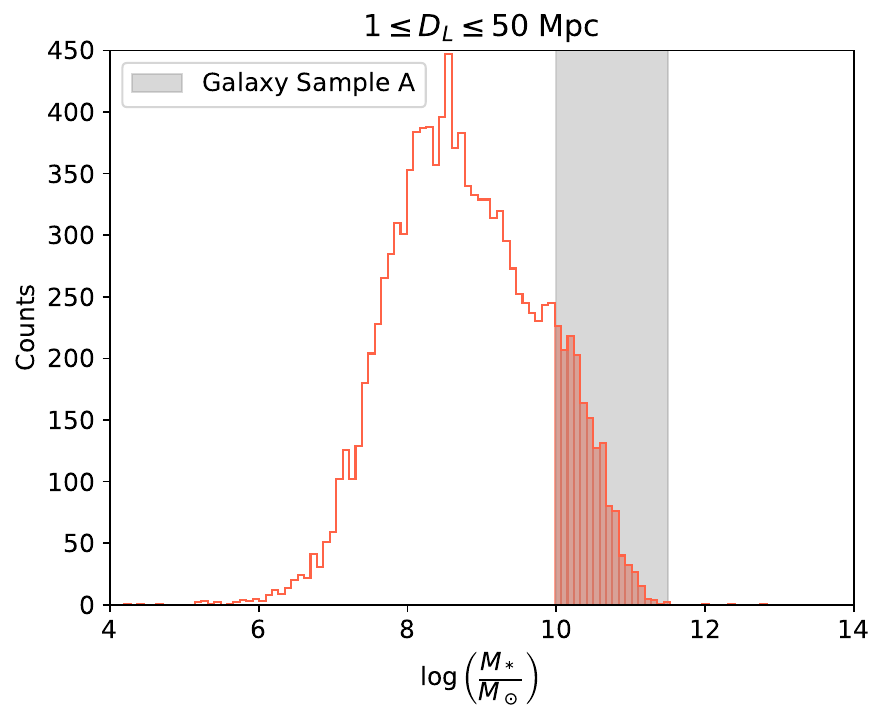}
    \caption{The stellar mass distributions of the galaxies in the 50MGC catalog displayed as the red outline, along with Galaxy Sample A highlighted in gray, with which we cross-match FRB sightlines to search for a signature of the DM contribution from intervening galaxy CGMs.}
    \label{fig:stellar1}
\end{figure}

We also exclude galaxies that have a luminosity distance less than 1 Mpc. This removes nearby galaxies with large angular extents, such as M31, which covers $>1600$\,sq~deg \citep[][]{2026arXiv260223749K}, corresponding to $\sim 1.3\%$ of CHIME's survey footprint \citep[][]{2022ApJS..261...29C}. This angular size is far larger than that of any other galaxy in our catalog and would heavily dominate the results of a stacking study. Once we place the mass restrictions along with the minimum distance restriction, the fiducial sample of galaxies in our catalog with which we cross-correlate FRB DMs is reduced to 1592. Hereafter, we denote this sample of galaxies to be `Galaxy Sample A'. We present the stellar mass distribution of the galaxies in Galaxy Sample A in Figure~\ref{fig:stellar1}, along with stellar masses from the full 50MGC catalog. Notably, Galaxy Sample A largely encompasses the high tail-end of this distribution; the majority of galaxies with $D_L \leq 50$\,Mpc are dwarf galaxies ($\log(M_*/M_\odot) \lesssim 9$). 

\subsubsection{Spatially mapping galaxy CGMs with FRB DMs}
\label{sec:mass_bins}
The electron density distribution of galaxy halos is known to decline radially, and this effect would be imprinted in the DMs of background FRBs as a function of impact parameter to foreground halos. We constrain the electron density distribution of halos using this signature. To measure this radial DM decline, we minimize the uncertainty on the impact parameter of the FRB to the foreground halo; as demonstrated in Figure~\ref{fig:1}, the median error on impact parameters becomes $\sim10\%$ of $r_\mathrm{vir}$ for MW-sized galaxies at a luminosity distance of 150\,Mpc \citep[][]{2019A&A...621A..56P}. We go to a higher luminosity distance than in Galaxy Sample A, to have a larger sample of galaxies to build statistical power by increasing the number of FRB-galaxy intersections. Going to higher luminosity distances than this will cause our sample to be predominantly polluted by data points that have $b_\mathrm{err} > 10\%~r_\mathrm{vir}$. Thus, for spatially mapping the DMs as a function of foreground galaxy impact parameter, we restrict ourselves to a maximum luminosity distance of 150\,Mpc ($z\approx 0.0329$). {At this maximum $D_L$, we estimate that $\sim99.5\%$ of the bursts in our sample are likely to be background to our galaxies using the FRB redshift probability distribution for CHIME/FRB from  Jain et al. (2026, submitted)}. {For this component of our analysis, we note that the completeness of the galaxy survey is not the limiting factor as we are only concerned with FRBs that we detect to be intersecting galaxy halos, and not with FRBs that we might erroneously misclassify to be non-intersecting because the galaxy they intersect is not visible in the catalog. } 

For spatially mapping the CGMs of intervening halos, we supplement the 1592 galaxies that are obtained from the 50MGC catalog with galaxies obtained from the REGALADE \citep[][]{2026A&A...706A.284T} galaxy catalog. We stress here that we do not include galaxies with $D_L \leq 50$\,Mpc from REGALADE to minimize potential double-counting. REGALADE combines data from multiple different galaxy catalogs including HECATE \citep[][]{2021MNRAS.506.1896K}, GLADE and GLADE+ \citep[][]{2018MNRAS.479.2374D, 2022MNRAS.514.1403D}, the NASA Extragalactic Database Local Volume Sample \citep[][]{2023ApJS..268...14C, 2025TNSAN..55....1C}, and the Sienna Galaxy Atlas \citep[][]{2023ApJS..269....3M}. In addition to this, \citet[][]{2026A&A...706A.284T} also queried and combined galaxy information derived from catalogs from deep imaging surveys such as the DESI Legacy Surveys \citep[][]{2019AJ....157..168D}, SDSS \citep[][]{2000AJ....120.1579Y, 2022ApJS..259...35A}, Pan-STARRS \citep[][]{2016arXiv161205560C}, and DELVE \citep[][]{2022ApJS..261...38D} to construct REGALADE. Distances for galaxies in the REGALADE catalog were obtained by selecting best distance estimates from the variety of photometric and spectroscopic redshifts available in the various component catalogs. 

We follow a similar approach to that described in Section~\ref{sec:cgm_det} by placing a lower limit of $D_L \geq 1$\, Mpc, but here we place an upper limit of $D_L\leq 150$\,Mpc. We also use the same maximum mass limit of $M_* \leq 10^{11.5}\,M_\odot$ to prevent probing galaxy cluster-scale objects. However, we reduce the mass lower-limit and now stack with galaxies down to $M_* \geq 10^{9.5}\,M_\odot$ (corresponding to $r_\mathrm{CGM} \gtrsim 138$\,kpc). This is primarily to search for the presence of ionized gas in these halos with the larger sample obtained by going out to a higher distance. These restrictions on the catalogs result in a sample of 92063 galaxies for this component of our analysis. Hereafter, we denote this full collection of galaxies to be `Galaxy Sample B'.

We note that different mass galaxies have intrinsically different physics involved: high-mass galaxies have strong AGN feedback that effectively pushes gas outward \citep[e.g.,][]{2022ApJ...932...18P, 2023MNRAS.524.5391A}, and the lack of strong AGN jets in less massive galaxies leads to the gas having shorter cooling times and therefore higher densities in the inner CGM \citep[e.g.,][]{2020MNRAS.499..768Z}. Because of this, the electron density distribution in different mass galaxies is intrinsically dissimilar; to accurately produce radial profiles, we need to treat these galaxies separately. For this reason, we divide the mass range of $9.5 \leq \log(M_*/M_\odot) \leq 11.5$ into 4 bins of 0.5\,dex each, numbered from Bin~0 to Bin~3 (in order of increasing mass).

\begin{figure}[!htb]
    \centering
    \includegraphics[width=0.47\textwidth]{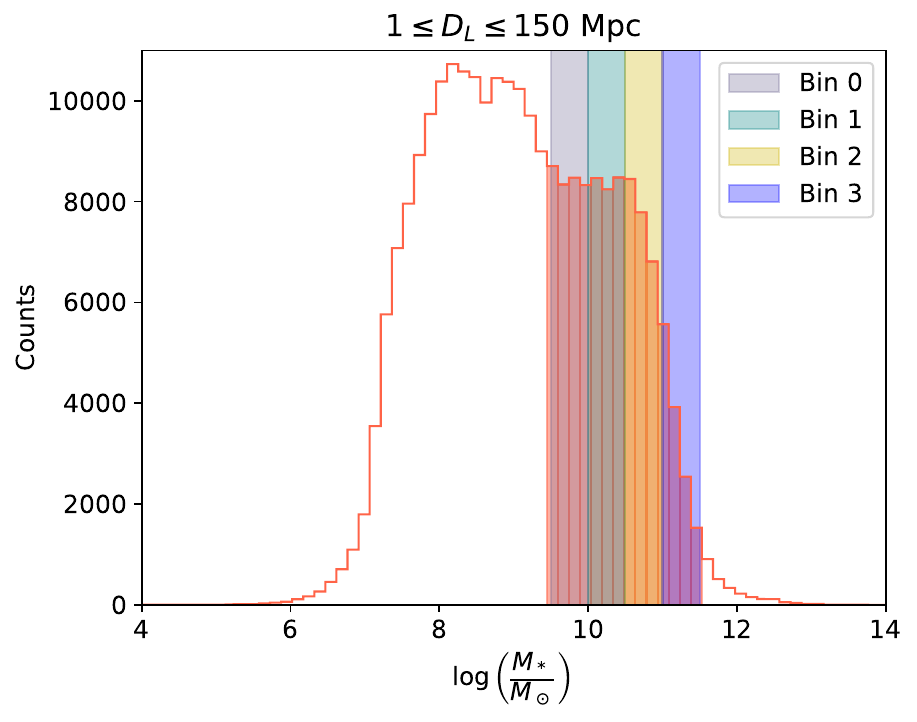}
    \caption{Stellar mass distribution for the galaxy sample obtained from supplementing 50MGC galaxies with REGALADE galaxies at a distance beyond 50\,Mpc, up to a maximum distance of 150\,Mpc. The highlighted Bins 0, 1, 2 and 3 correspond to stellar mass bins with widths of 0.5 dex, in the mass range $9.5 \leq \log(M_*/M_\odot)\leq 11.5$, in ascending order.}
    \label{fig:stellar2}
\end{figure}

Figure~\ref{fig:stellar2} displays the distribution of stellar masses in our sample, along with all four mass bins highlighted. In this sample, the fraction of massive galaxies is far higher than that in Figure~\ref{fig:stellar1}, likely because the completeness of the catalog to dimmer ($\log(L/L_\odot) < 9$) galaxies drops beyond a distance of 50\,Mpc. Despite this discrepancy in the relative abundances of high mass galaxies between Galaxy Sample~A and B, the width and the position of the peak of the stellar mass distributions in both our galaxy samples are similar, both peaking at $M_* \approx 10^{9}$ and having full-width at half-maximum (FWHM) values of $\sim4$\,dex. 

\subsection{Cross-matching FRB positions with galaxy halos}
\begin{figure*}[!htb]
    \centering
    \subfigure[]{\includegraphics[width = 0.9\textwidth]{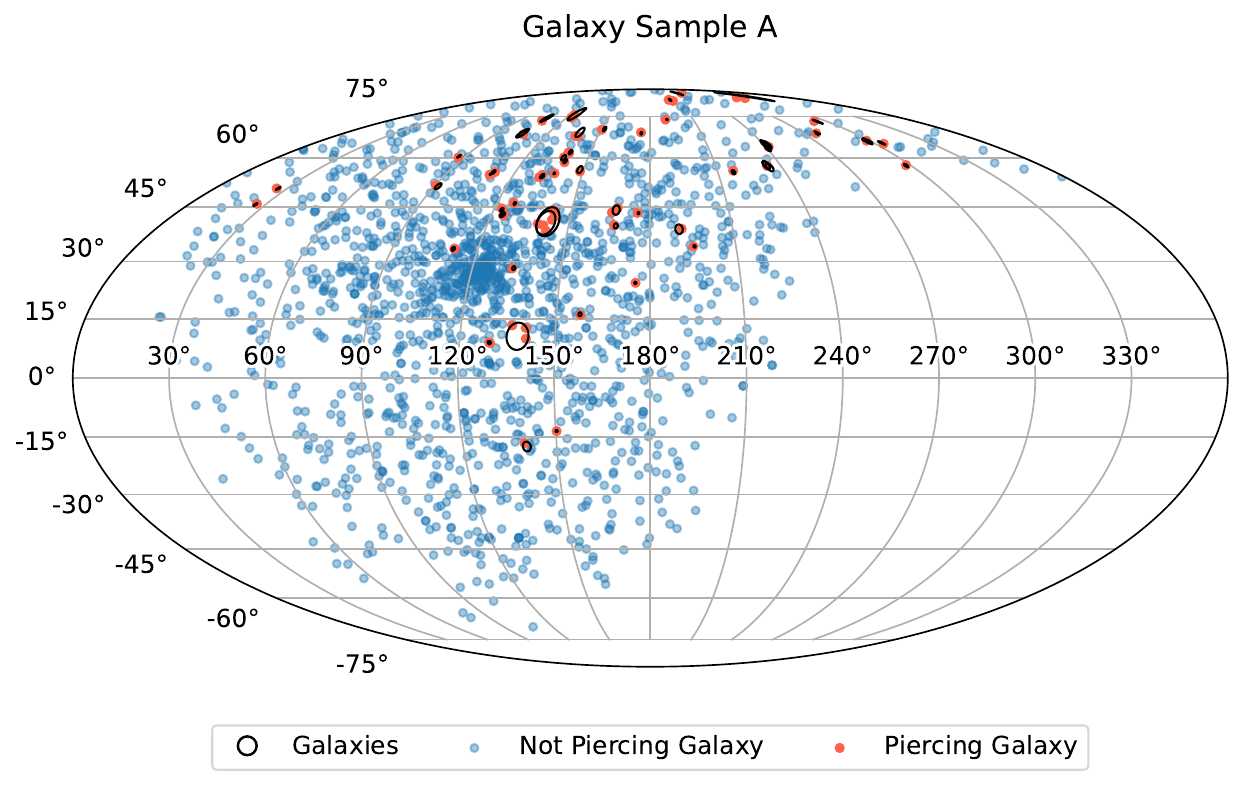}\label{fig:chime_all_sky1}}
    \subfigure[]{\includegraphics[width=0.9\textwidth]{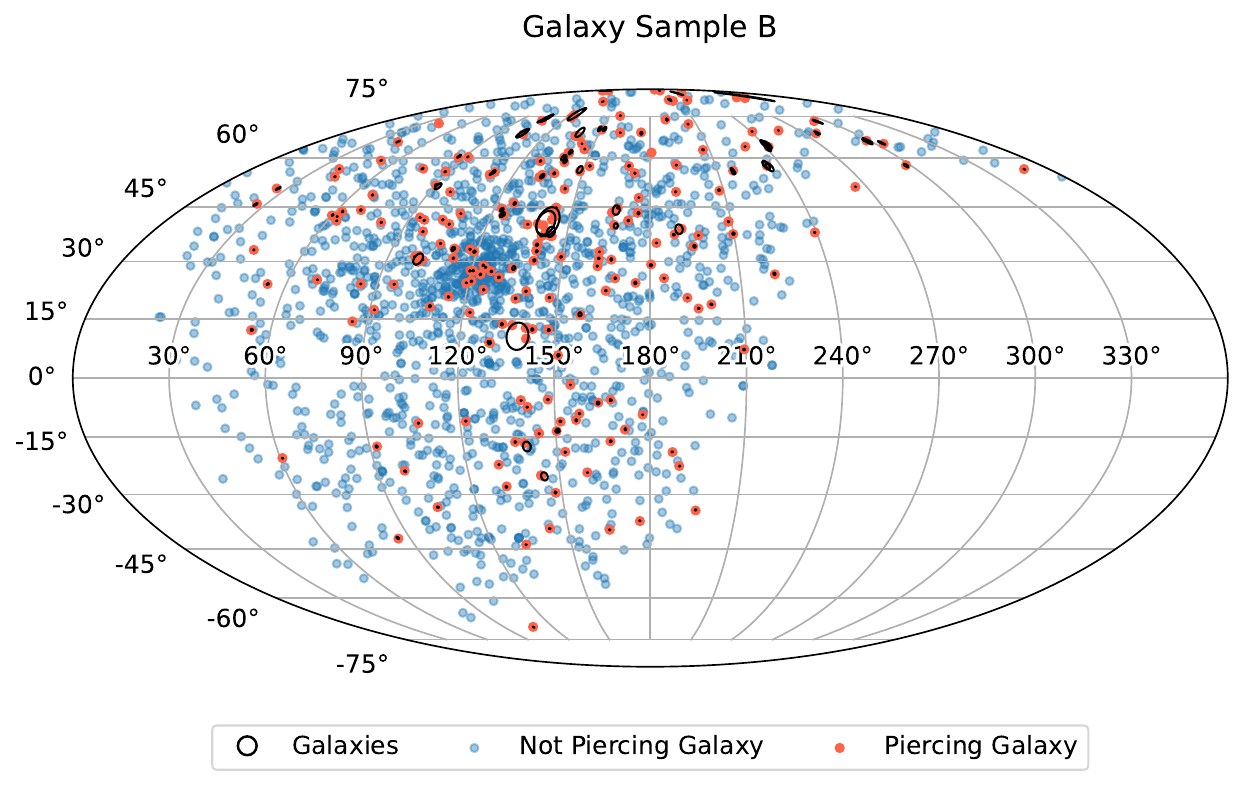}\label{fig:chime_all_sky2}}
    \caption{All-sky maps showing the positions of BaseCat2 FRBs that pierce galaxy halos within $1r_\mathrm{CGM}$ (in red) and FRBs that do not pierce galaxy halos (in blue), with the $1r_\mathrm{CGM}$ radii of galaxies indicated with black ellipses on the sky for our two galaxy samples. One of the most noticeable details are the high density clumps of FRBs at a position of $(l, b) \approx (120^\circ, 30^\circ)$; this corresponds to the North Celestial Pole, where CHIME/FRB has its highest exposure time and, therefore, detects more bursts \citep[][]{2026ApJS..283...34C}. We note that in both figures, the sizes of the data points for the FRBs were made large for visualization purposes, which can cause some to appear to be overlapping.}
\end{figure*}
For each FRB sightline in BaseCat2, we cross-match the FRB position with that of galaxies from our two galaxy samples and we only consider the FRB to be piercing a galaxy halo if the $1\sigma$ localization of the burst is fully encompassed within the $r_\mathrm{CGM}$ of an intervening galaxy \footnote{We do attempt a more conservative cross-matching by expanding the error region to the $2\sigma$ and $3\sigma$ levels for our analysis with the very nearby Galaxy Sample A and find no significant differences in our results; see Section \ref{sec:dm_excess}.}. Note that ideally, we would budget the DMs of the FRBs, with their redshift information, to ensure that they are foreground. However, since these FRBs are largely unlocalized to host galaxies, we cannot conduct an accurate budgeting of their DMs; additionally, our galaxies are extremely nearby so the vast majority of our FRB sample is likely background ($>99\%$) as previously noted. 

Additionally, for every FRB, we record the number of galaxy halos that it intersects and also the impact parameters at which it intersects these galaxies. In the case that an FRB intersects multiple halos along the LOS, we associate it with the halo to which it has the smallest impact parameter; {this applies to $\approx21\%$ and $\approx 26\%$ of the galaxy-piercing population in Galaxy Samples A and B, respectively}. This is primarily motivated by the fact that at a smaller impact parameter, the FRB passes through a larger column of the CGM and also through a denser part of the CGM and, therefore, will accumulate a larger DM contribution. We note that there is some degeneracy between the number of intersections, halo sizes, and the exact distribution of ionized gas in halos in producing a DM signal of a given strength. An example of a situation where this assumption fails is an FRB that passes through the outskirts of an incredibly massive galaxy's CGM but through the center of a smaller satellite galaxy. In this case, it would be associated with the smaller galaxy, even though it might contribute a smaller amount of DM than its massive host. Therefore, to precisely associate each FRB with the greatest contributor to its DM along the LOS, we would need to have information about the exact radial distribution of ionized gas in the CGM of the galaxies along the LOS, which is what we constrain. Therefore, as a simplifying assumption, we match each FRB with the closest galaxy along its LOS.

Figures~\ref{fig:chime_all_sky1} and \ref{fig:chime_all_sky2} display the positions of FRBs piercing galaxy halos for Galaxy Sample A and Galaxy Sample B, respectively. We note here that the size of the blue/red points is not representative of the localization regions of the FRBs (which are far smaller than the size of the points), but that the size of the black ellipses is equal to the angular size of $1r_\mathrm{CGM}$ of the galaxy halos. From these plots, it is evident that most of the galaxies with which we associate an FRB intersection are only probed by one FRB ($\approx 66\%$ and $\approx89\%$ for Galaxy Samples A and B, respectively); however, there are some large nearby galaxies such as the noticeable pair of M81 (with 5 background FRBs) and M82 (with 2 background FRBs) that do have multiple FRB sightlines that pass through their CGMs. 

In Galaxy Sample A, shown in Figure~\ref{fig:chime_all_sky1}, there are 66 FRBs probing galaxies within $1r_\mathrm{CGM}$ (that are indicated to be the piercing population in red), and 202 FRBs within $2r_\mathrm{CGM}$ (which are not highlighted in Figure \ref{fig:chime_all_sky1}){; there are no FRBs in the piercing population that are within $5^\circ$ of the Galactic plane}. In Galaxy Sample B, shown in Figure~\ref{fig:chime_all_sky2}, there are 223 FRBs that are probing galaxies within $1r_\mathrm{CGM}$ and 628 FRBs within $2r_\mathrm{CGM}${; there is one FRB in the piercing population that is within $5^\circ$ of the Galactic plane}. These numbers represent an order of magnitude higher galaxy halo-FRB intersections compared to previous stacking studies of galaxy CGMs with FRBs \citep[e.g.,][]{2022NatAs...6.1035C, 2025ApJ...993L..27H}.

\begin{figure}[!htb]
    \centering
    \subfigure[]{\includegraphics[width = 0.47\textwidth]{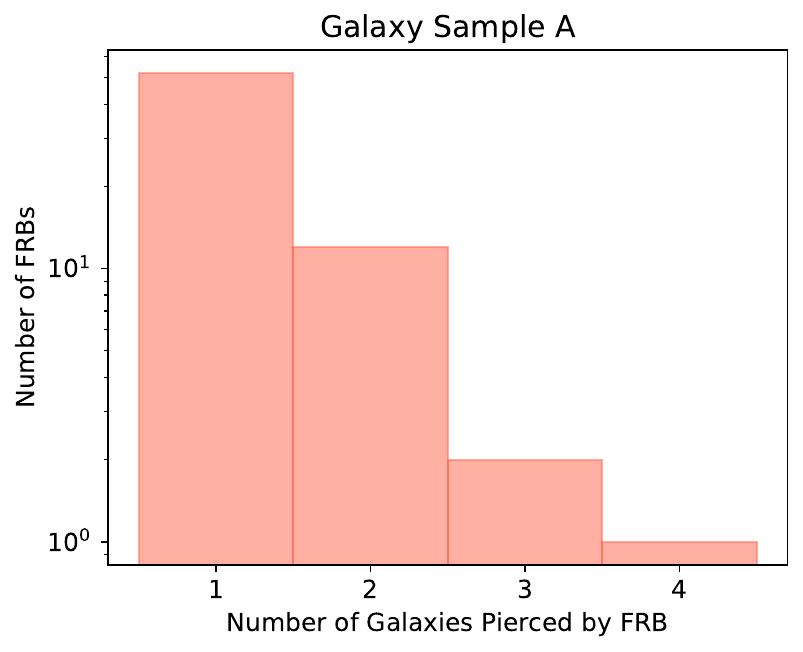} \label{fig:gal_hist1}}
    \subfigure[]{\includegraphics[width = 0.47\textwidth]{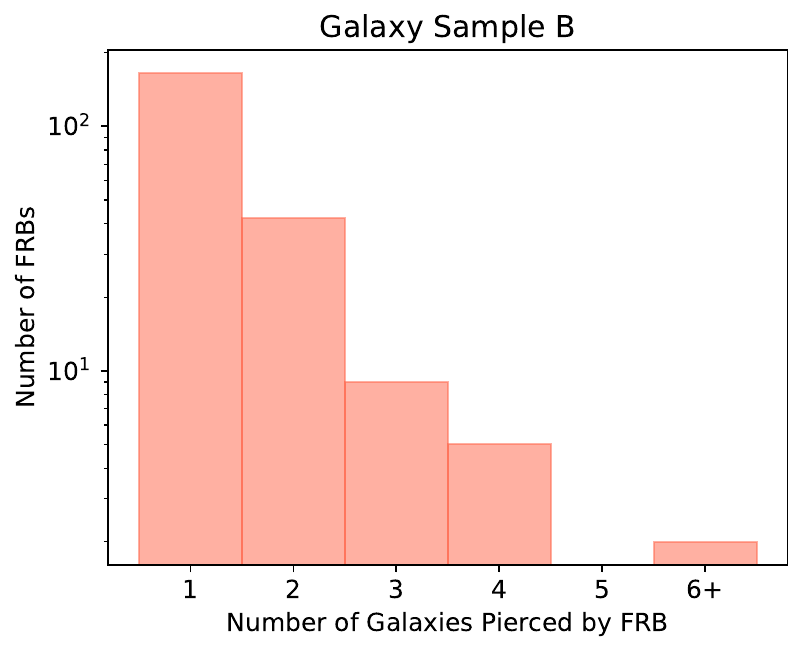} \label{fig:gal_hist2}}
    \caption{Histograms of the number of galaxies pierced by FRB sightlines for Galaxy Sample A (used for searching for a detection of the CGM; shown in a), and Galaxy Sample B (used for spatially mapping the distribution of gas in the CGM; shown in b). Non-intersecting FRBs (i.e., 0 galaxies pierced) have been excluded for visualization purposes.}
\end{figure}

Figures~\ref{fig:gal_hist1} and \ref{fig:gal_hist2} portray histograms of the number of galaxies pierced by FRBs for Galaxy Samples A and B. The vast majority of FRBs probe a single galaxy in both our samples, with just a few intersecting multiple galaxies. The most notable outlier is an FRB intersecting 17 galaxies within 1$r_\mathrm{CGM}$ in Galaxy Sample B, and also coinciding within $1r_\mathrm{vir}$ of the Perseus cluster. Because of the extreme number of halos that this FRB traverses, we exclude it in our radial CGM mapping analysis; the constraints on the properties of the baryons foreground to this FRB will be explored in a future work. 

\section{Results}
\label{sec:results}

\subsection{{DM contributed by galaxy halos}}
\label{sec:dm_excess}

\begin{figure}[!htb]
    \centering
    \includegraphics[width=0.47\textwidth]{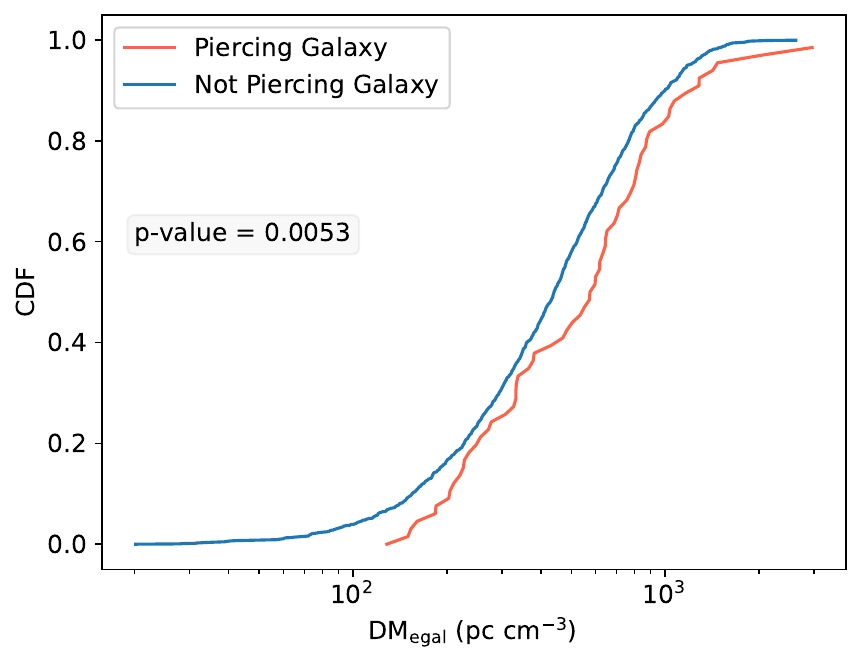}
    \caption{Cumulative distribution functions of the extragalactic DMs of BaseCat2 FRBs piercing galaxies (in red) and ones that do not (in blue).}
    \label{fig:cdf}
\end{figure}

Using Galaxy Sample A, we search for the presence of greater extragalactic DMs toward galaxy halos. We show the cumulative distribution functions (CDFs) of $\mathrm{DM_{egal}}$ that are piercing nearby galaxy halos and ones that are not in Figure~\ref{fig:cdf}. Notably, the CDF of the piercing population is significantly shifted to higher $\mathrm{DM_{egal}}$, with the greatest difference from the background occurring at $\mathrm{DM_{egal}}$ values of around 500\,pc~cm$^{-3}$. We calculate the excess mean $\mathrm{DM_{egal}}$ of the piercing sample to the non-piercing sample. Particularly, we note that variance due to high DM outliers may skew results. Recent studies of galaxy halos have utilized a DM-dependent exponential down-weighting to mitigate variance from high-DM outliers \citep[e.g.,][]{2023ApJ...945...87W, 2026arXiv260223749K}. Due to down-weighting high-DM FRBs, this method assigns more value to low-DM FRBs, which are more likely to be at low-$z$ and originating from galaxies within the galaxy sample. The number of these low-$z$ FRBs remains unconstrained, so we might bias ourselves to this contaminated FRB sample by following this weighting (see Section \ref{sec:contact} for more details). Additionally, the threshold DM scales used in these studies for down-weighting ($\alpha \sim 300 \, \mathrm{pc~cm}^{-3}$) heavily penalize (by a factor of $1/e^2$) typical piercing FRBs in our sample, which have DMs of $\sim 600\, \mathrm{pc~cm^{-3}}$. Because the largest difference between our piercing and non-piercing populations exists at DMs of $500-1000\,\mathrm{pc~cm}^{-3}$ (see the CDFs in Figure \ref{fig:cdf}), an exponential down-weighting suppresses the underlying physical signal. In addition to this,  \citet[][]{2023ApJ...945...87W} introduced this to counteract the effects of high DM outliers in a very small data sample ($N_\mathrm{FRB}\sim$20) that have large error regions; our analysis contains a much larger sample size and also has localizations with error regions that are $500\times$ smaller, which improves our stability to single high DM outliers. For these reasons, we choose to calculate the DM excess with a uniformly weighted mean and calculate our error bars using permutations of the full catalog to account for non-Gaussian cosmic variance that might lead to high DM outliers. 

We find a mean excess extragalactic DM of $119\pm 44$\,pc~cm$^{-3}$. We also compute the difference in the medians between the DMs of the FRBs piercing halos and ones that are not to be $132\pm 48$\,pc~cm$^{-3}$, as it is more robust to outliers. The two results are consistent within $1\sigma$ of each other, indicating that the excess seen in the means is not purely driven by outliers. We obtain the error bars on these measurements using permutations of the full catalog, which we also employ to quantify the significance of the excess by calculating the percentage of random iterations that have a greater excess than our observations. In the following, we present our approach to generate these permutations. From the full population of BaseCat2 FRBs, we randomly select a sample of 66 FRBs to be the piercing population (without replacement). Then, for each permutation, we calculate the mean difference between the random piercing sample and the remaining bursts, which are taken to be the non-piercing galaxy sample. We repeat this for $10^6$ iterations, and determine how many of the random iterations produce an excess extragalactic DM as high as in our true observations. Our random permutations have a mean excess extragalctic DM of $0.01\,\mathrm{pc~cm}^{-3}$ with a standard deviation of $43.80\,\mathrm{pc~cm}^{-3}$. 

We find that only 0.53\% of the random permutations have a DM excess as high as our observations, indicating that the observed DM excess toward galaxy halos is at the 2.6$\sigma$ threshold\footnote{We found a similar significance when using a more conservative classification, where we only considered FRBs to be intersecting halos if their $3\sigma$ localizations are encompassed by the galaxy halo}; the Kolmogorov-Smirnov (KS) or Anderson-Darling (AD) tests also find these distribution inconsistent with arising from the same underlying population ($p < 0.05$). Thus, we conclude that we find tentative evidence for the effect of the ionized gas in galaxy CGMs on the DMs of background FRBs.

\subsection{Radial distribution of baryons in the CGM}

\label{sec:full_stack_sample}
Here, we outline the results of mapping the radial distribution for FRBs intersecting galaxies from Galaxy Sample B, without any binning by mass. As previously mentioned, we first compute the impact parameter of the FRB to the foreground galaxy halo with the smallest angular separation along the LOS ($b_\mathrm{min}$). {We calculate $b_\mathrm{min}$ by first calculating the angular separation between the FRB position and the central position of galaxy halos, which is then translated to physical distances $(b)$ at the luminosity distance of the halos; finally, we take $b_\mathrm{min}$ to be the impact parameter that minimizes the normalized impact parameter $b/r_\mathrm{CGM}$ across all the galaxy halos}. We obtain error bars on $b_\mathrm{min}$, by propagating the errors in the FRB localization. {We also note here that we do not take into account the errors in the galaxy position, as these are typically far smaller than the uncertainty in FRB localizations}. First, we obtain an overall angular error by adding the errors in RA ($\alpha$) and Dec ($\delta$) in quadrature: 
\begin{align}
    \theta_\mathrm{error} = \sqrt{(\sigma_\alpha \cdot \cos(\delta))^2 + (\sigma_\delta)^2}. 
\end{align}
Then, we convert this angular error into a physical error at the luminosity distance of the galaxy that the FRB intersects.  

By not binning in mass, we assume that galaxy halos are self-similar \citep[e.g.,][]{2013ApJ...763L..42C}, i.e., the radial decline in electron density relies strictly on the normalized impact parameter ($b_\mathrm{min}/r_\mathrm{CGM}$), independent of the absolute halo mass. In Section~\ref{sec:mass_bins}, we explore the effect of this assumption by calculating the DM profile in stellar mass bins. Using this assumption of self-similarity, we explore the correlation between $\mathrm{DM_{egal}}$ and the fractional minimal impact parameter $b_\mathrm{min}/r_\mathrm{CGM}$. The unbinned extragalactic DMs as a function of fractional minimal impact parameter for the full Galaxy Sample B, as well as for the different mass bins is presented in Appendix \ref{app:dmvb}. Using these data, we compute the mean DM (in radial bins) as a function of impact parameter.

We calculate the mean DM profile by dividing the data into 10 bins, each of width $0.2~b_\mathrm{min}/r_\mathrm{CGM}$. Additionally, we note that due to the error in impact parameter, there is uncertainty regarding which bin a given FRB lies in. Because of this, we take the following binning approach: {an FRB is considered to be within a bin if its 1$\sigma$ error bar overlaps with the bin width}. Once we have determined which points lie within a given bin, we carry out a weighted mean in each bin. We describe the procedure taken to compute the weighted mean in Appendix \ref{app:weighted_mean}. 

We calculate the weighted mean over every radial bin to compute the averaged DM profile as a function of $b_\mathrm{min}/r_\mathrm{CGM}$. Once this is done, we correct for the contribution of the IGM, which we assume to be the mean $\mathrm{DM_{egal}}$ for $1.5 < b_\mathrm{min}/r_\mathrm{CGM} < 2$, where we expect the DM contribution of the CGM to be vanishingly small, and subtract this from the mean DM in each radial bin to obtain the background-corrected extragalactic mean DM profile, $\Delta\langle \mathrm{DM_{egal}}\rangle$. For BaseCat2 FRBs that we cross-match with galaxy halos in Galaxy Sample B, we find the mean IGM contribution to the DM to be 508\,pc~cm$^{-3}$.

We quantify two important facets of this mean DM profile: the robustness to outliers and statistical significance. We find the robustness by randomly excluding $\sim 10\%$ of the FRBs, recomputing the DM profile and calculating what percentage are similar to the original profile over $10^4$ random iterations. We determine a randomized profile to be similar to the observed profile if its reduced chi-squared with respect to the observation is less than 1 (i.e., it is within 1-$\sigma$ of the observed profile). 

To determine the statistical significance of the profile, we randomize the sky positions of the galaxies in Galaxy Sample B. We then cross-match BaseCat2 FRBs with this randomized sky and compute the mean DM profile. We repeat this process $10^4$ times and the significance of the observation is determined to be the fraction of randomized profiles that produce a stronger signal than the observation. We consider a profile to have a stronger decline than the observation if its Spearman rank correlation coefficient \citep[][]{spearman} is greater than that of the observation and if its DM excess in the interior ($0.2< b_\mathrm{min}/r_\mathrm{CGM}<1$) has a higher mean DM than the exterior ($b_\mathrm{min}/r_\mathrm{CGM}>1$). For this test, we ignore the innermost radial bin ($b_\mathrm{min}/r_\mathrm{CGM} < 0.2$) for all the profiles, as it is dominated by the FRB-galaxy ``contact term" (for more details regarding this, see Section \ref{sec:contact}).  

\begin{figure}[!htb]
    \centering
    \includegraphics[width=0.47\textwidth]{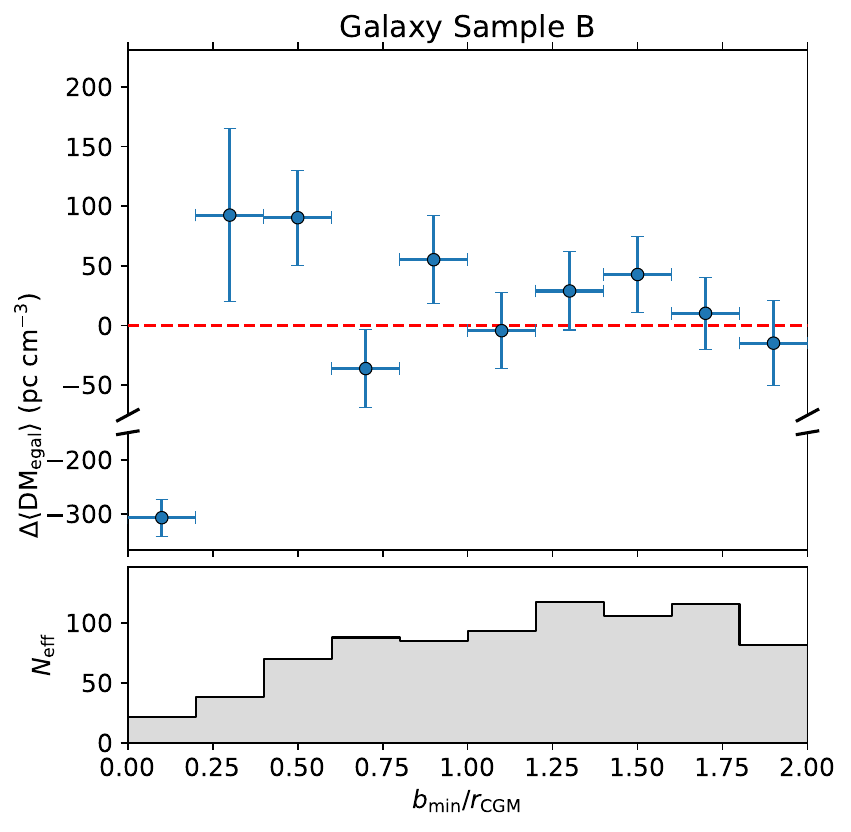}
    \caption{The background-corrected mean DM profile for FRBs that intersect CGMs within $2r_\mathrm{CGM}$ as a function of the fractional minimal impact parameter, indicated as blue dots. The error bars indicate standard error in the mean, with the red dotted line displaying the 0 $\mathrm{DM_{egal}}$ threshold. The bottom panel displays the effective number of data points per radial bin.}
    \label{fig:stack_profile}
\end{figure}

We present the observed mean DM profile for FRB DMs cross-correlated with Galaxy Sample B in Figure~\ref{fig:stack_profile}, along with the effective number of data points after our binning, $N_\mathrm{eff}$, as a function of minimal impact parameter. A distinct deficit in the mean extragalactic DM is seen at low impact parameters, which is likely caused by cross-matching FRBs with their own host galaxies (see Section~\ref{sec:contact} for more details). We observe a peak excess extragalactic DM of $\sim 90\pm75$\,pc~cm$^{-3}$ at $0.2 \lesssim b_\mathrm{min}/ r_\mathrm{CGM}\lesssim 0.4$ that diminishes to the background at higher impact parameters. We compute the robustness of this DM profile at $b_\mathrm{min}/r_\mathrm{CGM}  > 0.2$ using our aforementioned $10\%$ jackknifing to be 98.92\%, with a p-value of 0.0046. 

\subsection{Mass-binned DM profiles}
In this subsection, we present the mean DM profiles as a function of $b_\mathrm{min}/r_\mathrm{CGM}$ for the different mass bins in Galaxy Sample B that have been defined in Section~\ref{sec:mass_bins}. The number of FRBs that are cross-matched with Bins 0, 1, 2 and 3 are 91, 210, 217 and 110, respectively. Despite Bin~0 having a comparable number of galaxies to Bins 1 and 2 (see Figure~\ref{fig:stellar2}), it has the lowest number of FRB intersections because of their smaller physical size. The galaxies in Bin~3 are massive but less numerous (see Figure~\ref{fig:stellar2}), leading to fewer total FRB intersections. Bins 1 and 2 contain galaxies that are relatively abundant (see Figure \ref{fig:stellar2}), and also large enough for us to successfully associate many FRBs with them. 

\begin{figure*}
\includegraphics[width = 0.47\textwidth]{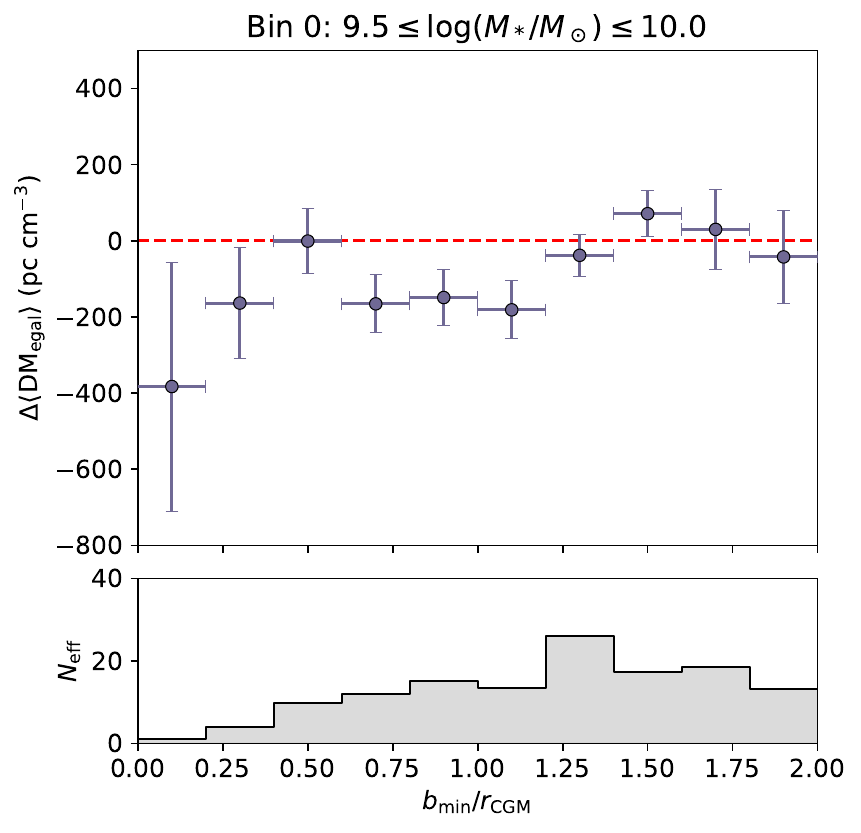}
\includegraphics[width = 0.47 \textwidth]{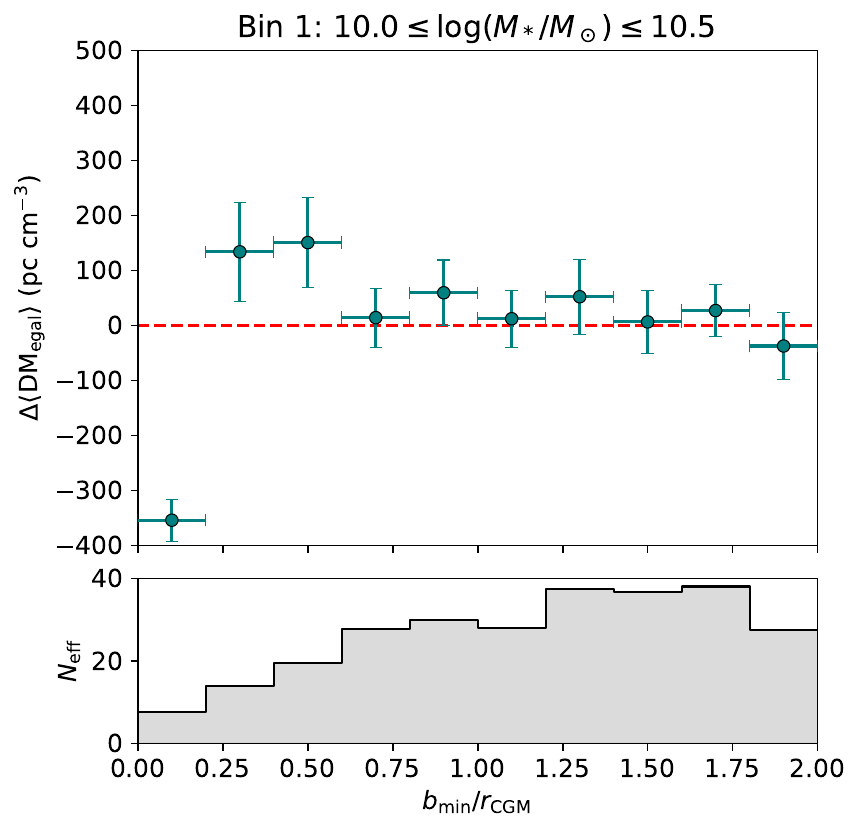}\\
\includegraphics[width = 0.47 \textwidth]{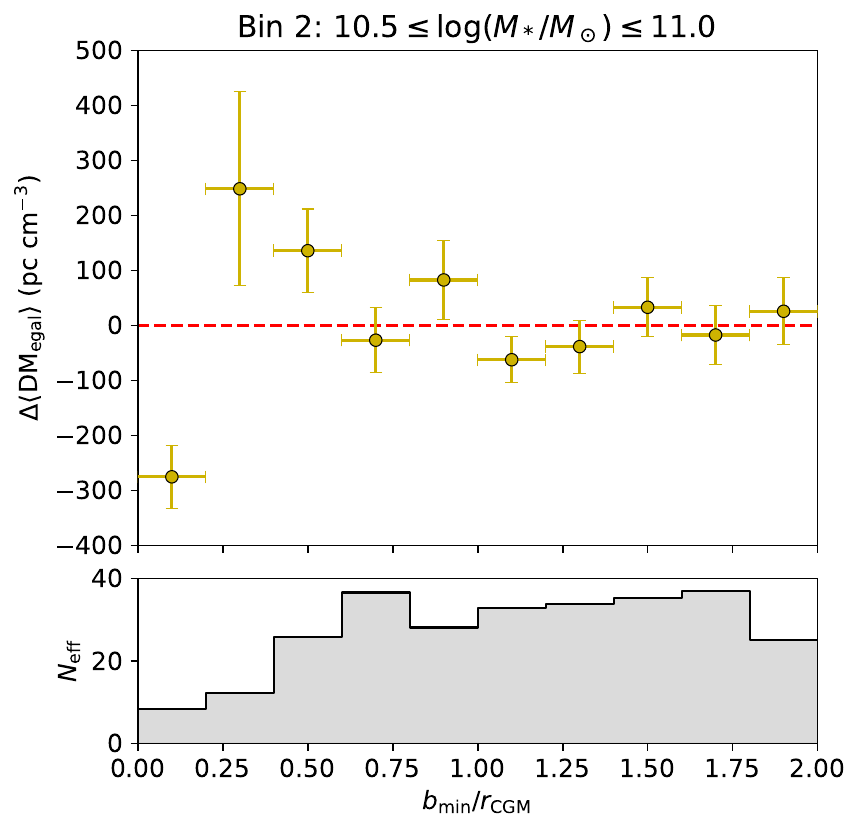}
\includegraphics[width = 0.47\textwidth]{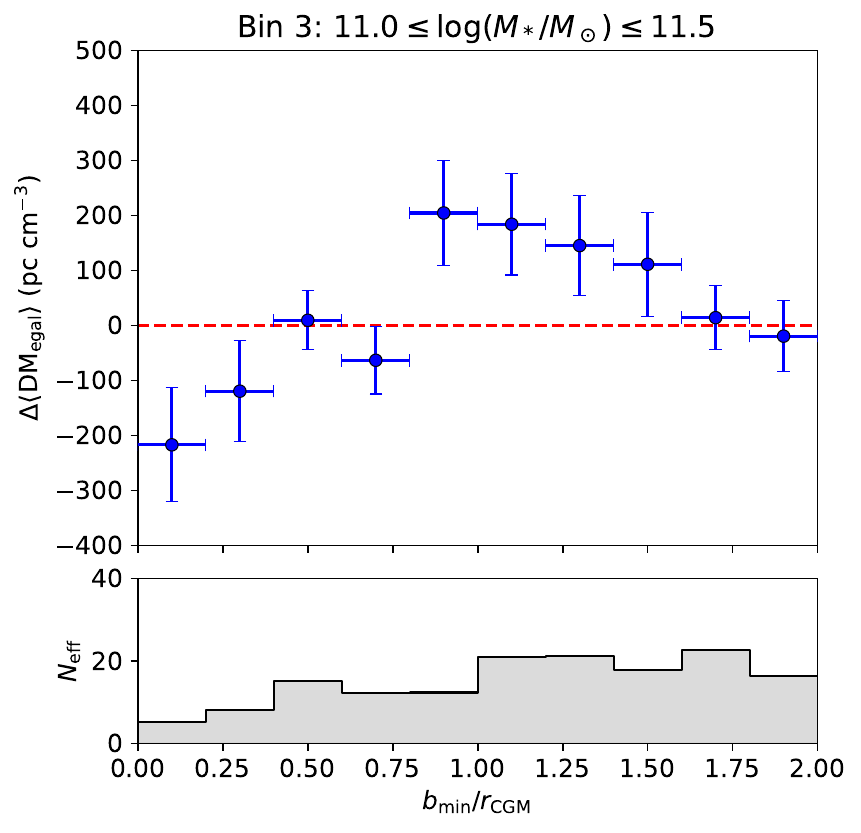}
    \caption{The background-corrected mean DM profiles (indicated with scatter points) as a function of fractional minimal impact parameter for the different mass bins, with the error bars representing the standard error on the mean, with the red dotted line displaying the 0 $\mathrm{DM_{egal}}$ threshold. The bottom panel displays the effective number of data points per radial bin.}
    \label{fig:bin_dm_profile}
\end{figure*}

We compute the mean DM profiles in each of the mass bins using the procedure described in Appendix~\ref{app:weighted_mean}, along with the robustness and the p-value for each mass bin. We present the DM profiles in Figure~\ref{fig:bin_dm_profile}, along with the effective number of FRBs per radial bin as a function of $b_\mathrm{min}/r_\mathrm{CGM}$. The trials-corrected p-values (by carrying out a Bonferroni correction with a multiplication by 4) and robustness for all the mass bins are displayed in Table \ref{tab:p-value}. Bin~2 is our most statistically significant mass bin, having the lowest p-value. 

\begin{table}[!htb]
    \centering
    \caption{{The trials-corrected p-values and robustness measures of the mass-binned DM profiles }}
    \begin{tabular}{|c|c|c|}
    \hline
        \textbf{Bin Number (and mass) }& \textbf{p-value} & \textbf{Robustness} \\
        \hline 
        \hline
         \textbf{0} ($9.5 \leq \log(M_*/M_\odot)\leq 10$) &  {0.8056} & $99.93\%$ \\
        \textbf{1} ($10\leq \log(M_*/M_\odot)\leq 10.5$)& {$0.1956$} & $71.11\%$\\
         \textbf{2} ($10.5\leq \log(M_*/M_\odot)\leq 11$)& {0.0092} & 97.78\%\\
         \textbf{3} ($11\leq \log(M_*/M_\odot)\leq 11.5$)& {1.000} & 97.88\%\\
         \hline
    \end{tabular}
    \label{tab:p-value}
\end{table}

A common theme among all the stellar mass bins is the deficit in mean DM in the lowest impact parameter bin; we discuss this in Section~\ref{sec:contact}. Notably, we see the most pronounced hints of an excess in Bins~1 and 2. Therefore, from this figure, we conclude that most of the apparent observed DM excess is largely driven by $L^*$ galaxies in Bins 1 and 2, which have massive enough halos ($\gtrsim10^{10}\, M_\odot$) to have a measurable impact on an FRB's DM, and are also numerous enough to provide us with high statistical power, allowing us to constrain the magnitude of the DM excess. Past the first radial bin, the DM profiles in both the stellar mass Bins 1 and 2 present an excess DM in the inner CGM and lower mean DMs in the outer CGM. Bin~0 has only one data point at small impact parameters and the DM profile appears to be relatively flat and consistent with no excess at all radii. Bin~3 also displays a profile that has a cavity near the center, is increasing monotonically, and plateauing to the background around $b_\mathrm{min} \sim r_\mathrm{CGM}$. We discuss some possible causes for the apparent inconsistency of these DM profiles compared to theoretical expectations, in Section~\ref{sec:theory}. 

\subsection{Modeling the observed DM profiles}
\label{sec:model}

While the overall significances of our observed profiles fall short of a formal detection ($< 3\sigma$), certain subsamples approach this threshold much more closely than others. Nevertheless, our results provide the most compelling indication to date for a radially declining electron density distribution in galaxies obtained using FRBs (see Section~\ref{sec:comparison}). Thus, we attempt to use our data to place constraints on a model of the electron density distribution for Galaxy Sample B, as well as for the different stellar mass bins. We model the electron density using a simple beta model \citep[e.g.,][]{1976A&A....49..137C, 2009A&A...500..103A}: 
\begin{equation}
    n_e^\mathrm{CGM}(r) = n_{e, 0} \left[ 1 + \left(\frac{r}{r_c}\right)^2\right]^{-3\beta/2},
    \label{eq:profile}
\end{equation}
where $n_{e,0}$ is the central electron density, $r_c$ is the core radius, and $\beta$ is the power-law index, which represents the ratio of the kinetic energy to the thermal energy of the gas \citep[][]{1994ApJ...426..513B}. Conventionally, this model has been used to study galaxy clusters \citep[e.g.,][]{2017ApJ...843...76A, 2024A&A...685A.106B} but recent work has showed that it is also applicable to galaxy CGMs \citep[e.g.,][]{2020NatAs...4.1072K, 2020ApJ...897...63D, 2024A&A...690A.267Z}. We use the beta model in comparison to more complicated models \citep[such as the modified NFW profile introduced by][]{2019MNRAS.485..648P} because of the limited number of model parameters. We also note that we exclude the innermost contact-term dominated radial bin from our modeling, constraining ourself to bins that satisfy $b_\mathrm{min}/r_\mathrm{CGM}>0.2$. 

Another crucial factor in our analysis is the fact that there are many FRBs that are probing more than a single galaxy halo along the LOS (as demonstrated in Figure~\ref{fig:gal_hist2}). Thus, we need to take into account the DM contributions of all galaxies along the LOS; {we note here that since we do not know the redshifts of the FRBs we cannot do a full accounting like that done by the FLIMFLAM survey \citep[][]{2024ApJ...973..151K, 2025ApJS..277...64H}}. For this reason, we model the CGM contribution for each FRB sightline by integrating Equation~\ref{eq:profile} for each galaxy halo (within 2$r_\mathrm{CGM}$) along the LOS. Additionally, we note here that $r_c$ is highly dependent on the virial radius of the galaxy, so, instead we explore the fractional core radius: 
\begin{equation}
    r_c^\mathrm{frac} = \frac{r_c}{r_\mathrm{CGM}}. 
\end{equation}
Thus, we explore the $n_{e,0}, r_c^\mathrm{frac},\beta$ parameter space through a  Markov Chain Monte Carlo (MCMC) approach. We restrict ourselves to the following region of parameter space:  
\begin{align}
    \{-3 \leq \log(n_{e,0}) \leq -1;~0.2 \leq \beta \leq 0.9;~\\ \nonumber 0.01\leq r_c^\mathrm{frac} \leq 0.15\}, 
\end{align}
\citep{2024A&A...690A.267Z} and maximize the Gaussian log-likelihood: 
\begin{equation}
    \ln \mathcal{L}(\theta) = -\frac{1}{2} \sum_i \left[\frac{\Delta \langle \mathrm{DM_{egal}}\rangle_i - \langle\mathrm{DM_{model} (\theta)}\rangle_i}{\sigma^\mathrm{DM_{egal}}_i}\right]^2, 
\end{equation}
where $\theta = (\log(n_{e,0}), r_c^\mathrm{frac},\beta)$ is our parameter vector, $\langle\mathrm{DM_{model} (\theta)}_i\rangle$ is the model DM profile at the $i$-th radial bin, and the sum is taken over all impact parameter bins that are within $1r_\mathrm{CGM}$. We apply this restriction to limit the effects of any external noise terms that might dominate at higher impact parameters (see Section~\ref{sec:theory} for further details).

\begin{figure*}
    \centering
    \subfigure[]{
        \includegraphics[width=0.47\textwidth]{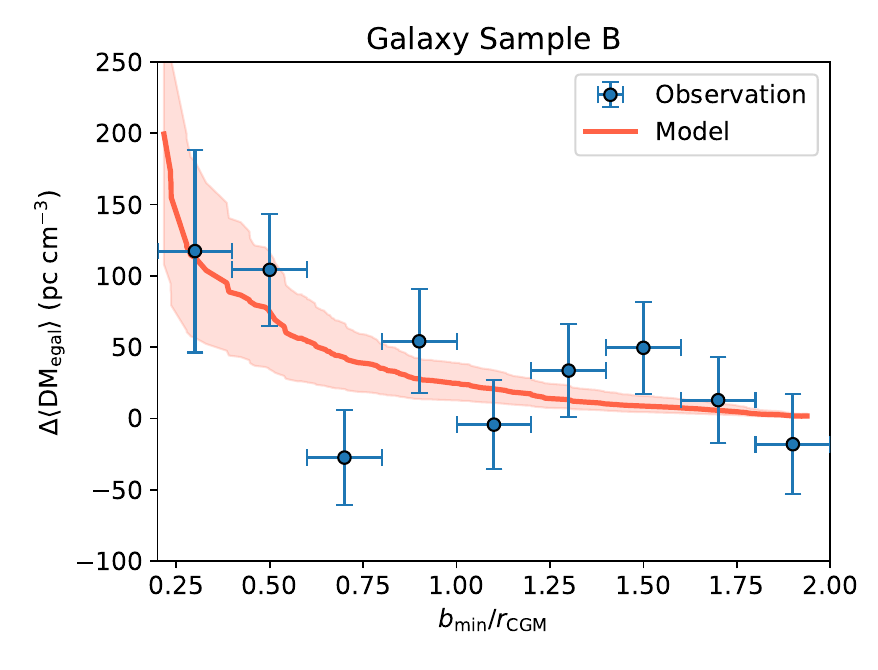}\label{fig:mcmc_fullstack}}\subfigure[]{\includegraphics[width=0.47\textwidth]{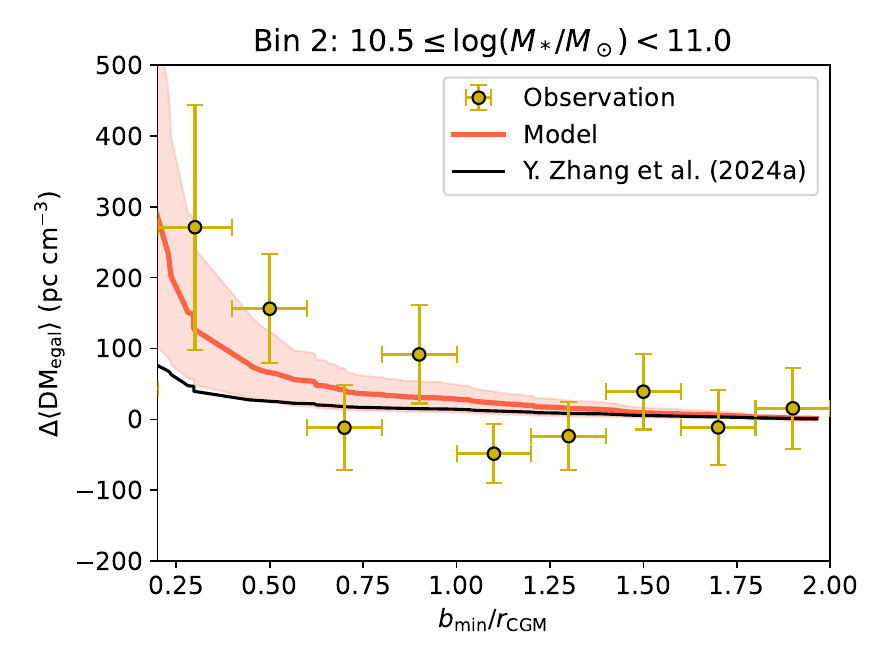}\label{fig:bestfit_model}}
    \caption{(a) The median MCMC model DM profile (in red) with the shaded region encompassing the 16th and 84th percentiles; the observed background-corrected mean DM profile (in blue) for Galaxy Sample~B with the error bars indicating the $1\sigma$ error. (b) The median MCMC model DM profile (in red), with the shaded region encompassing the 16th and 84th percentiles; the observed profile (in mustard) for the Bin~2 galaxies, with the error bars indicating the $1\sigma$ error; and the mean DM profile obtained from the fit results of \citet[][]{2024A&A...690A.267Z} (in black).}
\end{figure*}

We present the results of our MCMC for Galaxy Sample~B in Figure \ref{fig:mcmc_fullstack}. Additionally, we also present the results from our MCMC for Bin~2 in Figure \ref{fig:bestfit_model}, as it is the only stellar mass bin that has measurements of the hot plasma from another similar statistical stacking study; here, we compare our results to the mean DM profile obtained from the best-fit parameters of \citet[][]{2024A&A...690A.267Z}. We obtain $n_{e,0}$ from their central X-ray surface brightness by assuming a virial temperature of $T_\mathrm{vir} = 0.22\,$keV and a CGM metallicity of 0.3$Z_\odot$ \citep[which is typical of MW-like galaxies; e.g.,][]{2017ApJ...837..169P}, and evaluating the X-ray emissivity $\epsilon_{X}$ at $0.3 Z_\odot, T_\mathrm{vir}$, using the Astrophysical Plasma Emission Code models \citep[][]{2001ApJ...556L..91S}. We evaluate the emissivity at these values as they correspond to the ones that \citet[][]{2024A&A...690A.267Z} found to produce a total galactic baryon fraction that is consistent with the cosmological baryon fraction for galaxies in this stellar mass bin. The MCMC radial profiles for the remaining stellar mass bins are presented in Appendix \ref{app:stellar_mass}, along with the posteriors for all our MCMC runs. We present the constraints obtained on parameters from the MCMC for all our galaxy samples in Table \ref{tab:mcmc}, along with the reduced chi-squared values for the models evaluated at the maximum a posteriori best-fit parameter values. 

\begin{table*}[]
    \begin{center}
    \renewcommand{\arraystretch}{1.5}
    \caption{MCMC results from fitting with a $\beta$ model}
    \begin{tabular}{|c|c|c|c|c|}
    \hline
         {Galaxy Sample} & $n_{e,0}$ ($10^{-3}$\,cm$^{-3}$) & $\beta$ & $r_c^\mathrm{frac}$ & $\chi^2/\mathrm{d.o.f}$ (Best-Fit)\\
    \hline 
    \hline 
        Galaxy Sample B & $1.7-21.1$\tablenotemark{a}& $0.68^{+0.15}_{-0.16}$& $0.074^{+0.048}_{-0.039}$ & 1.35 \\ 
        Bin 0 ($9.5\leq \log(M_*/M_\odot)\leq 10$) & $<13.0$ & $-$ & $-$ & 3.16\tablenotemark{b}\\
        Bin 1 ($10 \leq \log(M_*/M_\odot)\leq 10.5$) & $2.1-26.87$ & $0.64^{+0.16}_{-0.17}$& $0.083_{-0.044}^{+0.045}$&0.52 \\
        Bin 2 ($10.5\leq \log(M_*/M_\odot) \leq 11$) & $1.8-26.3$ & $0.75^{+0.11}_{-0.17}$& $0.071^{+0.049}_{-0.040}$ & 1.05\\
        Bin 3 ($11\leq \log(M_*/M_\odot)\leq 11.5$) & $<12.3$ & $-$ & $-$  & 2.03\tablenotemark{b}\\
    \hline
    \end{tabular}
    \end{center}
    \tablenotetext{a}{We report the 16th--84th percentile range as the MCMC posteriors are relatively flat at the lower limit and remain unconstrained due to the large error bars on our data.}
    \tablenotetext{b}{For MCMC runs that yield a poor goodness-of-fit ($\chi^2$/d.o.f. $>1.5$) we only report the 84th percentile upper limit on $n_{e,0}$ and do not report the shape parameters, $\beta$ and $r_c^\mathrm{frac}$, that remain unknown since the amplitude is unconstrained. }
    \label{tab:mcmc}
\end{table*}

\section{Discussion}
\label{sec:discussion}

\subsection{Other potential source of the observed excess extragalactic DM}
\label{sec:environment}
In Section~\ref{sec:dm_excess}, we established that there is tentative evidence ($\sim 2.6\sigma$) for higher extragalactic DMs in the direction of galaxy halos. {The magnitude of this excess is consistent (within error) with the expected DM contribution from MW and M31-like galaxy halos \citep[e.g.,][]{2020ApJ...888..105Y, 2015MNRAS.451.4277D, 2023ApJ...946...58C, 2026arXiv260223749K}. } However, we note that galaxies preferentially reside in large-scale matter overdensities, such as cluster and group environments \citep[e.g.,][]{1996MNRAS.282..347M}. Most of the baryonic matter in galaxy clusters and groups resides as hot \citep[$T\sim 10^7 - 10^8$\,K;][]{2024A&A...691A.340R}, ionized gas in the intracluster medium (ICM) \citep[][]{1993Natur.366..429W}. At the center of a galaxy cluster, the ICM can have densities as high as $10^{-1}$\,cm$^{-3}$ \citep[e.g.,][]{2002MNRAS.335..256A, 2024A&A...685A.106B}. Thus, as an FRB traverses through the ICM, it undergoes dispersion due to the ICM's plasma. Therefore, one might argue that the excess extragalactic DM we observe toward galaxy CGMs is not caused by the CGM itself but by the cluster environment within which the CGM resides. We have attempted to exclude FRBs that are affected by DM contamination from clusters by masking the 26 FRBs determined by \citet[][]{2026arXiv260810358S}  to be embedded or background to clusters, but we note that the method therein lies on large DM contribution from the clusters to identify these bursts and therefore some low level contamination from clusters may persist. 

To test this possibility, we produce simulated DMs for each FRB sightline. For each simulated realization, first, we draw a host DM, $\mathrm{DM_{host}}$, for each BaseCat2 FRB sightline from the probability distribution of Jain et al. (2026, submitted) which is based on the work of \citet[][]{2023ApJ...944..105S}, from which we also draw the FRB's redshift. Then, we assume the following IGM contribution:
\begin{equation}
    \mathrm{DM_{IGM}} = 892\mathrm{\,pc~cm}^{-3}z,
\end{equation}
following the findings of \citet[][]{2021ApJ...906...49Z}. We model the simulated DM, $\mathrm{DM_{sim}}$, as follows:
\begin{equation}
    \mathrm{DM_{sim}} = \frac{\mathrm{DM_{host}}}{1+z}  + 892 \mathrm{\,pc~cm}^{-3}z + \mathrm{DM_{LSS}},
    \label{eq:sim}
\end{equation}
where $\mathrm{DM_{LSS}}$ is the DM contribution from the large-scale structure (LSS) local to the galaxy halos, which includes clusters, groups, and filaments.

To obtain this $\mathrm{DM_{LSS}}$, we make use of the SLOW magnetohydrodynamic (MHD) cosmological simulation \citep[][]{2023A&A...677A.169D, 2024A&A...692A.232B}. SLOW is a constrained cosmological simulation of the local cosmic web, with a box size of 500$h^{-1}$\,Mpc (going out to $z \approx 0.058$), centered on the MW. The SLOW simulation uses the positions and motions of galaxies from CosmicFlows2 \citep[][]{2013AJ....146...86T} to constrain the LSS and build the initial conditions \citep[][]{sorce18, sorce+21, sorce+24}. The identified structures and clusters obtained from these initial conditions are described by \citet[][]{2023A&A...677A.169D}, \citet[][]{hernandezmartinez+24} and \citet[][]{seidel+24}. We note here that SLOW is a purely radiative MHD simulation without any star formation or feedback physics \citep[][]{2024A&A...692A.232B, 2026arXiv260706665S}. 

By doing this, SLOW is able to reproduce the positions of more than 40 local galaxy clusters and groups. We integrate the electron density along the LOS in SLOW throughout the full simulation volume (out to $z = 0.058$), following Equation~\ref{eq:dm} to obtain the DM contribution from the local LSS. The simulation box we use has $2\times 3072^3$ particles (both dark matter and gas particles). Particularly, the gas particle mass resolution ($5.4\times 10^7 h^{-1}\,$Mpc) of SLOW sets a minimum resolvable halo mass of $10^{11}\, M_\odot$. Consequently, the simulation accurately captures the density profiles of the ICM of clusters and groups, but it lacks the resolution to produce CGM density profiles of MW-like ($M_\mathrm{vir} \sim 10^{12}\,M_\odot$) intervening galaxies. It also contains no galaxies; only galaxy-sized dark matter and baryon halos. For these reasons, SLOW is ideal to test if contributions from the LSS (excluding CGM contributions) are sufficient to account for the observed extragalactic DM excess toward galaxy CGMs. 

For a given simulated realization, we obtain the $\mathrm{DM_{LSS}}$ for each BaseCat2 FRB by perturbing the position of the FRB within a radius of $0.74^{\circ}$, which corresponds to a physical radius of $\sim$1\,Mpc \citep[the virial radius of average galaxy clusters;][]{1983ApJ...271..446A} at $D_L = 75$\,Mpc (which roughly corresponds to the average luminosity distance of a galaxy in our sample). This is done to account for variability in $\mathrm{DM_{LSS}}$ that arises from uncertainties in the exact electron density distribution in galaxy clusters. Once we obtain the $\mathrm{DM_{LSS}}$, it is added to the host and IGM contributions to produce the simulated DM, as in Equation \ref{eq:sim}. This process is repeated to produce $10^{5}$ realizations of BaseCat2 FRBs that do not have any CGM contribution to their DM. 

\begin{figure}[]
    \centering
    \subfigure[]{\includegraphics[width=0.47\textwidth]{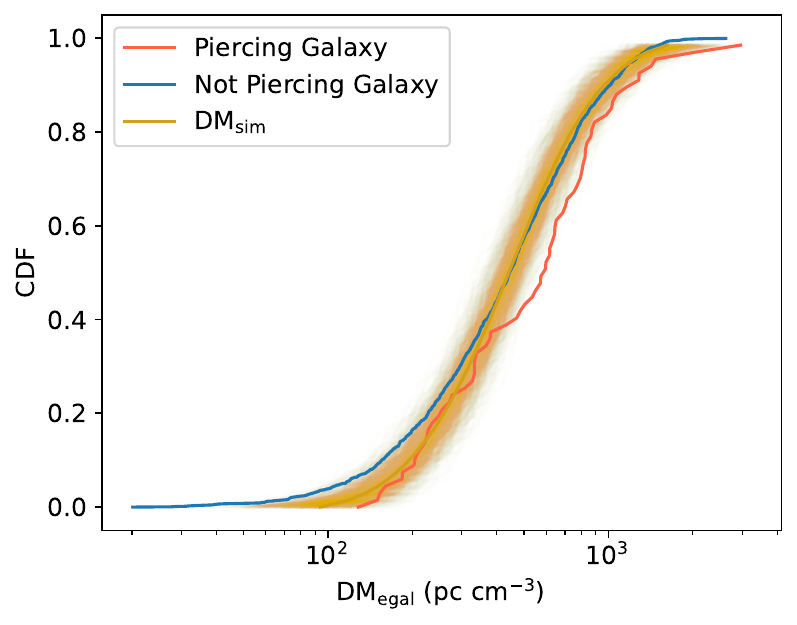}\label{fig:cdf_sim}}
    \subfigure[]{\includegraphics[width=0.47\textwidth]{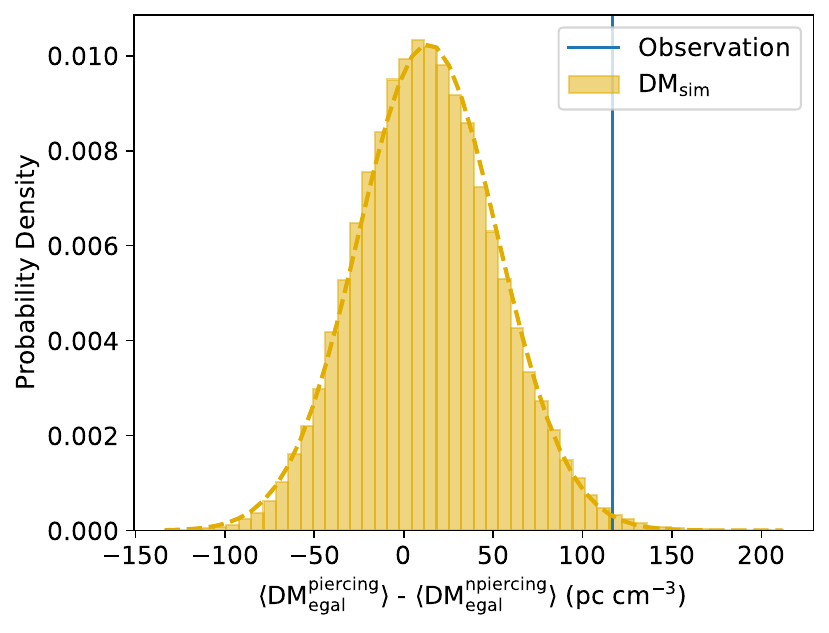}\label{fig:diff_lss_sim}}
    \caption{(a) CDFs for 1000 simulated realizations of BaseCat2 FRBs without any CGM contribution, over-plotted on the CDFs for the observed piercing and non-piercing samples, with the median of all the realizations plotted with a dark solid line. (b) The mean difference between the simulated DMs for BaseCat2 FRBs that intersect galaxy halos and ones that do not. The dotted line represents a Gaussian fit to the distribution. We indicate the observed mean excess.}
\end{figure}

To determine if the LSS contribution (along with the host and IGM contribution) can account for the observed DM excess of $119\pm44\,\mathrm{pc~cm}^{-3}$ presented in Section \ref{sec:dm_excess}, we compare the CDFs of the DM of 1000 randomly selected realizations to our observations in Figure~\ref{fig:cdf_sim}, each of which has 1748 unique bursts. 

We also calculate the mean differences between the simulated DMs for the BaseCat2 FRBs that intersect galaxy halos and ones that do not to explore the impact that the environment of galaxy halos has on the observed DM. We plot this distribution for all simulated realizations in Figure~\ref{fig:diff_lss_sim}. We find that the simulations are only able to reproduce the excess seen in the observations 0.62\% of the time, {providing tentative evidence that the full extragalactic DM excess we observe cannot be {solely} explained by cluster and group contributions}. We also note that the distribution for the simulations is not centered on 0; in particular, it indicates that the simulated DMs for the piercing population are slightly higher than that of the not piercing population. We fit this distribution with a Gaussian and find a mean of 14\,pc~cm$^{-3}$ with a standard deviation of 39\,pc~cm$^{-3}$. This indicates that the contribution from the LSS is insufficient to account for the excess extragalactic DM seen toward galaxy halos. In addition, since the {full} distribution in Figure~\ref{fig:diff_lss_sim} is shifted to the right, we conclude that, on average, the environment (the plasma of groups, clusters and filaments) of galaxy halos in the local Universe contributes 14\,pc~cm$^{-3}$ to the DMs of background FRBs, conditioned on the astrophysics and cosmology of the simulation. We note that our measurement in a constrained simulation is consistent with the findings from random cosmological simulations such as IllustrisTNG. For example, \citet[][]{2024A&A...683A..71W} measured  a mean ($1\sigma$) DM contribution of $14 ~(48)$\,pc~cm$^{-3}$ from halos in the IllustrisTNG simulation for low redshift ($z = 0.1$) FRBs, which trace the local Universe that is simulated in SLOW.
\subsection{{Radial DM profiles of CGM halos}}
\label{sec:theory}
{Here, we discuss the radial DM profiles of the galaxies in our full Galaxy Sample~B, as well as the DM profiles binned by stellar masses. }
\subsubsection{The contact term}
\label{sec:contact}
A notable feature present at the several $\sigma$ level across all our stellar mass bins is the deficit in DM at the smallest radial bin. We note that this is likely a manifestation of the ``contact term", presented in Equation 6 of \citet[][]{2026PhRvL.137d1001W}, which arises when the FRB sources originate from the same redshift as the galaxy sample with which they are being cross-matched. In particular, in this case, it indicates that some of these FRB sightlines at low impact parameters are associated with their own host galaxies. This contact term therefore occurs due to these low$-z$ FRBs having a much lower mean DM than that of the average background FRBs which originate from cosmic distances and have a larger $\mathrm{DM_{IGM}}$. This results in us over-correcting the IGM contribution (which we assume to be the mean DM of FRBs at $1.5 < b_\mathrm{min}/r_\mathrm{CGM} <2)$ of these low-$z$ FRBs, which manifests as a DM deficit. A full exploration of the association of BaseCat2 FRBs to host galaxies is left to future work. However, here we derive the theoretical form of this contact term effect to the full signal that we observe. We do this by starting with the DM-galaxy angular cross power spectrum obtained by \citet[][]{2026PhRvL.137d1001W} and taking Hankel transforms \citep[][]{hankel1869cylinderfunctionen} under the flat-sky and Limber approximations \citep[][]{1953ApJ...117..134L} to obtain the DM-galaxy cross-correlation function for a galaxy at redshift $z_g$: 

\begin{align}
\label{eq:contact}
    \xi_b^{dg}(z_g) &= \frac{f_f(z_g)}{(1+z_g)^2}\int_{-\infty}^{\infty}n_e^\mathrm{CGM} \left(\sqrt{b^2 + l_g^2}\right)dl_g\\ \nonumber +& \Delta \mathrm{DM}_g p_f(z_g) \frac{\mathcal{P}_\mathrm{FRB\times gal}(z_g)}{2\pi \sigma_b^2}e^{-b^2/2\sigma_b^2}.
\end{align}
where $f_f$ is the fraction of FRBs at redshifts higher than $z_g$, $n_e^\mathrm{CGM}$ is the electron number density distribution, $l_g$ is the distance along the LOS from the center of the galaxy halo, $\Delta \mathrm{DM}_g$ is the DM deficit that an FRB originating from a host at $z_g$ has (in comparison to the average DM of the background), $\sigma_b$ is the error in the impact parameter, and $\mathcal{P}_{\mathrm{FRB\times gal}}$ is the probability of an FRB originating from a galaxy at redshift $z_g$ within the mass range in our galaxy sample. The full derivation of Equation \ref{eq:contact} is presented in Appendix \ref{app:cross}. The first term in this equation represents the DM contribution due to the CGM of an intervening halo, and the second term is the contact term.

\begin{figure}
    \centering
    \includegraphics[width=0.47\textwidth]{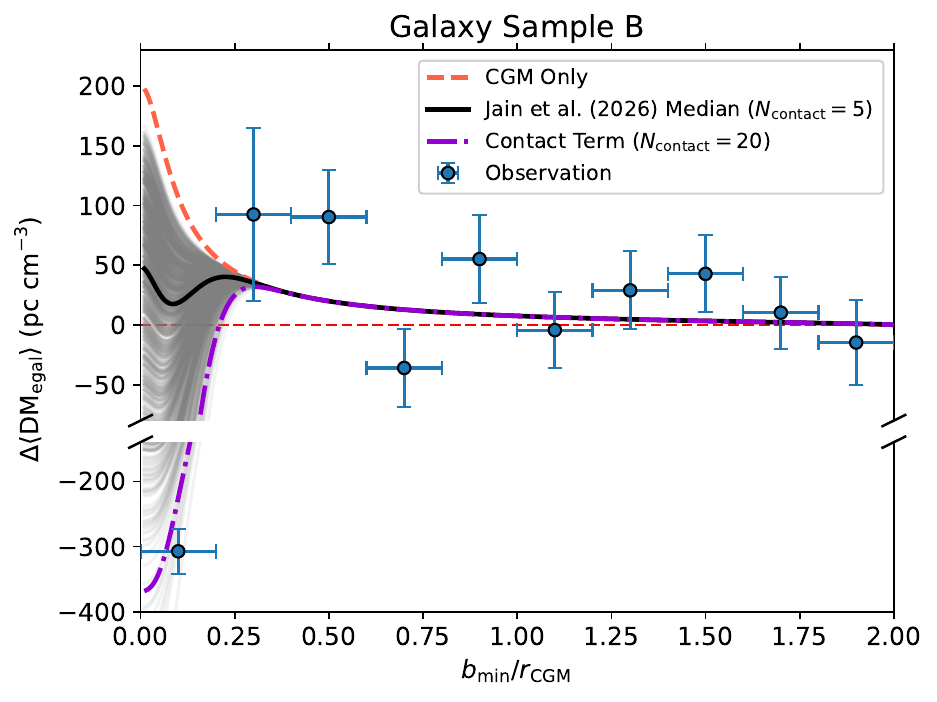}
    \caption{The theoretical DM profile obtained from the DM-galaxy cross-correlation function that only displays the DM contribution from a single halo for the CGM, without accounting for multi-halo intersection (in red), and 20 FRBs contributing to the contact term (in purple), in comparison to the mean background-corrected DM profile (in blue). We also show the theoretical DM profile using the probability distribution of Jain et al. (2026, submitted); we show their median prediction of 5 contact FRBs in black and show the full distribution of DM profiles (in gray), using their MCMC sampling chains.}
    \label{fig:contact}
\end{figure}

In our measurement of Galaxy Sample B, we find that there are $\sim 20$ FRBs that are in the innermost radial bin, indicating that they are possible candidates for FRBs that are cross-matched with their own hosts. Andersen et al. (in prep.) have run a PATH \citep[][]{2021ApJ...911...95A} analysis on a subset of BaseCat2 FRBs to associate FRBs with their hosts. We have cross-matched our 20 contact term candidate FRBs with theirs and find that 8 of these 20 have a PATH probability association to the galaxy that we cross-match them with of $>90\%$, with 3 additional being probable candidates with probabilities $>80\%$, indicating that a significant fraction of these FRBs have been cross-matched to intersect with their own hosts. We present the predicted theoretical DM profiles as a function of impact parameter for the full Galaxy Sample~B using our DM-galaxy cross-correlation function and our $n_e^\mathrm{CGM}$ best-fit MCMC results from Section \ref{sec:model} for various values of the number of contact term FRBs, $N_\mathrm{contact}$, in Figure \ref{fig:contact}\footnote{The theoretical models here only contain contributions from a single halo, and are not forward-modeled along multiple halo intersections as in Section \ref{sec:model}, so the amplitude of the DM at larger impact parameters in our models may be slightly different from those in Figure \ref{fig:mcmc_fullstack}.}. Particularly, we show $N_\mathrm{contact} = 20$ (corresponding to the number of contact term candidates in our data), $N_\mathrm{contact} = 5$ (demonstrating the median results of Jain et al. 2026, submitted), along with the full allowed range of $N_\mathrm{contact}$ from Jain et al. (2026, submitted). From this figure, we see that based on our theoretical predictions, $\sim 20$ FRBs contributing to the contact term are expected (exactly the number of FRBs observed at $b_\mathrm{min}/r_\mathrm{CGM}< 0.2$ ), and only 5 FRBs under predicts the deficit due to the contact term. Overall, our results are consistent with the allowed range of Jain et al. (2026, submitted), and they suggest that theoretical predictions best match the observations when there are $\sim 20$ FRBs within 150 Mpc. The redshift distribution of Jain et al. (2026, submitted) allows a large range of the number of FRBs at $D_L < 150$\,Mpc and is poorly constrained ($1\leq N_\mathrm{contact}\leq 28$, with a median of $N_\mathrm{contact} = 5$); our results provide independent evidence that the number of FRBs at these low redshifts might be at the higher end of this range. Thus, these results indicate that CHIME/FRB detects FRBs at lower redshifts than expected. 


\subsubsection{The DM profile at larger radii}
{Besides this interior feature, Bins 1 and 2, along with the full Galaxy Sample B, are qualitatively in line with the theoretical expectation of a radially declining electron density that fluctuates about some background value at large impact parameters. The radial profile of Bin~0 appears to be completely flat (excluding the contact term-dominated innermost radial bin) and is consistent with the null hypothesis (based on the high p-value of {0.81}) of no detection. } 
{The best-fit electron density in Bin~2 is consistent with the gas densities of $10^{-2} - 10^{-3}$\,cm$^{-3}$ in the inner CGM for galaxies in this mass range from simulations such as TNG300 \citep[e.g.,][]{2020MNRAS.499..768Z}.  Additionally, our findings are consistent with those of \citet[][]{2015ApJ...800...14M} for the MW's \citep[which has a stellar mass of $\log(M_*/M_\odot) = 10^{10.7}$;][]{2020MNRAS.494.4291C} hot halo using \ion{O}{6} and \ion{O}{7} emission lines. The fractional core radius of 0.07 (equivalent to $r_c = 8$\,kpc for a $\log(M_*/M_\odot) = 10.5$ galaxy) that we find is also consistent within error of the values found using the eROSITA All-Sky Survey by \citet[][]{2024A&A...690A.267Z} for galaxies in this mass range. Our measured values of $\beta$ are also within error of those found for MW-like galaxies by \citet[][]{2013ApJ...770..118M}, \citet[][]{2015ApJ...800...14M}, and \citet[][]{2024A&A...690A.267Z}, who found $\beta = 0.71^{+0.13}_{-0.14}, ~0.50\pm 0.03, ~0.43^{+0.10}_{-0.06}$, respectively. Overall, we note that, as displayed in Figure \ref{fig:bestfit_model}, the best-fit electron density models from \citet[][]{2024A&A...690A.267Z} are consistent within error with our observations and our models. 


Using our median MCMC models, we also compute the baryon fraction for a typical MW-like galaxy in Bin~2 to compare our baryon fractions with those found by \citet[][]{2024A&A...690A.267Z} for galaxies in this stellar mass bin. We assume that a typical galaxy in this stellar mass bin has a stellar mass of $\log(M_*/M_\odot) = 10.75$, and a halo mass of $\log(M_\mathrm{halo}/M_\odot)  = 12.128$ \citep[obtained from the stellar-to-halo mass relation of ][]{2020A&A...634A.135G}. We assume a Navarro-Frenk-White \citep[NFW;][]{1997ApJ...490..493N} profile for the underlying dark matter halo (not the baryonic halo that we constrain in Section \ref{sec:model}) with a concentration parameter of $c = 13.3$, following the measurements of \citet[][]{2020MNRAS.494.4291C} for the MW. We compute the cumulative baryon fraction enclosed in various multiples of the virial radii of such a galaxy, and present these fractions in Figure \ref{fig:baryon_frac}. From these plots, we see that it is likely that the closure radius \citep[the radius at which the baryon fraction reaches the cosmic mean;][]{2023MNRAS.524.5391A} for these galaxies might be as low as $\sim 1r_\mathrm{vir}$, indicating that it is possible that baryons are tracing the underlying dark matter distribution well and that there maybe suppressed feedback in these galaxies. Our results in this mass range are in-line with simulations; \citet[][]{2023MNRAS.524.5391A} found that the closure radius for galaxies from TNG50 MHD simulation \citep[][]{2019MNRAS.490.3196P} in this halo mass range can go lower than $1r_\mathrm{vir}$. However, we cannot rule out higher closure radii of $>3r_\mathrm{vir}$ \citep[which have been found by][]{2024A&A...690A.267Z}, as this is within our $1\sigma$ error region. 
\begin{figure}[!htb]
    \centering
    \includegraphics[width=0.47\textwidth]{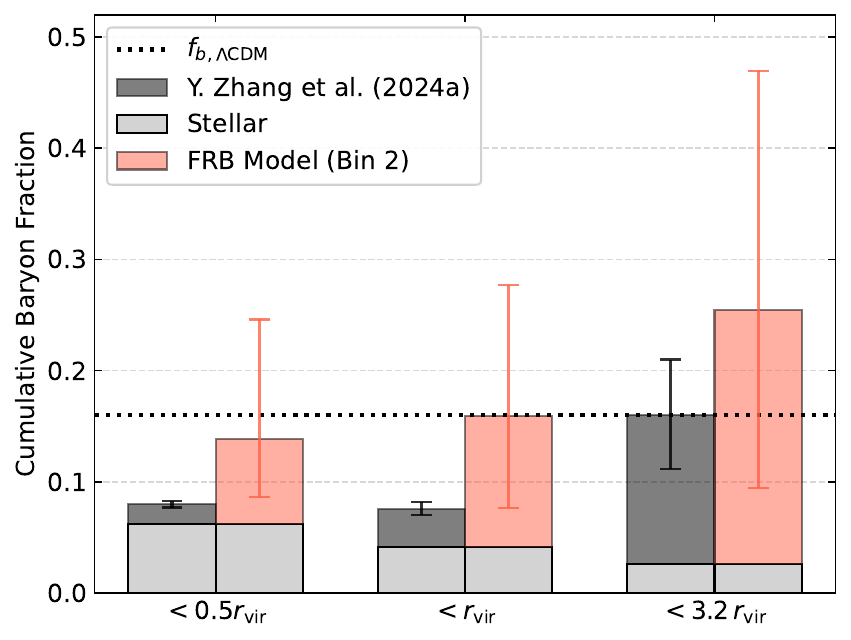}
    \caption{The cumulative baryon fraction enclosed in various virial radii for a MW-like galaxy. We present the hot gas baryon fraction from our best-fit models, along with the stellar mass fraction and the hot gas predictions from the work of \citet[][]{2024A&A...690A.267Z}. The error bars represent the $1\sigma$ error of the baryon fraction as determined with the 16th and 84th percentiles from the posterior of our MCMC samples. The cosmic mean baryon fraction as predicted by $\Lambda$CDM \citep[][]{2020A&A...641A...6P} is indicated with a black dotted line. }
    \label{fig:baryon_frac}
\end{figure}

{However, we note that a radially declining electron density model is unable to capture the intricacies of the real data for all the profiles, most importantly for Bin~3, where a radial decline is heavily disfavored. We cannot rule out the null hypothesis of being consistent with a flat line; however, we do see tentative hints of a deficit in DM in the CGM ($b_\mathrm{min}/r_\mathrm{CGM} < 1$). Nevertheless, we offer below some possibilities that might explain the data.} In Bin~3, we see that there is a collection of low extragalactic DMs that is accompanied with the absence of high DMs, and that this trend extends out to distances of  $b_\mathrm{min}\sim 0.8 r_\mathrm{CGM}$ (see Figure \ref{fig:raw_dm}). Our assumed background IGM contribution is assumed to be the mean DM at $1.5r_\mathrm{CGM} \leq b_\mathrm{min} \leq 2r_\mathrm{CGM}$, and for Bin~3, this exterior mean is higher than the mean DM at low impact parameters. Thus, our background subtraction manifests as the large central DM deficit seen in Figure~\ref{fig:bin_dm_profile}. Given the low statistical significance of this stellar mass bin, we only provide speculative interpretation that this apparent deficit may be due to scattering away of high DM sightlines through the center of the CGMs of these galaxies, or lower observed DMs due to stronger AGN feedback at these higher mass ranges \citep[e.g.,][]{2013MNRAS.429.1827P}. It might also be possible that due to a high DM contribution from the CGM, these bursts might have a lower S/N and thus harder to detect, as CHIME/FRB's detectability depends on the DMs of the FRBs \citep[][]{2026arXiv260626334M}. This might bias these high DM bursts to be closer on average, which might lower their IGM contribution, resulting in an over-subtraction of $\mathrm{DM_{IGM}}$. Many more FRBs are needed to investigate this feature further (see Appendix \ref{sec:forecast}) and to constrain models in this stellar mass bin.

{\subsection{Comparison to \textsc{Simba}}}
{Feedback processes are strong enough to dictate the large-scale distribution of baryons; for example, \citet[][]{2026MNRAS.548ag699D} have shown that feedback has a large effect on the distribution of baryons in the cosmic web, with the gas mass fraction increasing by more than 2\% in knots when AGN feedback is turned off. Hence, with less powerful feedback, this excess baryonic mass in the knots is more likely to condense into the CGM of galaxies, and therefore alter the observed DM profiles of galaxies. Since we restrict ourselves to local neighborhood galaxies, and we have FRB localization errors of $<1'$, we are able to probe the inner CGM with more accurate impact parameters. Thus, our study is uniquely suited to probing the inner CGM $(b/r_\mathrm{CGM} < 0.5)$, where feedback processes have the greatest effect on the CGM gas \citep[e.g.,][]{2017MNRAS.466.3810F}. Thus, we use our DM profiles to attempt to determine the predominant feedback processes in each of the stellar mass bins. To investigate this, we use the \textsc{Simba} suite of cosmological MHD simulations \citep[][]{2019MNRAS.486.2827D}, as they contain a combination of simulations with various feedback mechanisms involved. We refrain from using simulations like FLAMINGO \citep[][]{2023MNRAS.526.4978S}, as they have much larger box sizes ($\sim1\,$Gpc) with lower mass resolution ($\sim 10^{8-9}\,M_\odot$) that are more suited for cosmological studies. We primarily compare our observations to the \textsc{Simba}$-50$ simulation that has the following feedback mechanisms: stellar, AGN winds, AGN jets, AGN X-ray emission; the \textsc{Simba}$-$nojet that has only stellar feedback and radiative AGN winds; and the \textsc{Simba}$-$nofb, where all feedback mechanisms are turned off. These \textsc{Simba} feedback variant simulations are limited in length at $50\,h^{-1}\mathrm{Mpc}$. Therefore,  we mimic our observations by simulating an FRB intersecting galaxies along a distance of $\sim150$\,Mpc, which roughly corresponds to traveling through two simulation boxes. A proper treatment of this would require ray-tracing through a box that has periodic boundary conditions. However, we only conduct a qualitative comparison so instead we randomly select two pixels and add their DM values to imitate the effect of passing through the simulation volume twice. Furthermore, in our comparisons we decide to exclude the contact term dominated innermost radial bin in the observations, as in the simulations we assume that all FRBs originate background to galaxy halos. }

We present an overlay of the mean DM profiles as a function of impact parameter for galaxies in Galaxy Sample~B from the various \textsc{Simba} variants in Figure \ref{fig:simba_full}. Our full stacked Galaxy Sample B DM profile qualitatively aligns most closely with the DM profile in the \textsc{Simba}$-$nojet simulation. However, we note that in the individual stellar mass bins, the observations show more deviation from the simulation, particularly, in our most statistically significant stellar mass bin, Bin~2. This suggests that while some simulation variants are able to capture physics globally across galaxies of various stellar masses, they might not be able to do so at certain stellar masses. However, we note that more FRBs are needed to further explore these mass-dependent deviations. 

\begin{figure}
    \centering
    \includegraphics[width=0.47\textwidth]{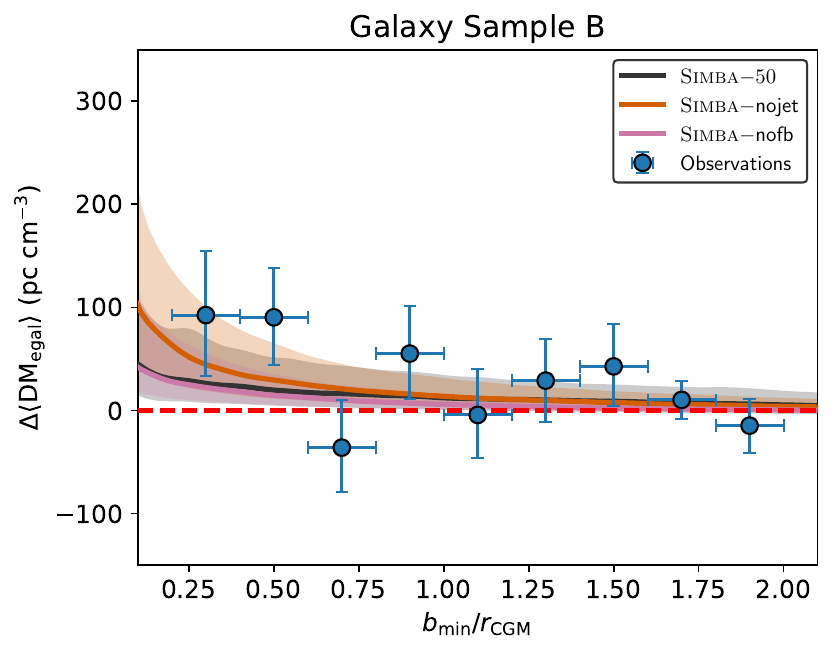}
    \caption{Overlays of simulated DM profiles for CGMs from the \textsc{Simba} simulation suite for Galaxy Sample B. We present the simulated mean DMs for \textsc{Simba}$-50$, \textsc{Simba}$-$nojet, \textsc{Simba}$-$nofb.}
    \label{fig:simba_full}
\end{figure}

\subsection{Comparison to rotation measure studies of CGMs}
There have been several works that have attempted to study the hot CGM through tracing the magnetoionic plasma using the Faraday rotation measure (RM) of background polarized radio sources \citep[e.g.,][]{2008Natur.454..302B, 2014ApJ...795...63F}, including FRBs \citep[e.g.,][]{2019Sci...366..231P, 2026A&A...706A..11K}. However, there have largely been limited studies that attempt to study the radial correlation between RMs and impact parameters. Using $\sim 1000$ RMs, \citet[][]{2020MNRAS.496.3142L} placed upper limits on the magnitude of the CGM contribution to background RMs as a function of impact parameter. Following this, \citet[][]{2023A&A...670L..23H} used 2461 RMs from the LOw Frequency ARray \citep[][]{2013A&A...556A...2V} to study the CGM and found no significant excess in the RM at low impact parameters. However, they see an excess of $\sim 8\,\mathrm{rad~m}^{-2}$ at $b < 200\,$kpc, for RMs along the minor axes of galactic disks, followed by a steep drop off at higher impact parameters. In addition to this, \citet[][]{2026ApJ...997..214K}, attempted to radially map the CGMs of cluster member galaxies in the nearby galaxy cluster Abell 3581 but do not see any correlation. Most recently, \citet[][]{2026A&A...710L..32M} used 2483 RMs, some of which were cross-matched with \ion{Mg}{2}, and find a $4.5\sigma$ RM excess at $z \sim 1$. 

Overall, very few statistical RM studies have resolved a clear radial dependence for the CGM contribution. \citet[][]{2023A&A...670L..23H} remains a notable exception, finding that the excess is strictly confined to $b < 200\,$kpc ($\lesssim 1\,r_{\mathrm{CGM}}$). This trend is similar to what we observe, with DM excesses at $b < 1r_\mathrm{CGM}$ (in Bins~1 and 2) that steeply fall off to zero at higher impact parameters. However, future FRB DM and RM studies of the CGM will be able to further illuminate any trend between CGM DM and RM profiles.

\subsection{Comparison to previous FRB studies of the CGM}
\label{sec:comparison}
The first study to attempt to probe the inner CGM using FRBs, as we have in this work, was conducted by \citet[][]{2022NatAs...6.1035C} who used 474 FRBs from CHIME/FRB's First Catalog to find 26 FRBs intersecting galaxies with impact parameters less than 200 kpc, and found a DM excess of $>90\,$pc~cm$^{-3}$ with a p-value of $<0.01$. This study was limited by the large uncertainty regions of the available intensity catalog localizations, which have a median error in impact parameter of $\sim 100$\, kpc (see Figure \ref{fig:1}); therefore, it is possible that some fraction of their FRBs were incorrectly associated with foreground halos, making their results prone to being skewed by a small number of outliers. Following this, \citet[][]{2023ApJ...945...87W} showed that the work of \citet[][]{2022NatAs...6.1035C} is indeed heavily affected by two high DM outliers that artificially raise the significance of their results. In comparison to such works, our results use a sample of FRBs that is four times larger and also has localization uncertainties that have 500$\times$ smaller sky area. Due to these improvements in our sample, we are able to be more certain of FRB-galaxy intersections. Using this sample of FRB-galaxy intersections, which is less contaminated by false positives, we obtain the highest statistical indication for a DM excess due to the CGM with FRBs to date, by having detected a DM excess that is $119\pm 44\,$pc~cm$^{-3}$, representing a p-value of 0.0053. 

The only studies to attempt to radially map the DM contribution of the inner CGM were conducted by \citet[][]{2023ApJ...945...87W} and \citet[][]{2026arXiv260223749K}. These studies only used the less precise localizations from CHIME/FRB intensity catalogs with fewer FRBs; particularly, \citet[][]{2023ApJ...945...87W} used bursts from the CHIME/FRB Intensity Catalog 1 and obtained a marginal ($<2\sigma$) detection of a radial decline in galaxies in the halo mass range of ($10^{11-13}\,M_\odot$) using only 26 FRBs. \citet[][]{2026arXiv260223749K} attempted the first radial study of the inner CGM of a single galaxy using 171 FRBs but find a low significance with a p-value of 0.25. We improve on these results by obtaining the first constraints on the electron density distribution in the inner CGM using $>600$ FRBs with sub-arcminute localization errors, by stacking over many galaxy halos and obtain the strongest statistical indication of a radially declining DM profile in galaxy halos, with a p-value of 0.0046 across our global sample.


This work places some of the first statistical constraints on the baryon distribution in the inner CGM ($<200\,$kpc) using a large sample of FRBs. \citet[][]{2026PhRvL.137d1001W} produce an almost entirely complementary measurement of baryons using CHIME/FRB Intensity Catalog 2 \citep{2026ApJS..283...34C}, constraining the baryon distributions at physical scales of $>200$ kpc (corresponding to the smallest angular scales in their lowest redshift bin). Following this work, \citet[][]{McCarty_2026_arXiv}, have also produced a similar measurements of baryons using 130 localized FRBs. In combination, the measurements from those works and the ones from this work have provided some of the first constraints on baryons in the cosmic web using a large sample of FRBs across all spatial scales, as displayed in Figure \ref{fig:cosmic_web}. Additionally, this work is also the first to detect and place constraints on the `contact term' predicted by \citet{2026PhRvL.137d1001W} due to low-$z$ FRBs by deriving the DM-galaxy cross-correlation function at a high significance.

\begin{figure}
    \centering
    \includegraphics[width=0.49\textwidth]{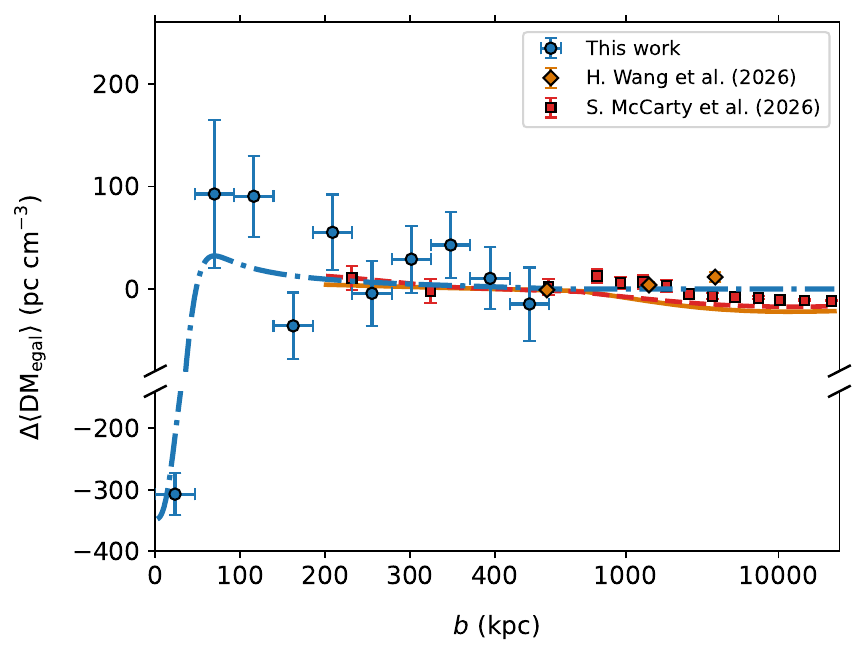}
    \caption{The DM excess as a function of impact parameter, as measured in this work, with comparisons to the data (presented as points) and model fits (presented as lines) from \citet{2026PhRvL.137d1001W} and \citet[][]{McCarty_2026_arXiv}. Here, we only display their models to the minimum spatial scale of $\sim 200\,$kpc that can be resolved by their data. We also apply the same background correction as ours for their data and models by subtracting the mean DM in the range of $1.5r_\mathrm{vir}-2r_\mathrm{vir}$ ($350-460\,$kpc) for MW-like halos.}
    \label{fig:cosmic_web}
\end{figure}

In Appendix \ref{sec:forecast}, we conduct a detailed forecast of a future CGM study with a catalog of 10000 FRBs with baseband localizations. We simulate the CGM contributions to the DM from models of galaxy halos \citep[][]{2025MNRAS.543.2649O}, where we assume that the baryon distribution follows the underlying dark matter distribution. We cross-match this catalog of FRBs with our galaxy samples and find that it results in a $\sim5\sigma$ detection ($p<10^{-7}$) of the CGM impact to a background FRB. CHIME/FRB is expected to obtain this large sample of baseband-localized bursts in the coming years, so similar future stacking studies with baseband data are likely to find much stronger statistical evidence of the hot halo of galaxies.

\subsection{{Ideal redshift FRBs for studying intervening CGMs}}
Upcoming FRB survey instruments (like CHORD, the DSA and the SKA), will have increased detection rates and will be more sensitive than CHIME/FRB, thus detecting many FRBs from much larger redshifts \citep[][]{2009IEEEP..97.1482D, 2019BAAS...51g.255H, 2019clrp.2020...28V}. However, at higher redshifts, there is larger DM contamination due to the IGM; thus, we investigate the ideal redshift FRBs to maximize our constraining potential on the CGM. Below, we describe a simulation in which we estimate the fractionl DM contribution of the CGM to an FRB's DM as a function of redshift of the FRB's host. 


The steps of this simulation are as follows. First, we begin by placing an FRB source at a given redshift. Then, we populate the LOS along this FRB with galaxy halos, drawing them from the halo mass function of \citet[][]{2013MNRAS.433.1230W}, where the number of halos in a given comoving volume cell and mass interval is given by:
\begin{equation}
    N_{\mathrm{halo}} = \int_{z_1}^{z_2} \frac{dV_\mathrm{comoving}}{dz}dz \int_{\ln(M_1)}^{\ln(M_2)} \frac{dn}{d\ln (M)}d\ln (M), 
    \label{eq:big_theory}
\end{equation}
where $dn/d\ln (M)$ is the halo mass function, $M_1$ and $M_2$ are the minimum and maximum halo masses (in solar masses), $z_1$ and $z_2$ are the minimum and maximum redshifts of the cell, and $dV_\mathrm{comoving}/dz$ is the comoving volume element. The minimum halo mass that we explore is $10^{10}M_\odot$ and the maximum halo mass is $10^{15}M_\odot$, and the minimum and maximum redshift that we consider are 0.005 and 2.5, respectively. We divide our masses into 1000 logarithmically spaced bins and work in redshift slices of 0.005. For each redshift slice in the sightline of the FRB, we consider a volume cell that has a radial size equivalent to three times the radius of a $10^{15}M_\odot$ dark matter halo, assuming spherical collapse using the semi-empirical fits of \citet[][]{1998ApJ...495...80B}:

\begin{equation}
    r_\mathrm{vir} = \left(\frac{3M_\mathrm{halo}}{4\pi \Delta_c\rho_c(z)}\right)^{1/3}, 
\end{equation}
where $\rho_c(z)$ is the critical density of the Universe at that redshift, and $\Delta_c$ is given by:
\begin{align}
    \Delta_c = 18\pi^2 + 82x - 39x^2, 
\end{align}
where $x = \Omega_m(z) - 1$, with $\Omega_m(z)$ being the cosmological matter density parameter. 

We then randomly populate each volume cell with galaxy halos, probabilistically drawing the number of galaxies in each mass bin from Equation~\ref{eq:big_theory}. We assume the galaxy halo electron densities to follow Model~1 of the Descriptive Parameter Halo (DPM) models introduced by \citet[][]{2025MNRAS.543.2649O}. Their Model~1 assumes that the baryons perfectly trace the underlying dark matter distribution without any evacuation of gas due to feedback; we choose to use this to demonstrate the maximum theoretical detectability for future surveys. We also explicitly avoid using static single-halo models (such as models of the MW halo; e.g., YT2020) as these are calibrated to a single halo. Our simulated sightlines must integrate DMs across a variety of halos masses and redshifts; thus, it is challenging to properly scale these models with mass. Due to these constraints, we require a parametric framework such as DPM halo.


Then, we integrate the electron density along the LOS using Equation~\ref{eq:dm} to obtain $\mathrm{DM_{CGM}}$ for the FRB sightline. We repeat this by placing an FRB at the edge of each redshift cell to build $\mathrm{DM_{CGM}}$ as a function of redshift. We redo all the above steps for $10^4$ iterations to obtain a probability density distribution of $\mathrm{DM_{CGM}}$ as a function of redshift. We compute the fractional contribution of the CGM to the cosmic DM (the IGM contribution, along with the contribution of intervening halos) from \citet[][]{2025NatAs...9.1226C} as follows:
\begin{equation}
    \eta_\mathrm{CGM} = \frac{\mathrm{DM_{CGM}}}{\mathrm{DM_{cosmic}}}.
\end{equation}
We then calculate the joint probability distribution function for $\eta_\mathrm{CGM}$ as:
\begin{align}
\mathcal{P}(\eta_\mathrm{CGM}|z)=  \int \mathrm{DM_{cosmic}} &\mathcal{P}(\eta_\mathrm{CGM}\mathrm{DM_{cosmic}}|z)\\ & \nonumber \mathcal{P}(\mathrm{DM_{cosmic}}|z) d \mathrm{DM_{cosmic}}.
\end{align}
We note here that we compute this joint probability distribution under the assumption that $\mathrm{DM_{cosmic}}$ and $\mathrm{DM_{CGM}}$ are independent of each other. Because these quantities are drawn from two different simulations, our approach assumes that they are statistically independent from each other. A sightline that intersects a massive halo in one of the simulations might pass through a void in the other, uncoupling the two quantities. We do acknowledge that a real world sightline passing through a halo would experience an increase in the DM of both quantities. Nevertheless, applying this simplifying assumption allows us to get a statistical estimate of the CGM's fractional contribution to the DM.  

We present the probability density distribution for $\eta_\mathrm{CGM}$ in Figure~\ref{fig:f_cgm}. We note that there are regions where $\eta_\mathrm{CGM} > 1$ has a non-zero probability. This is because the two quantities are drawn from two different simulations and Poisson noise can cause the CGM contribution to the DM to be more than the cosmic DM obtained by \citet[][]{2025NatAs...9.1226C}. However, the probabilities where $f_\mathrm{CGM} > 1$ are comparatively low, confirming the validity of our approach and simulation. At very low redshifts, we note that $\eta_\mathrm{CGM}$ is incredibly low; this is likely because the FRB is so close that it has only passed through very few galaxy halos and therefore the IGM dominates its total DM. At intermediate redshifts, we see that $\mathrm{DM_{CGM}}$ begins to dominate; here, there are sufficiently many galaxies and we are at a low enough cosmic distance that the IGM contribution is relatively low. However at $z \gtrsim 0.75$, we see that the IGM contribution begins to be more prevalent and $\eta_\mathrm{CGM}$ steadily declines to $f_\mathrm{CGM} \lesssim 0.4$ at FRB redshifts of 2.5. Beyond this redshift, we expect the IGM to dominate even more increasingly as there are fewer and fewer massive galaxy halos. 

Thus, we conclude that surveys hoping to target the CGMs of intervening galaxies should prioritize FRBs at redshifts of $0.4 \lesssim z_\mathrm{FRB}\lesssim 0.75$ to study the CGMs of intervening galaxies, as this is the region where the DM contribution of the CGM is likely to be maximized. To date, the highest redshift FRB localized to a host galaxy has been at a redshift of 2.148 \citep[][]{2025arXiv250801648C}. We expect that this sample of high-redshift FRBs will expand greatly with the increased sensitivity of CHORD, the DSA, and the SKA. Due to this, it will be more challenging to conduct a similar stacking of the CGM using FRBs unlocalized to host galaxies with these observatories, as the odds of having a high-redshift FRB with minimal $\mathrm{DM_{CGM}}$ increases. However, this redshift range is where CHIME/FRB detects most of its FRBs. In addition to this, the CHIME/FRB system will soon undergo a hardware upgrade that will increase the event rate by a factor of 2-5. Thus, CHIME/FRB will continue to be a uniquely well-suited instrument for studying foreground galaxy halos.

\begin{figure}[!htb]
    \centering
    \includegraphics[width=0.47\textwidth]{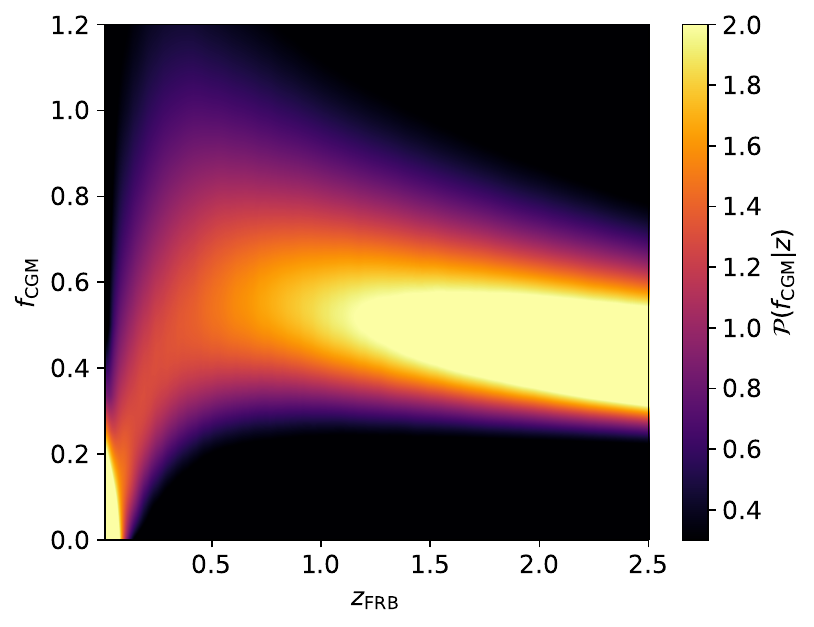}
    \caption{The probability density of the fractional contribution to the cosmic DM from intervening galaxy halos.}
    \label{fig:f_cgm}
\end{figure}
\section{Conclusion}
We have conducted a statistical stacking study of the circumgalactic media (CGM) of galaxies with $D_L$ less than 150\,Mpc using the DMs of more than 1600 unique fast radio burst (FRB) sources from the CHIME/FRB Baseband Catalog~2 that have 1-$\sigma$ localization errors on the order of $\sim 10^{\prime\prime}$. The results of this work are summarized as follows: 

\begin{enumerate}
    \item We cross-matched our FRBs with the positions of 1592 nearby ($1\mathrm{\,Mpc}\leq D_L \leq 50$\,Mpc) massive ($10\leq \log(M_*/M_\odot) \leq 11.5$) galaxies and found 66 FRBs intersecting halos within 1 CGM radius ($r_\mathrm{CGM}$), with tentative evidence of an excess DM of $119\pm 44$\,pc~cm$^{-3}$ in this piercing population compared to the population of FRBs not found to be piercing galaxies, with a p-value of 0.0053 ($\sim 2.6 \sigma$), which was measured using a non-parametric permutation test, representing the highest statistical indication for a DM excess due to the CGM with FRBs to date. 
    
    \item We quantified the fraction of this excess that might arise from the environments of the galaxy CGMs (such as the intracluster medium or cosmic filaments) by using simulated DMs from the SLOW cosmological simulation \citep[][]{2023A&A...677A.169D}, which is a constrained magnetohydrodynamic simulation. We found that a DM excess greater than the observation only occurs in 0.62\% of $10^5$ random realizations. However, we have measured that, on average, the environment of galaxies contributes 14\,pc~cm$^{-3}$ to the DM of background FRBs, with a standard deviation of 39\,pc~cm$^{-3}$. 
    
    \item We attempted to radially map the DM profile (the mean DM as a function of normalized impact parameter) in the stacked CGM of many galaxies by cross-matching FRB positions with intervening galaxies that have $1\mathrm{\,Mpc}\leq D_L\leq 150$\,Mpc and $9.5\leq\log(M_*/M_\odot)\leq 11.5$ and found a global trend of a radial decline in DM as a function of impact parameter with a p-value of 0.0046 ($\sim 2.6\sigma$), most of which is driven by $L^*$ galaxies in the range $10\leq \log(M_*/M_\odot) \leq 11$. We also detect a significant DM deficit (deemed the ``contact term'') at low impact parameters likely due to cross-matching FRBs with their own host galaxies.

    \item{By developing an analytical DM-galaxy cross-correlation model, we demonstrated that the observed deficit can be reproduced with $\sim20$ FRBs contribute to the contact term. This is the same number of FRBs in our catalog that exist at impact parameters of $b/r_\mathrm{CGM} < 0.2$. These numbers also indicate that CHIME/FRB might have a higher fraction of local ($D_L < 150$\,Mpc) FRBs than previously thought.}
    
    \item We used the observed DM profiles to constrain $\beta$-models for the electron density, finding the central electron density to be in the range of $(1.7-21.1)\times 10^{-3}\mathrm{\,cm}^{-3}$, with a power-law index of $\beta = 0.68^{+0.15}_{-0.16}$ and a fractional core radius of $0.07^{+0.05}_{-0.04}$. Our model parameters are consistent within error with those found from X-ray stacking studies \citep[][]{2024A&A...690A.267Z} for MW-like galaxies. 
    
    \item We forecasted a catalog of FRBs with 10,000 bursts with CHIME/FRB baseband localizations and determined that it is sufficient to provide a $\sim5$-$\sigma$ detection of the CGMs of intervening halos, with the expectation of $\sim 400$ FRB intersections with foreground galaxies. 
    \item We produced a probability density distribution for the expected theoretical contribution to the DM from intervening halos as a function of redshift. We compared this to the total cosmic DM-redshift probability density distribution from \citet[][]{2025NatAs...9.1226C}, and found that the CGM's fractional contribution to a background FRB's DM is maximized for FRB's in the redshift range $0.4\lesssim z_\mathrm{FRB}\lesssim0.75$. 
\end{enumerate}
In all, FRBs present an exciting opportunity to serve as precision probes of the distribution, content and properties of the baryonic matter in the Universe, particularly with the upcoming observatories that are expected to detect tens of thousands of FRBs \citep[][]{2009IEEEP..97.1482D, 2019clrp.2020...28V, 2019BAAS...51g.255H}. In this study, we have used the largest catalog of well-localized FRBs to date to demonstrate the constraining power of a large statistical sample of FRBs in studying the baryonic matter of the Universe down to the scale of galaxy halos.

\begin{acknowledgments}
We acknowledge that CHIME is located on the traditional, ancestral, and unceded territory of the Syilx/Okanagan people. We are grateful to the staff of the Dominion Radio Astrophysical Observatory, which is operated by the National Research Council of Canada. CHIME is funded by a grant from the Canada Foundation for Innovation (CFI) 2012 Leading Edge Fund (Project 31170) and by contributions from the provinces of British Columbia, Québec and Ontario. The CHIME/FRB Project is funded by a grant from the CFI 2015 Innovation Fund (Project 33213) and by contributions from the provinces of British Columbia and Québec, and by the Dunlap Institute for Astronomy and Astrophysics at the University of Toronto. Additional support was provided by the Canadian Institute for Advanced Research (CIFAR), McGill University and the Trottier Space Institute at McGill thanks to the Trottier Family Foundation, and the University of British Columbia. The baseband recording system for CHIME/FRB is funded in part by a CFI John R. Evans Leaders Fund award to IHS. This research was enabled in part by software provided by the Digital Research Alliance of Canada (\href{https://alliancecan.ca/}{alliance.can.ca}). 

A.K. acknowledges the support of a Bourses de ma\^{\i}trise en recherche grant (`2002729' or \url{https://doi.org/10.69777/2002729}) from the Fonds de recherche du Qu\'ebec. The AstroFlash research group at McGill University, University of Amsterdam, ASTRON, and JIVE is supported by: a Canada Excellence Research Chair in Transient Astrophysics (CERC-2022-00009); an Advanced Grant from the European Research Council (ERC) under the European Union's Horizon 2020 research and innovation programme (`EuroFlash'; Grant agreement No. 101098079); an NWO-Vici grant (`AstroFlash'; VI.C.192.045); an NSERC Discovery Grant (RGPIN-2025-06681); an ERC Starting Grant (`EnviroFlash'; Grant agreement No. 101223057); and an NWO-Veni grant (VI.Veni.222.295). A.P. is a Trottier Space Institute postdoctoral fellow and a Fonds de recherche du Québec - Nature et technologies (FRQNT) postdoctoral fellow. C.L. acknowledges support from the Miller Institute for Basic Research at UC Berkeley. L.M.B. is partially supported by NASA through grant 80NSSC24K0173 and NSF through grant AST-2510951. A.M.C. is a Banting Postdoctoral Fellow. A.P.C. is a Canadian SKA Scientist and is funded by the Government of Canada / est financé par le gouvernement du Canada. K.D. acknowledges support by the Deutsche Forschungsgemeinschaft (DFG, German Research Foundation) under Germany’s Excellence Strategy - EXC-2094 - 390783311. The SLOW simulation was supported by the grant agreements ANR-21-CE31-0019 / 490702358 from 
the French Agence Nationale de la Recherche / DFG for the LOCALIZATION project and carried out at the 
Leibniz Supercomputer Center (LRZ) under the project pn68na. V M.K. holds the Lorne Trottier Chair in Astrophysics \& Cosmology, a Distinguished James McGill Professorship, and receives support from an NSERC Discovery grant (RGPIN 228738-13). K.W.M. is supported by NSF Grant No. 2510771. D.M.~acknowledges support from the French government under the France 2030 investment plan, as part of the Initiative d'Excellence d'Aix-Marseille Universit\'e -- A*MIDEX (AMX-23-CEI-088). M.N. is a Fonds de Recherche du Quebec - Nature et Technologies (FRQNT) postdoctoral fellow. A.B.P.~acknowledges support by NASA through the NASA Hubble Fellowship grant \mbox{HST-HF2-51584.001-A} awarded by the Space Telescope Science Institute, which is operated by the Association of Universities for Research in Astronomy, Inc., under NASA contract \mbox{NAS5-26555}. A.B.P.~also acknowledges prior support from a Banting Fellowship, a McGill Space Institute~(MSI) Fellowship, and a Fonds de Recherche du \mbox{Qu\'ebec -- Nature} et Technologies~(FRQNT) Postdoctoral Fellowship. A.Pd.~acknowledges support from the French government under the France 2030 investment plan, as part of the Initiative d'Excellence d'Aix-Marseille Universit\'e -- A*MIDEX (AMX-23-CEI-088). M.W.S. acknowledges support from the French government under the France 2030 investment plan, as part of the Initiative d’Excellence d’Aix-Marseille Université - AMIDEX (AMX-23-CEI- 088). P.S. acknowledges the support of an NSERC Discovery Grant (RGPIN-2024-06266). J.S. is supported by the University of Lille via the Welcoming Internationals to Lille initiative for the UNIVERSITWINS project.  

\end{acknowledgments}

\appendix
\counterwithin{figure}{section}
\counterwithin{table}{section}

\setcounter{figure}{0}
\renewcommand{\thefigure}{\thesection\arabic{figure}}

\setcounter{table}{0}

\setcounter{equation}{0}

\section{Unbinned DM scatter plots }
\label{app:dmvb}

Figure~\ref{fig:dmvb} displays the extragalactic DMs as a function of fractional minimal impact parameter for the 628 FRBs that intersect halos within $2r_\mathrm{CGM}$ for the full Galaxy Sample B.  We display the unbinned extragalctic DMs as a function of impact parameter in Figure~\ref{fig:raw_dm}.
\begin{figure}[!htb]
    \centering
    \includegraphics[width=0.47\textwidth]{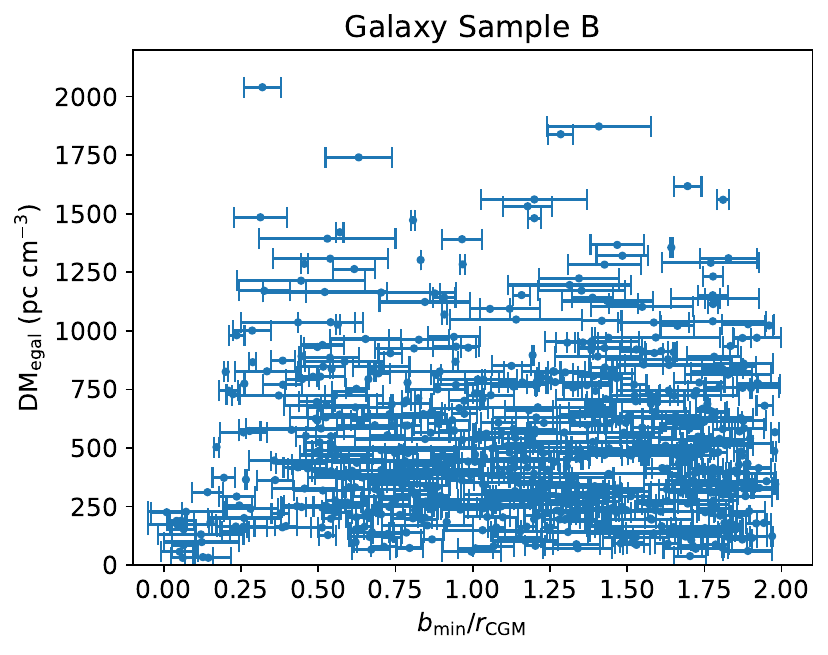}
    \caption{The extragalactic DMs as a function of fractional minimal impact parameter for FRBs cross-matched with foreground galaxies from Galaxy Sample B.}
    \label{fig:dmvb}
\end{figure}
\begin{figure*}[!htb]
    \includegraphics[width = 0.47\textwidth]{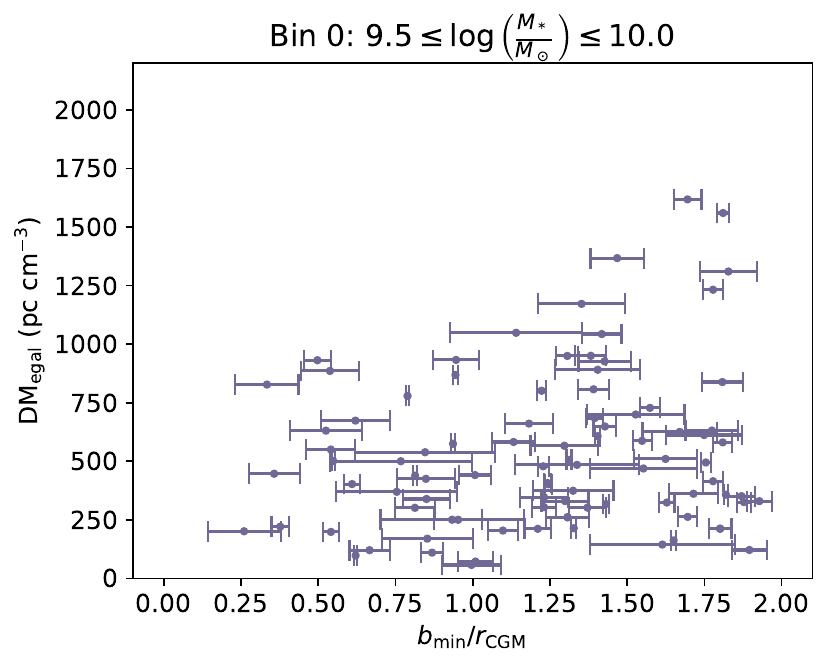} \includegraphics[width = 0.47\textwidth]{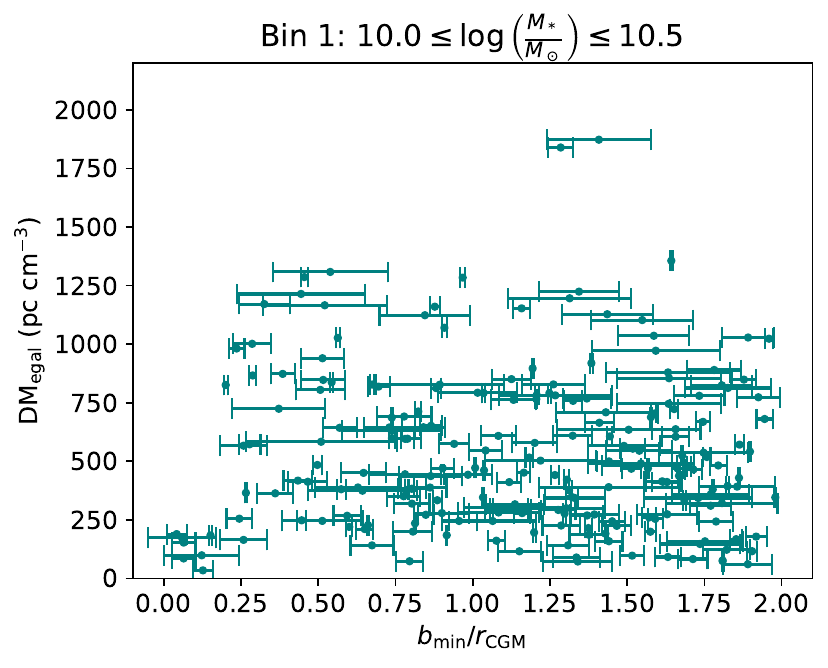}\\
    \includegraphics[width = 0.47\textwidth]{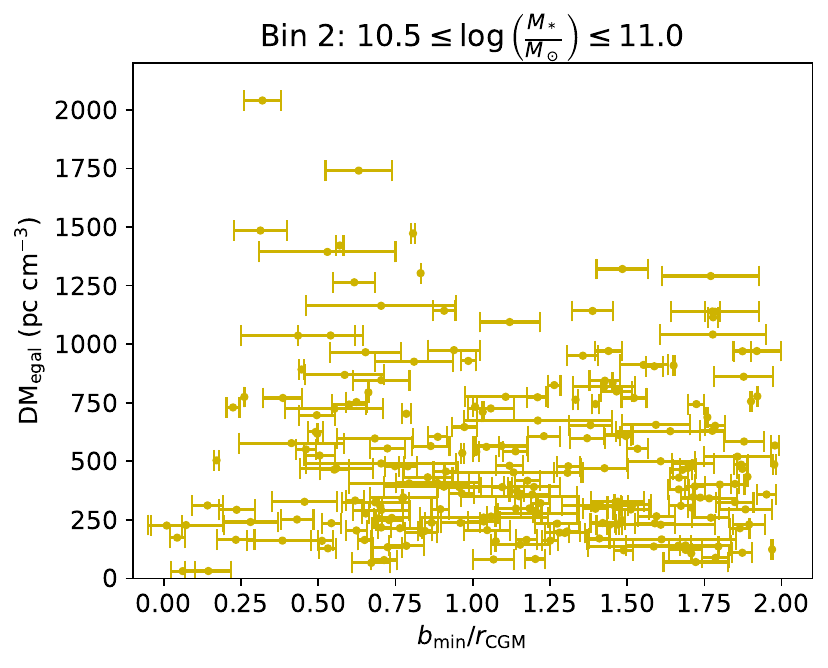} \includegraphics[width = 0.47\textwidth]{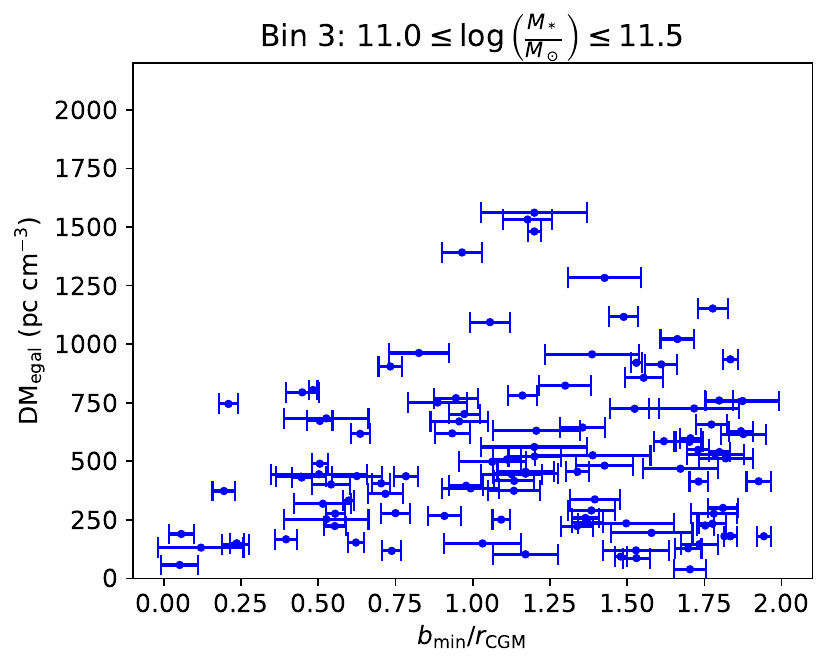}
    \caption{The extragalactic DM as a function of $b_\mathrm{min}/r_\mathrm{CGM}$ for mass Bins 0 to 3. In ascending order of bin index, the number of FRBs per bin is 91, 210, 217 and 110. }
    \label{fig:raw_dm}
\end{figure*}

\section{Weighted mean DM per radial bin}
\label{app:weighted_mean}

In the following, we describe, in full, our procedure to obtain the weighted mean. The weight of an FRB to the bin is determined by the probability that the FRB lies within the bin, which we term the `membership probability'. We calculate the membership probability assuming that the impact parameter of the FRB has a Gaussian probability distribution, with the mean being the impact parameter and the standard deviation being given by the positional uncertainty of the FRB. Under these assumptions, the membership probability for the $i-$th FRB in the bin is given by: 
\begin{align}
    p_i = \Phi\left(\frac{x_\mathrm{right} - x_i}{x_i^\mathrm{ err}}\right) - \Phi\left(\frac{x_\mathrm{left} - x_i}{x^\mathrm{ err}_i}\right),
\end{align}
where $x = b_{\mathrm{min}}/r_\mathrm{CGM}$ is the fractional minimal impact parameter, $\Phi$ is the CDF of the normal distribution, $x_\mathrm{right}$ is the fractional minimal impact parameter of the right-most data point in the bin, $x_\mathrm{left}$ is the fractional minimal impact parameter of the left-most data point, $x_i$ is the fractional minimal impact parameter of the $i-$th FRB, and $x^\mathrm{err}_i$ is the error in the fractional minimal impact parameter of the $i-$th FRB. For any given FRB within a particular bin, $0 < p_i < 1$ because the full probability distribution cannot be contained within a single bin, and it is bounded below because the FRB is only included in the bin because its 1-$\sigma$ error is within the boundaries of the bin. Then, the weighted mean DM per bin is computed to be: 
\begin{align}
    \langle \mathrm{DM_{egal}}\rangle_\mathrm{bin} = \frac{\sum_i p_i \mathrm{DM_{egal,}}_{i}}{W_1},
\end{align}
where $W_1 = \sum_{i}p_i$ is the sum of the membership probabilities. 


In addition to the mean, we calculate the standard error in each bin using the unbiased weighted standard deviation \citep[e.g.,][]{cochran1977sampling} in each bin,  $s^{\mathrm{DM_{egal}}}_\mathrm{bin}$, which we compute as:
\begin{align}
    s^{\mathrm{DM_{egal}}}_\mathrm{bin} = \sqrt{\frac{W_1}{W_1^2 - W_2} \sum_{i}p_i (\mathrm{DM_{egal,}}_{i} -\langle \mathrm{DM_{egal}}\rangle_\mathrm{bin})^2},
\end{align}
where $W_2 = \sum_i p_i^2$ is the sum of the squared membership probabilities. We note that in the limit that all data points in the bin are weighted equally, i.e., $p_i=1$ for all $i$, $s^{\mathrm{DM_{egal}}}_\mathrm{bin}$ reduces to Bessel's correction to the standard deviation \citep[][]{bessel}.  In addition to this, we calculate the effective number of data points in each bin to be $N^\mathrm{eff}_\mathrm{bin} = W_1^2/W_2$. Then, finally the standard error in each bin is computed to be:
\begin{align}
    \sigma^{\mathrm{DM_{egal}}}_\mathrm{bin} = \frac{s^\mathrm{DM_{egal}}_\mathrm{bin}}{\sqrt{N^\mathrm{eff}_\mathrm{bin}}}.
\end{align}

\section{The Dispersion Measure-Galaxy Cross-Correlation Function}
\label{app:cross}
We note that our observable $\Delta \langle \mathrm{DM_{egal}}(b)\rangle$ is a proxy of the underlying DM-galaxy cross-correlation function. Due to this, we derive the relationship between our observable and this underlying cross-correlation function. To do so, we first derive the cross-correlation function, which has two major components: the dispersion measure contribution from the CGM of intervening galaxies to the FRB, and the contact term due to contamination of FRBs that have host galaxies within our galaxy sample. Because of these two major contributions, we begin from the approach of \citet[][]{2026PhRvL.137d1001W}, who derived the DM-galaxy angular cross power spectrum and convert their models to real space by taking Hankel transforms \citep[][]{hankel1869cylinderfunctionen} to convert from spherical harmonic space to real space. The DM-galaxy angular cross power spectrum is of the form: 
\begin{align}
    C_\ell ^{dg} (z_g) = C_\ell^{eg}(z_g) + C_\ell^{fg}(z_g), 
\end{align}
where $\ell$ is the spherical harmonic, $z_g$ is the redshift of the intervening galaxy, $C_\ell^{eg}$ is the electron-galaxy angular cross power spectrum, and $C_\ell^{fg}$ is the FRB-galaxy angular cross power spectrum representing the contact term. Thus, the corresponding DM-galaxy cross-correlation function is of the form: 
\begin{align}
    \xi_\theta^{dg}(z_g) = \xi_\theta^{eg}(z_g) + \xi_\theta^{fg}(z_g),
\end{align}
where $\xi_\theta^{eg}$ and $\xi_\theta^{fg}$ are the cross-correlation functions corresponding to $C_\ell^{eg}$ and $C_\ell^{fg}$, respectively. We note that we can do so since the zeroth order Hankel transform is linear in nature. 

Thus, we begin by deriving $\xi_\theta^{eg}$, by taking the Hankel transform of $C_\ell^{eg}$:
\begin{align}
    \xi_\theta^{eg}(z_g) = \int_0^{\infty} d\ell \frac{\ell}{2\pi} C_\ell^{eg}(z_g) J_0(\ell\theta), 
\end{align}
where $J_0(\ell \theta)$ is the zeroth order Bessel function of the first kind \citep[][]{abramowitz1948handbook}. We now write out $C_\ell^{eg}(z_g)$ using the form derived by \citet[][]{2026PhRvL.137d1001W}: 
\begin{align}
    \xi_\theta^{eg}(z_g) = \int_0^{\infty}d\ell~\frac{\ell}{2\pi} \frac{\bar{n_e}(z_g)}{(1+z_g)^2} \frac{f_f(z_g)}{\chi^2_g}P_{eg}(k = \ell/\chi_g, z_g)J_0(\ell \theta), 
\end{align}
where $\bar{n_e}(z_g)$ is the average electron density at the galaxy's redshift, $f_f(z_g)$ is the fraction of FRBs background to the galaxy (i.e., $z_\mathrm{FRB} > z_g$) obtained from the inferred redshift distribution of FRBs (\citealp{2023ApJ...944..105S}; Jain et al. 2026, submitted), $P_{eg}$ is the electron-galaxy cross power spectrum, and $\chi_g$ is the comoving distance to the galaxy. Working under Limber's approximation \citep[][]{1953ApJ...117..134L}, we note that $\ell \approx k\chi_g$, and we transform from spherical harmonic space to wavenumber space: 
\begin{align}
    \xi_\theta^{eg}(z_g) = \int_0^\infty dk~\frac{k}{2\pi} \frac{\bar{n_e}(z_g)}{(1+z_g)^2} f_f(z_g) P_{eg}(k, z_g) J_0(k\chi_g\theta).
\end{align}
We note that $f_f(z_g)$ and $(1+z_g)^2$ are independent of wave number so we can write this as: 
\begin{align}
    \xi_\theta^{eg}(z_g)= \frac{f_f(z_g)}{(1+z_g)^2} \int_0^\infty dk ~ \frac{k\bar{n_e}(z_g)}{2\pi} P_{eg}(k, z_g) J_0 (k\chi_g \theta). 
\end{align}
Now, to convert this integral from Fourier space to real space, we once again apply Limber's approximation: 
\begin{align}
    \int_0^\infty dk~ \frac{k\bar{n_e}(z_g)}{2\pi} P_{eg}(k, z_g) J_0(k\chi_g\theta ) = \int_{-\infty}^\infty dl_g~\bar{n_e}(z_g) \xi^{\mathrm{CGM}} (\sqrt{b^2 + l_g^2}), 
\end{align}
where $l_g$ is the distance along the LOS from the center of the galaxy halo, $b$ is the impact parameter of the FRB sightline to the galaxy, and $\xi^{\mathrm{CGM}}$ is the cross-correlation function that describes the distribution of hot gas around the CGM. In particular, we note that multiplying the mean electron density at the redshift of the galaxy with the cross-correlation function describing the distribution of the hot plasma in the CGM simply provides us with the electron density distribution of the CGM: 
\begin{align}
    n_e^{\mathrm{CGM}} \left(\sqrt{b^2 + l_g^2}\right) = \bar{n_e}(z_g) \xi^\mathrm{CGM} \left(\sqrt{b^2 + l_g^2} \right). 
\end{align}
Thus, using this expression, we write down the electron-galaxy cross-correlation function as: 
\begin{align}
    \xi^{eg}_b(z_g) = \frac{f_f(z_g)}{(1+z_g)^2} \int_{-\infty}^{\infty} dl_g~ n_e^\mathrm{CGM} \left(\sqrt{b^2 + l_g^2}\right),
\end{align}
for an intervening galaxy at redshift $z_g$. 

Similarly, we also derive the FRB-galaxy cross-correlation function that quantifies the contact term, starting at the FRB-galaxy angular power spectrum
\begin{align}
    \xi_\theta^{fg} (z_g) &= \int_0^\infty d\ell~ \frac{\ell}{2\pi}C_\ell^{fg}(z_g)J_0(\ell \theta),
    \label{eq:c12}
\end{align}
where we once again use the formalism of \citet[][]{2026PhRvL.137d1001W} and assume that: 
\begin{align}
    C_\ell^{fg}(z_g) = \Delta \mathrm{DM}_g p_f(z_g) \frac{H(z_g)}{\chi_g^2} P_{fg}(k = \ell/\chi_g, z_g).
\end{align}
Here, $p_f(z_g)$ is the probability of an FRB existing at the galaxy's redshift obtained from the FRB redshift distribution (\citealp{2023ApJ...944..105S}; Jain et al. 2026, submitted), $H(z_g)$ is the Hubble expansion factor at the galaxy's redshift, $P_{fg}(k, z_g)$ is the FRB-galaxy cross-power spectrum, and $\Delta \mathrm{DM}_g$ is the DM deficit of an FRB originating at $z_g$ in comparison to the average FRBs in our catalog, which are mostly background to the galaxies in our sample and will therefore have higher DMs due to traveling through a greater path length in the IGM. Following, \citet[][]{2026PhRvL.137d1001W}, we model this as the difference between the expected DM contribution due to the IGM at $z_g$, along with expected host DM contribution, and the average DM of the background. \citet[][]{2026PhRvL.137d1001W} take this background DM to be the mean DM of all the FRBs in their sample; however, to be consistent with our analysis, we take this to be the average extragalactic DM of the FRB sightlines that have normalized impact parameters of $b/r_\mathrm{CGM} > 1.5$. Thus, our $\Delta \mathrm{DM}_g$ is given by:
\begin{align}
    \Delta \mathrm{DM}_g = \mathrm{DM_{IGM}} (z_g) + \frac{\mathrm{DM_{host}}}{1+z_g} - \langle \mathrm{DM_{egal}}(b/r_\mathrm{CGM} > 1.5)\rangle.
\end{align}
Once again, under Limber's approximation, we have that $\ell \approx k\xi_g$, so we write Equation \ref{eq:c12} using the wavenumber $k$ as follows: 
\begin{align}
\xi_\theta^{fg} (z_g) = \Delta \mathrm{DM}_gp_f(z_g) \int_0^\infty dk ~\frac{k}{2\pi} H(z_g)P_{fg}(k, z_g) J_0(k\chi_g\theta), 
\label{eq:c15}
\end{align}
where we have taken $\Delta \mathrm{DM}_g$ and $p_f(z_g)$ out of the integral, as they only depend on the galaxy redshift and not $k$. Here, we observe that the spatial separation between FRBs and their own host galaxies is effectively zero, so that the real-space analog of the FRB-galaxy cross power spectrum will appear to be a Dirac delta function \citep[][]{10.1098/rspa.1927.0012}. Due to this, $H(z_g)P_{fg}(k, z_g) = \kappa(z_g)$ will be constant in terms of $k$ and will only depend on $z_g$. However, this is only true in the limiting case of infinitely accurate FRB localizations. In the case of the BaseCat2 sample, the errors on the localizations of an FRB are on the order of 10s of arcseconds (CHIME/FRB Collaboration et al. 2026, in prep.), which translates to a physical transverse scale of $~\sim 23$\,kpc at the maximum luminosity distance of 150\,Mpc that we consider in our Galaxy Sample~B. This uncertainty in the location of the FRB can cause smearing of the contact term and cause potential contamination to the mean DM at radial bins at higher impact parameters. Therefore, we account for this smearing of the contact term by multiplying $\kappa(z_g)$ by a Gaussian window function: 
\begin{align}
    H(z_g)P_{fg}(k, z_g) = \kappa (z_g)e^{-k^2/2\sigma_k^2}.
\end{align}
Here, $\sigma_k$ represents the spread of the Gaussian window function and depends on the localization uncertainty of the FRB. Particularly, $\sigma_k = 1/\sigma_b$, where $\sigma_b$ is the uncertainty in the FRB's impact parameter, since wavenumbers are inversely proportional to physical scales. 

Using this form of the FRB-galaxy cross power spectrum, the integral in Equation \ref{eq:c15} can be written as:
\begin{align} 
    \int_0^\infty dk \frac{k}{2\pi} \kappa(z_g)e^{-k^2/2\sigma_k^2}J_0(k\chi_g\theta) = \frac{\kappa(z_g)}{2\pi \sigma_b^2}e^{-b^2/2\sigma_b^2},
\end{align}
since the zeroth order Hankel transform of a Gaussian function is Gaussian. Now, we attempt to model the relevant terms in $\kappa$. Note that $\kappa$ quantifies the probability that an FRB at a given redshift originates from a galaxy in the galaxy catalogs we use, and also within the stellar mass ranges that we limit our experiment to. Therefore, we cast this as follows: 
\begin{align}
    \kappa(z_g) = \mathcal{P}_{\mathrm{FRB\times gal}}(z_g) = f_\mathrm{overlap} \times f_\mathrm{completeness}(z_g) \times f_\mathrm{mass}(z_g),     
\end{align}
where $f_\mathrm{overlap}$ is the ratio of the overlap in the sky coverages of the footprint of the galaxy catalog to that of CHIME/FRB, $f_\mathrm{completeness}(z_g)$ quantifies the completeness of the galaxy catalog out to a redshift of $z_g$, and $f_\mathrm{mass}(z_g)$ is the fraction of FRBs that originate from galaxies of a certain stellar mass. We obtain the $f_\mathrm{overlap}$ from our FRB and galaxy catalogs; for $f_\mathrm{completeness}$, we take the redshift completeness fraction of \citet[][]{2020ApJ...895...32F}; and for $f_\mathrm{mass}$, we assume that FRBs trace star formation and stellar mass \citep[e.g.,][]{2023ApJ...954...80G, 2025ApJ...991...85L}, and that it is given by:
\begin{align}
    f_\mathrm{mass}(z_g) = \frac{\int_{M_\mathrm{min}}^{M_\mathrm{max}} dM_*~ M_* \phi_{z_g}(M_*)}{\int_0^{\infty} dM_*~ M_* \phi_{z_g}(M_*)},
\end{align}
where $M_\mathrm{min}$ and $M_\mathrm{max}$ correspond to the minimum and maximum stellar masses of galaxies in our sample, and $\phi_{z_g}(M_*)$ is the galaxy stellar mass function (GSMF), which we take be a double Schecter function. We use the best-fit parameters of \citet[][]{2012MNRAS.421..621B}, who constrained the GSMF at redshifts of less than 0.06 using properties of galaxies from the Galaxy And Mass Assembly survey \citep[][]{2011MNRAS.413..971D}. 

Under these assumptions, we write the FRB-galaxy cross-correlation function as: 
\begin{align}
    \xi_b^{fg}(z_g) = \Delta \mathrm{DM}_g p_f(z_g) \frac{\mathcal{P}_\mathrm{FRB\times gal}(z_g)}{2\pi \sigma_b^2}e^{-b^2/2\sigma_b^2}.
\end{align}
Thus, we derive the DM-galaxy cross-correlation function for an ensemble of galaxies at redshift $z_g$ being probed by an FRB at an impact parameter of $b$ to be: 
\begin{align}
    \xi_b^{dg}(z_g) = \frac{f_f(z_g)}{(1+z_g)^2}\int_{-\infty}^{\infty}n_e^\mathrm{CGM} \left(\sqrt{b^2 + l_g^2}\right)dl_g + \Delta \mathrm{DM}_g p_f(z_g) \frac{\mathcal{P}_\mathrm{FRB\times gal}(z_g)}{2\pi \sigma_b^2}e^{-b^2/2\sigma_b^2}. 
\end{align}
We note here that this is only for galaxies at a single redshift $z_g$; averaging over many redshifts we obtain our observable $\Delta \langle\mathrm{DM_{egal}}(b)\rangle$:
\begin{align}
    \langle \mathrm{DM_{egal}}(b)\rangle = \frac{\int d{z_g} \frac{dN_g}{dz_g}\xi^{dg}_b(z_g)}{\int dz_g \frac{dN_g}{dz_g}}, 
\end{align}
where $\frac{dN_g(z_g)}{dz_g}$ is the differential number of galaxies at redshift $z_g$. 

\section{MCMC Constraints on Radial DM Profiles}
\label{app:stellar_mass}
Here, we present the radial profiles of the various stellar mass bins, along with the MCMC constraints on the electron density distributions accompanied with the posteriors. Figure \ref{fig:mcmc_bins} displays the radial profiles for Bins 0, 1 and 3. Figure \ref{fig:posterior_full} displays the posteriors for Galaxy Sample B, and Figure \ref{fig:mcmc_posterior} displays the posterior distributions for all the stellar mass bins. From these posterior distributions, we note that significant degeneracy exists particularly between $r_c^\mathrm{frac}$ and $n_{e,0}$. Particularly, we note that $r_c^\mathrm{frac}$ is at very low CGM radii ($<0.1 r_\mathrm{CGM}$); therefore, given the current collection of FRBs we are unable to meaningfully constrain the core radius. To do so, we would require many FRBs that traverse intervening CGMs at very small impact parameters. 

\begin{figure}\begin{center}
    \includegraphics[width = 0.47\textwidth]{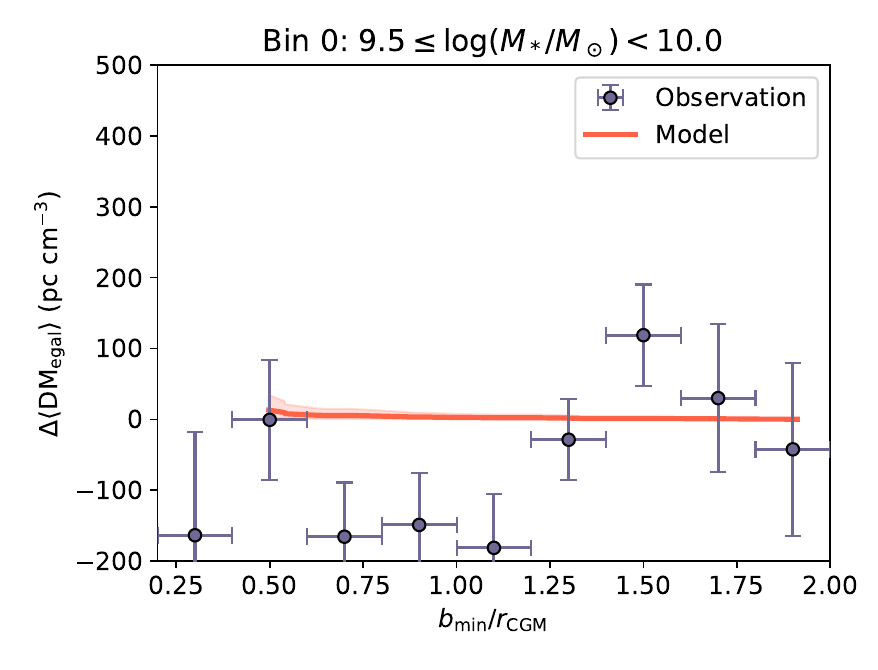}\includegraphics[width = 0.47\textwidth]{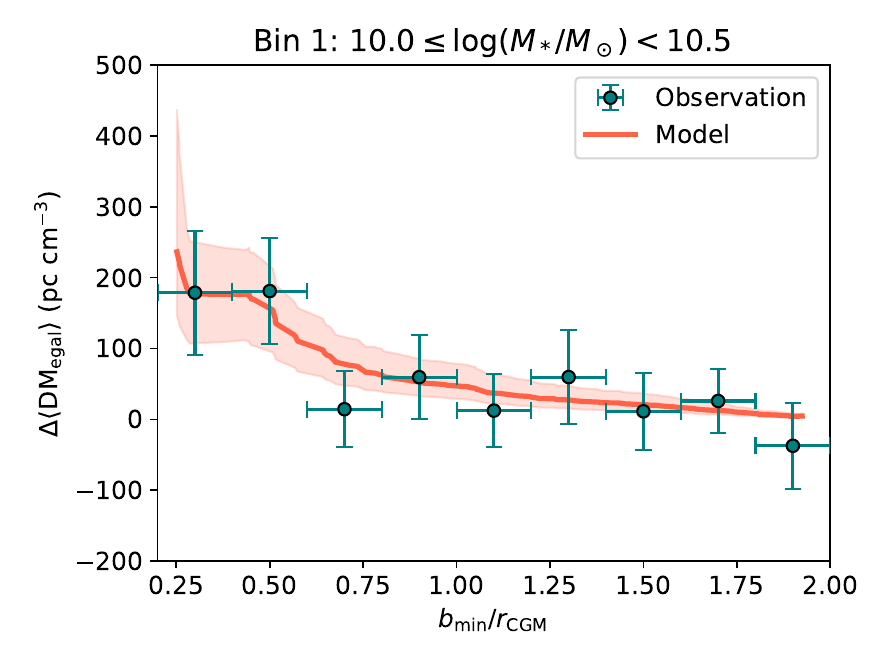}\\
    \includegraphics[width = 0.47\textwidth]{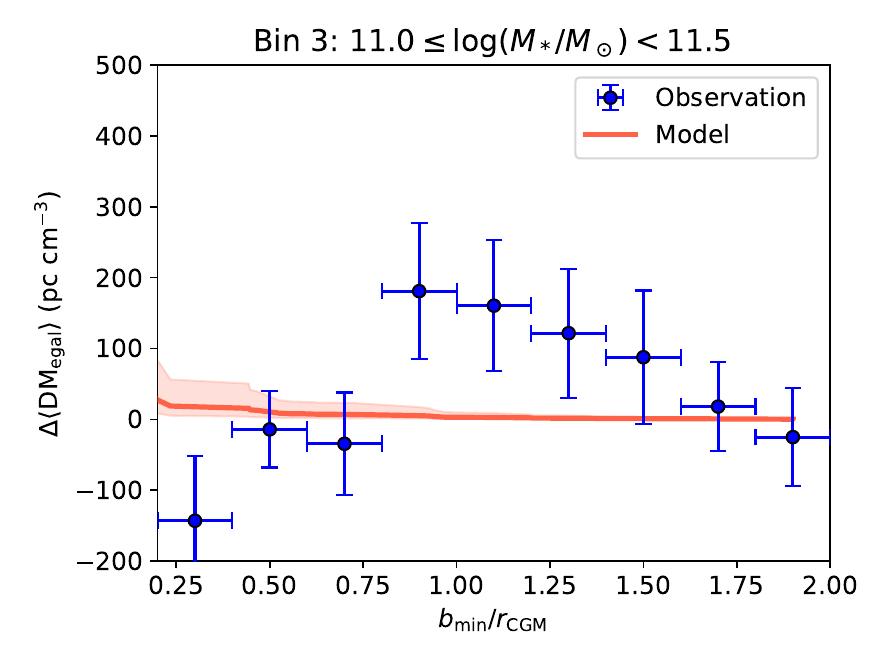}
    \end{center}
    \caption{The The median MCMC model DM profiles (in red) with the shaded region encompassing the 16th and 84th percentiles overlaid on the observed DM profiles for galaxies in the different stellar mass bins.}
    \label{fig:mcmc_bins}
\end{figure}

\begin{figure}
    \centering
    \includegraphics[width=0.47\textwidth]{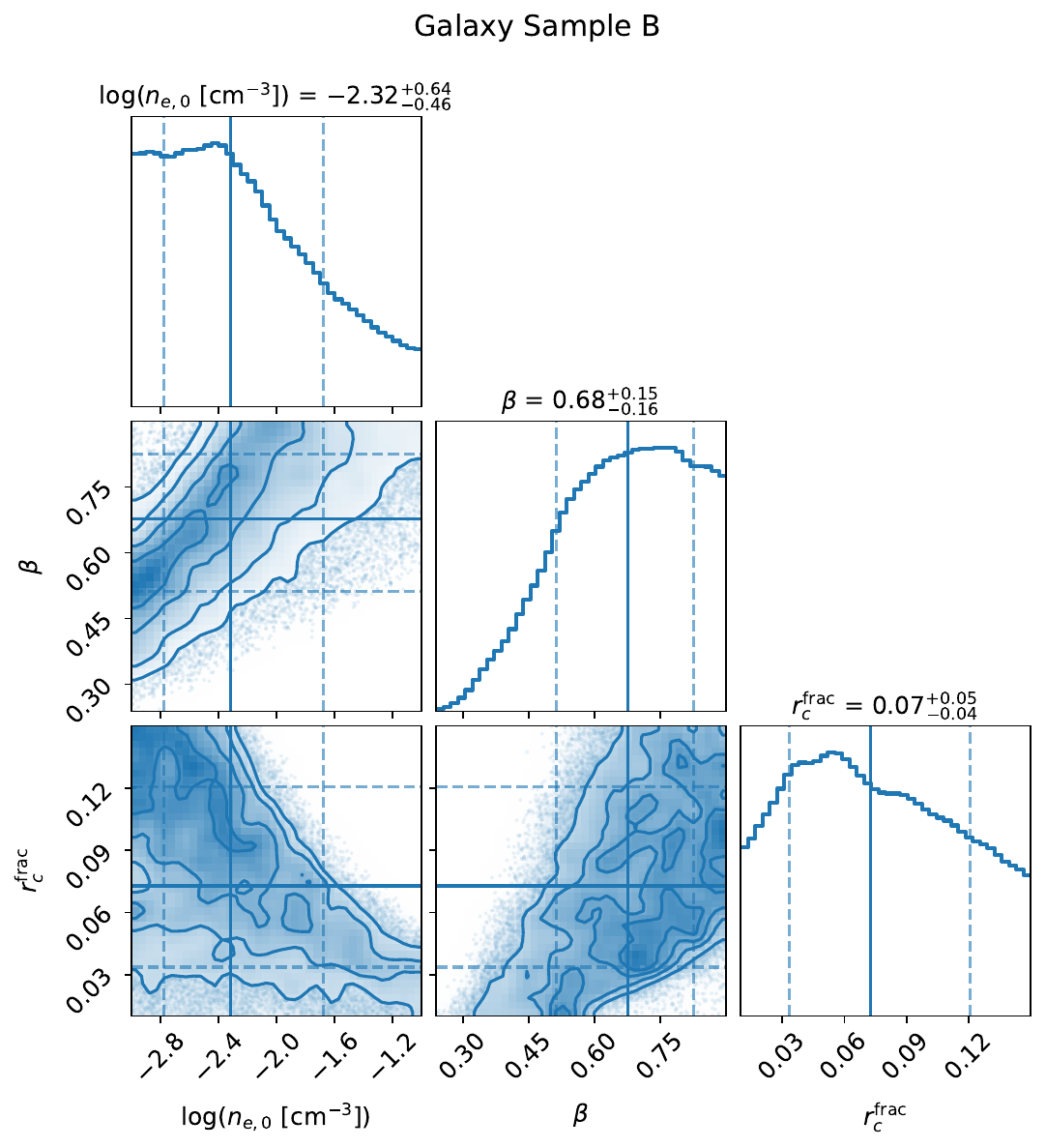}
    \caption{The posteriors from the MCMC for Galaxy Sample B. The contours indicate the $1\sigma$, $2\sigma$, and $3\sigma$ confidence intervals. The solid line indicates the median for each parameter, with the dotted lines indicating the $1\sigma$ regions.}
    \label{fig:posterior_full}
\end{figure}

\begin{figure*}
    \includegraphics[width = 0.47\textwidth]{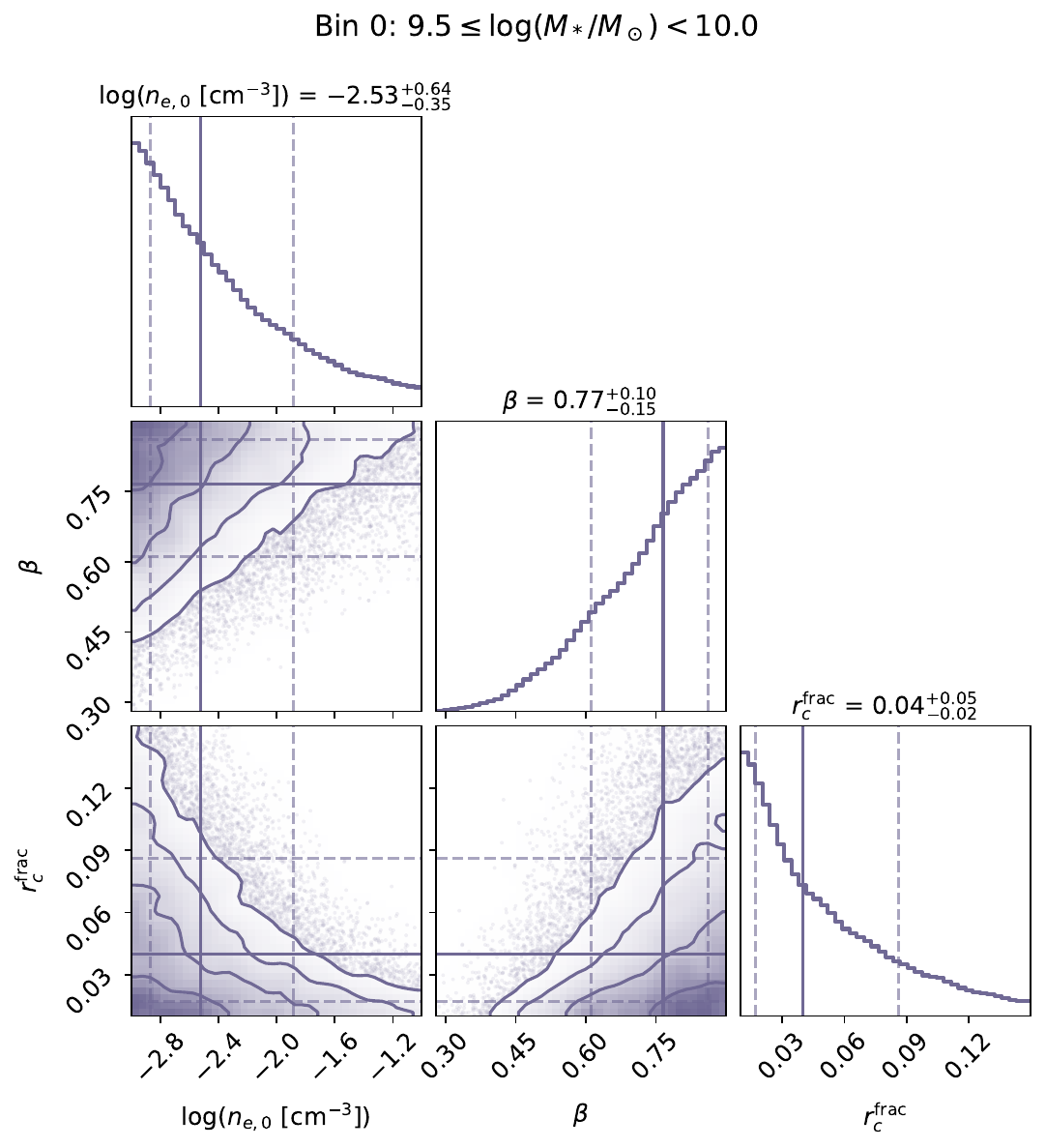}\includegraphics[width = 0.47\textwidth]{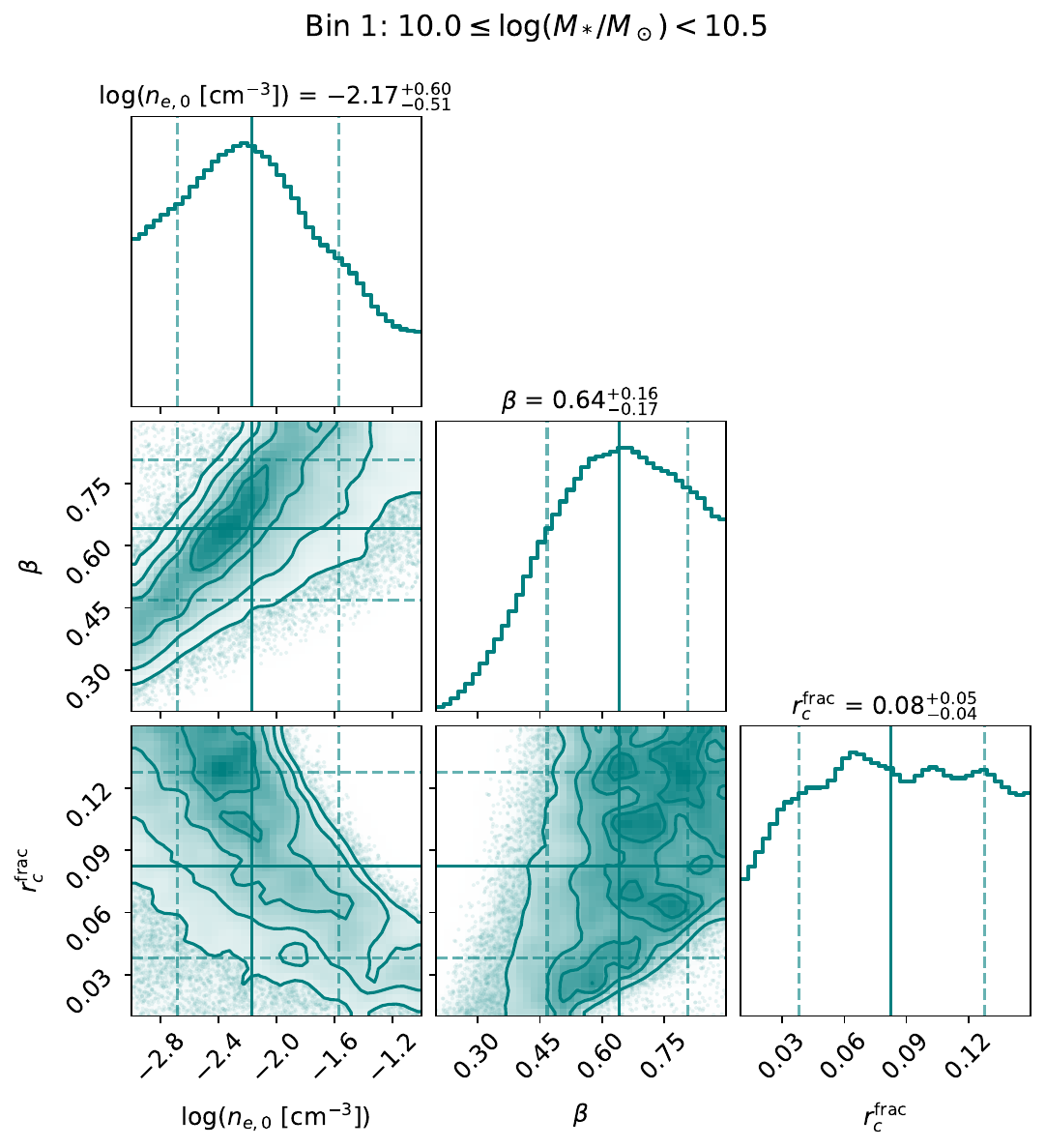}\\
    \includegraphics[width = 0.47\textwidth]{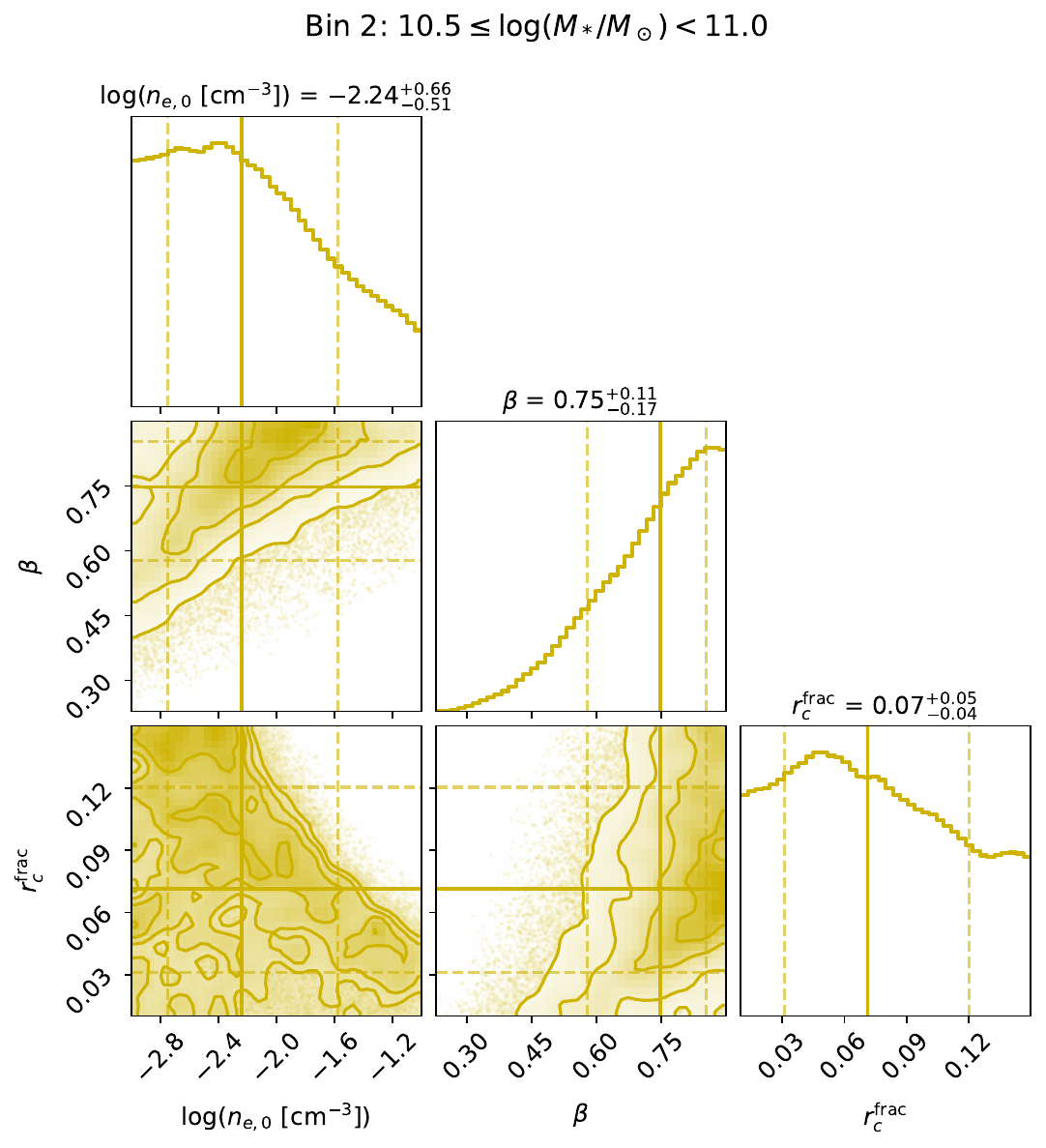}\includegraphics[width = 0.47\textwidth]{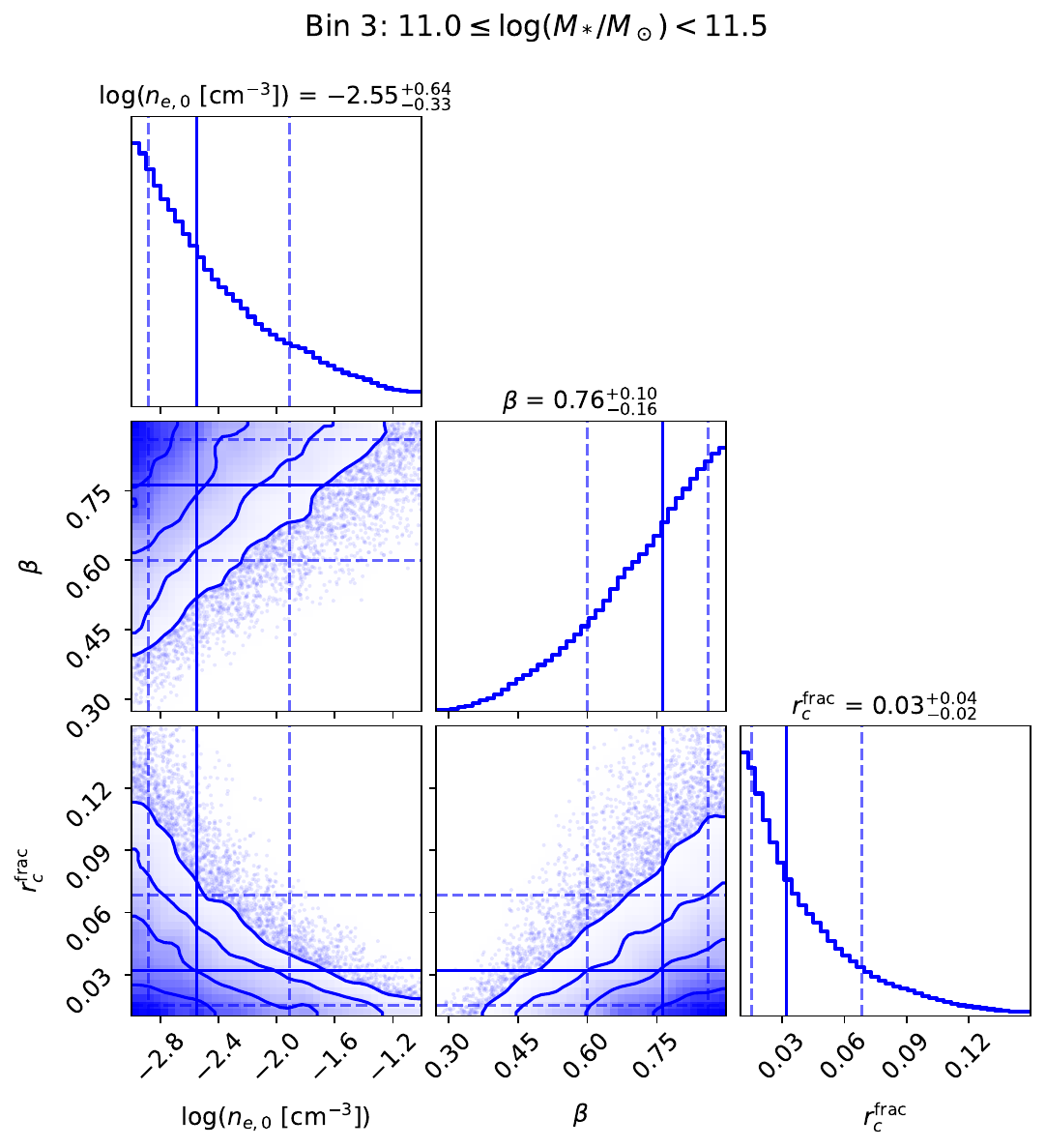}
    
    \caption{The posteriors from the MCMC for galaxies in all the stallar mass bins. The contours indicate the $1\sigma$, $2\sigma$, and $3\sigma$ confidence intervals. The solid line indicates the median for each parameter, with the dotted lines indicating the $1\sigma$ regions.}
    \label{fig:mcmc_posterior}
\end{figure*}

\section{Forecasting a CGM study with 10000 Baseband FRBs}
\label{sec:forecast}

In the future, we expect CHIME/FRB to record baseband data for many more FRBs. Here, we forecast a similar CGM stacking study that can be done using a catalog of $10^4$ FRBs with baseband localizations. We sample the sky positions and localizations of the FRBs throughout the survey footprint of CHIME/FRB with a Gaussian kernel density estimator that uses the sky positions and localizations of BaseCat2 bursts. Then, we simulate the DM contributions of these bursts from the host, the IGM and the LSS using the procedure described in Section \ref{sec:environment} (see also Equation \ref{eq:sim}). We cross-match the FRBs with foreground galaxies from our Galaxy Sample B and also add the CGM contribution to the DM for galaxies along the LOS using Model~1 from the DPM models introduced by \citet[][]{2025MNRAS.543.2649O}.

Using this simulated catalog of 10000 FRBs with baseband localizations, we conduct both of our analyses once again. First, we search for a DM excess toward very nearby galaxies by cross-matching with Galaxy Sample A. In doing so, we find 419 intersections of background FRBs with foreground halos within 1$r_\mathrm{CGM}$ ($\sim 5\times$ greater than our current number of intersections). Furthermore, we conduct the same statistical test as before and obtain a p-value of $<10^{-7}$ for the non-parametric permutation test. Thus, we conclude that with upcoming baseband catalogs, it will be possible to obtain a $\sim5$-$\sigma$ detection of the CGM of nearby galaxies. {We stress that 10000 is around the \textit{minimum} number of baseband FRBs that are needed to obtain a result at the 5-$\sigma$ detection threshold.} 

Furthermore, we also investigate the accuracy with which we can study the electron density profiles of galaxies in Sample B with such a high number of FRBs. For these purposes, we compute the DM profiles for the full stacked sample as well as the binned sample of galaxies. The total number of FRBs that are found to intersect halos in Galaxy Sample B is 3829, which is more than double the total number of unique FRBs that are currently present in BaseCat2. We find a central excess of 120\,pc~cm$^{-3}$, with a radial decline. However, we note that the fraction of bursts that intersect halos ($\sim 0.3829$) is still comparable to that in the observations ($\sim 0.391$); therefore, in general we expect roughly 40\% of all FRBs to be intersecting foreground halos at $D_L < 150$\,Mpc. We find that the p-value of this profile to be $<10^{-5}$.

We display the mean DM profile as a function of impact parameter for our forecasted FRB catalog in Figure~\ref{fig:forecast_fullstack} for Galaxy Sample B. The number of forecasted FRBs cross-matched with the Bins 0 to 3 in ascending order is: 652, 1201, 1366 and 610. The mean DM profiles for the forecasts across our stellar mass bins are displayed in Figure~\ref{fig:forecast_bin}. We also compute the p-values for these forecasts and find them to be: $0.0357, 0.00003, <10^{-5}, 0.00006$ for Bins 0, 1, 2 and 3, respectively. We note that the forecasted DM profile for Bin~3 is the one that is most dissimilar from our observed data; the reason for this is that the CGM model that we incorporate does not take into account any of the feedback mechanisms that might explain the central DM cavity observed in high-mass galaxies (as discussed in Section~\ref{sec:theory}). Furthermore, based on the p-values, we see that even with an incredibly large sample of FRBs, it is challenging to meaningfully study the CGM of the low-mass galaxies in Bin~0 primarily due to their low electron densities and small radii. Obtaining a detection of the CGM for galaxies of this mass using FRBs will likely be even more difficult in practice due to stellar feedback that may further evacuate the CGM. We conclude that FRB DMs from upcoming catalogs are most suited for studying galaxies with $\log(M_*/M_\odot) > 10$, as these have a large enough reservoir of baryons to impart a detectable signature on the FRB DMs.

\begin{figure}[!htb]
    \centering
    \includegraphics[width=0.47\textwidth]{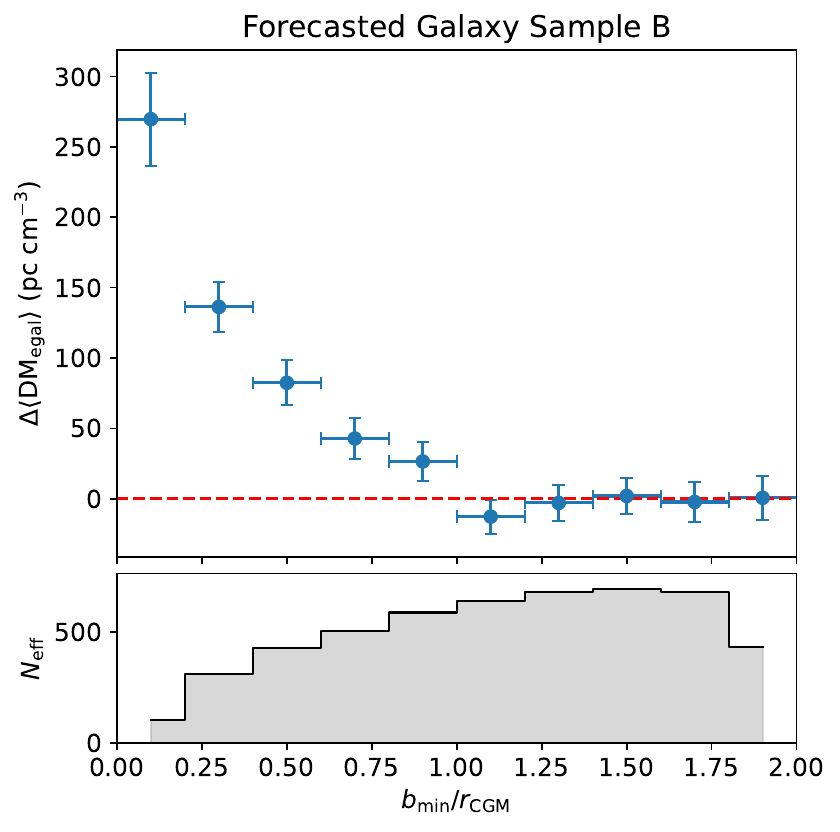}
    \caption{The mean DM profile (indicated with a solid line) as a function of fractional impact parameter for FRB DMs from a forecasted catalog of 10000 baseband bursts cross-matched with galaxies in Sample B. The shaded region indicates the 1-$\sigma$ error region, with the red dotted line displaying the 0 $\mathrm{DM_{egal}}$ threshold. The bottom panel displays the effective number of data points per radial bin.}
    \label{fig:forecast_fullstack}
\end{figure}

\begin{figure*}
    \centering
    \includegraphics[width=0.47\textwidth]{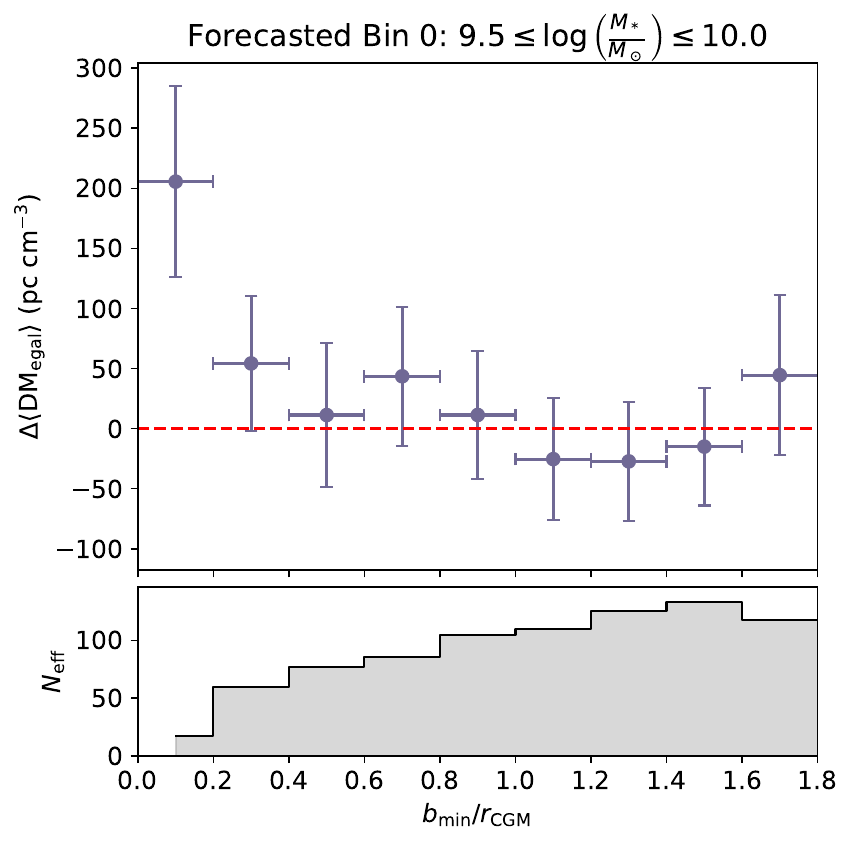}\includegraphics[width = 0.47\textwidth]{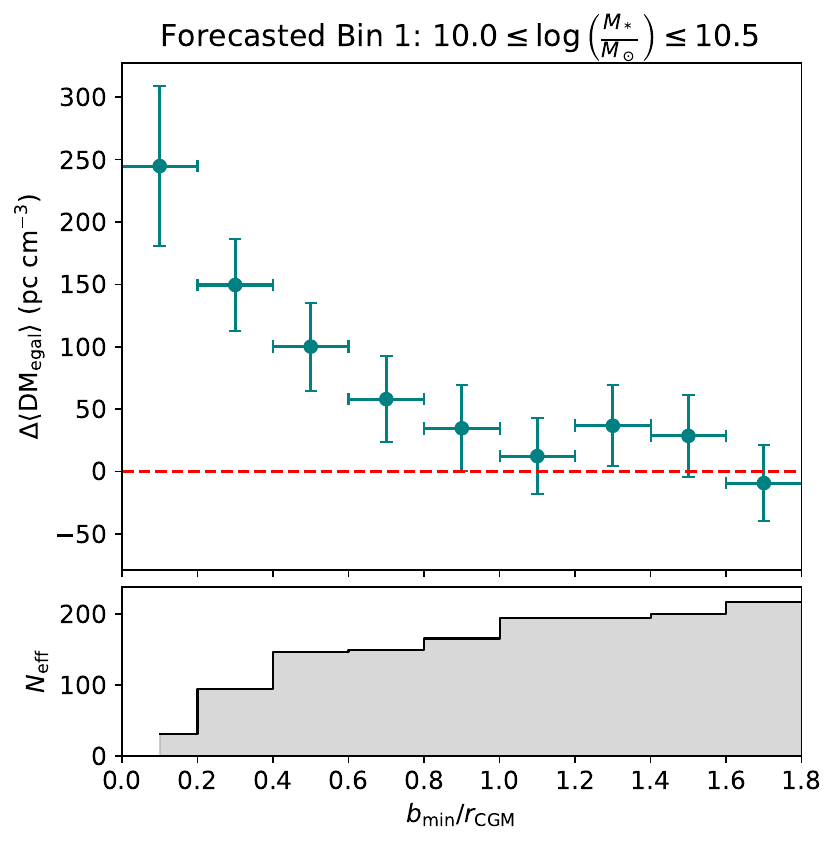}
    \includegraphics[width=0.47\textwidth]{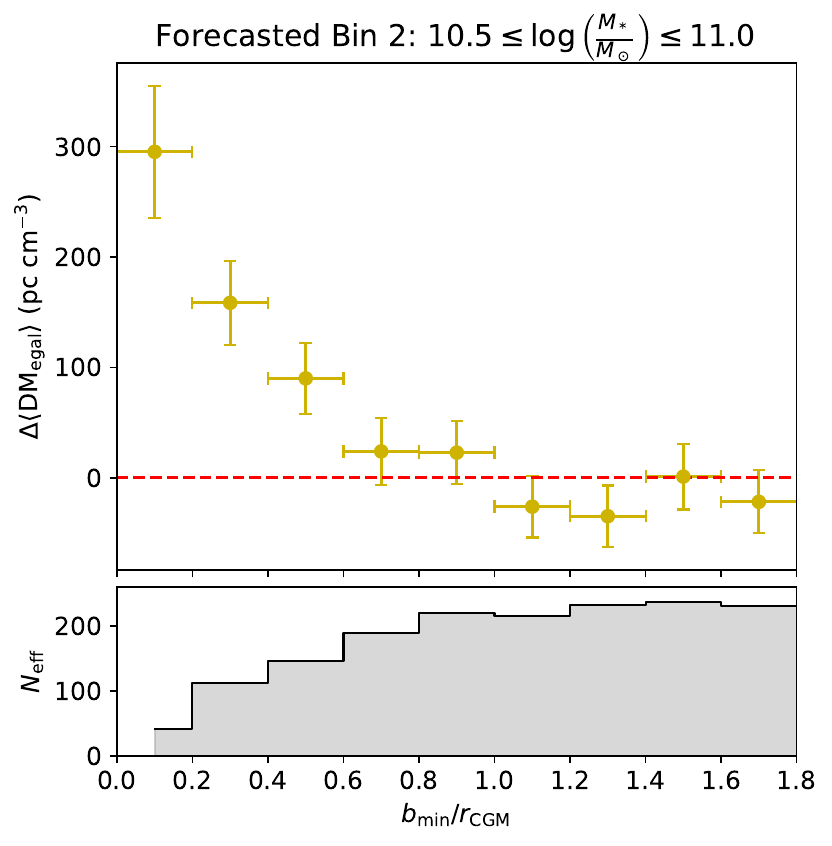}\includegraphics[width = 0.47\textwidth]{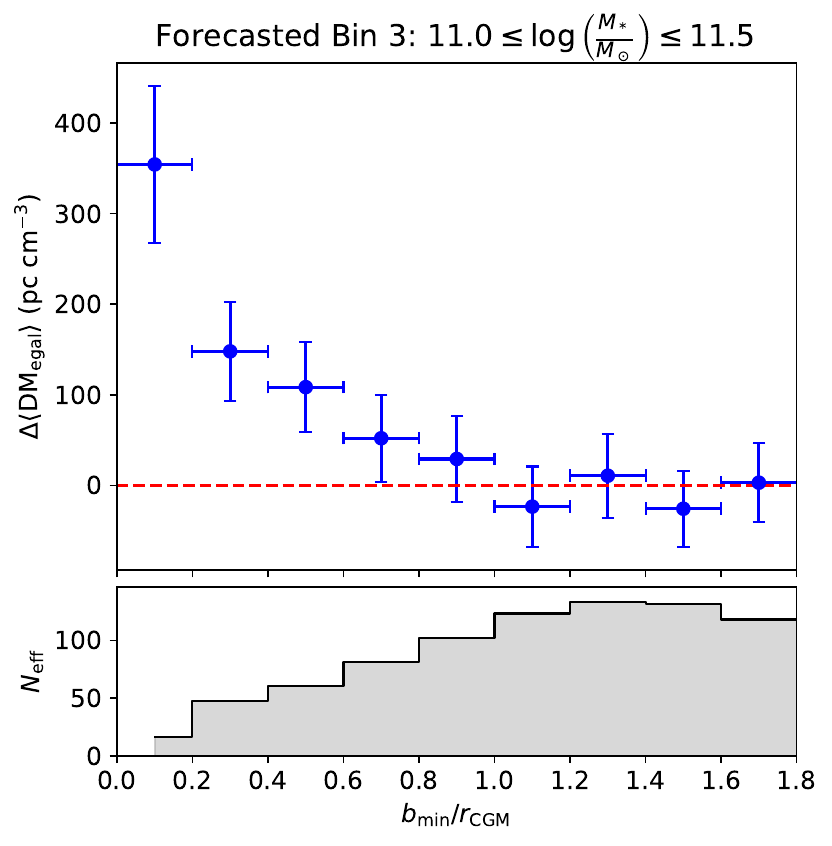}
    \caption{The mean DM profiles (indicated with solid lines) as a function of fractional minimal impact parameter for the forecasted observations with an FRB catalog that contains 10000 baseband bursts.}
    \label{fig:forecast_bin}
\end{figure*}
\newpage
\bibliography{sample701}{}
\bibliographystyle{aasjournalv7}



\end{document}